%% file: SG_dim_def.tex
\pdfoutput=1
\documentclass[twocolumn,prd,floatfix,nofootinbib,longbibliography,notitlepage,superscriptaddress]{revtex4-2}

\usepackage{amsmath}
\usepackage{amssymb}
\usepackage{bm}
\usepackage{subfigure}
\usepackage[usenames,svgnames,dvipsnames]{xcolor}
\usepackage{array}

\newcommand{\Figref}[1]{Fig.~\ref{#1}}
\newcommand{\figref}[1]{Fig.~\ref{#1}}
\newcommand{\Secref}[1]{Section~\ref{#1}}
\newcommand{\secref}[1]{section~\ref{#1}}
\newcommand{\appref}[1]{appendix~\ref{#1}}

\newcommand{\ie}{i.e.,\,}

\newcommand{\ud}{\ensuremath{{\rm d}}}
\newcommand{\pd}{\partial}

\newcommand{\phiScl}{\ensuremath{\phi_\star}}
\newcommand{\breatherSub}{\ensuremath{{\rm B}}}
\newcommand{\phiB}{\ensuremath{\phi_{\breatherSub}}}
\newcommand{\piB}{\ensuremath{\Pi_{\breatherSub}}}
\newcommand{\omegaB}{\ensuremath{\omega_{\rm B}}}

\newcommand{\omegamin}{\ensuremath{\omega_{\mathrm{min}}}}
\newcommand{\omegamax}{\ensuremath{\omega_{\rm max}}}
\newcommand{\Emax}{\ensuremath{E_{\rm max}}}
\newcommand{\Emin}{\ensuremath{E_{\rm min}}}
\newcommand{\omegaini}{\omega_{\mathrm{ini}}}

\newcommand{\omegaosc}{\omega_{\mathrm{osc}}}

\newcommand{\piDim}{\ensuremath{\bar{\Pi}}}

\newcommand{\lenPar}{\ensuremath{\ell}}

\newcommand{\head}[2]{\multicolumn{1}{>{\centering\arraybackslash}p{#1}|}{#2}}

\usepackage{graphicx}
\graphicspath{{../Figures/},{../scripts/}}

\usepackage[T1]{fontenc}
\usepackage[utf8]{inputenc}
\usepackage{lmodern}

\usepackage[unicode, colorlinks, allcolors=blue!70!black, linktocpage, pdfusetitle]{hyperref}
\hypersetup{
    colorlinks=true,
    urlcolor=SteelBlue,
    linkcolor=red,
    citecolor=blue,
}
\usepackage[all]{hypcap}
\usepackage{orcidlink}
\usepackage{microtype}

\newcommand{\CITA}{\affiliation{Canadian Institute for Theoretical Astrophysics, University of Toronto, 
60 St. George Street}}

\newcommand{\UTEC}{\affiliation{Universidad de Ingenieria y Tecnologia - UTEC, 
Jr. Medrano Silva 165, Lima - Per\'u}}

\input{figures.tex}

\begin{document}

\title{Dimensional deformation of sine-Gordon breathers into oscillons}

\author{Jos\'e T.\ \surname{G\'alvez Ghersi}\,\orcidlink{0000-0001-7289-3846}}
\email{jgalvezg@utec.edu.pe}
\CITA
\UTEC

\author{Jonathan Braden,\orcidlink{0000-0001-9338-2658}}
\email{jbraden@cita.utoronto.ca}
\CITA

\date{\today}

% Because hyperref only gets the *last* author, we need to be explicit.
\hypersetup{pdfauthor={G\'alvez Ghersi and Braden}}

\begin{abstract}
Oscillons are localized field configurations oscillating in time with lifetimes orders of magnitude longer than their oscillation period. In this paper, we simulate non-travelling 
oscillons produced by deforming the breather solutions of the sine-Gordon model.
Such a deformation treats the dimensionality of the model as a real parameter to produce spherically symmetric oscillons. After considering the post-transient oscillation 
frequency as a control parameter, we probe the initial parameter space to continuously connect breathers and oscillons at various dimensionalities. For sufficiently small 
dimensional deformations, we find that oscillons can be treated as perturbatively deformed breathers. In $D\gtrsim 2$ spatial dimensions, we observe solutions undergoing intermittent 
phases of contraction and expansion in their cores. Knowing that stable and unstable configurations can be mapped to disjoint regions of the breather parameter space, 
we find that amplitude modulated solutions are located in the middle of both stability regimes. These solutions display the dynamics of critical behavior around the stability limit.  
\end{abstract}
\maketitle

Oscillons~\cite{Bogolyubsky:1976yu}  are a remarkable set of long-lived localized states that oscillate in time, which emerge as solutions of nonlinear field theories. In this context, long-lived 
means that these have lifetimes orders of magnitude longer than their oscillation period. They are held together by a delicate balance between attractive forces and dispersion preventing them 
from dilution or collapse.In early universe cosmology, oscillons may have been produced at the end of inflation~\cite{Amin:2011hj}, leading to a number of 
potentially interesting consequences, including: possible connections to dark matter~\cite{Olle:2019kbo, Cyncynates:2021rtf}, and 
a variety of effects after their interactions with primordial scalar and tensor gravitational modes~\cite{Amin:2018xfe,Cotner:2018vug,Hiramatsu:2020obh, Kou:2021bij}.

Despite a significant body of existing work~\cite{Gleiser:1999tj, Graham:2006xs,Gleiser:2008ty,Saffin:2014yka, Zhang:2020bec, Olle:2020qqy, 
Cyncynates:2021rtf, vanDissel:2023zva}, oscillon longevity is not fully understood.
However, a similar class of objects, known as breathers, exist in the sine-Gordon model.
Breathers can be interpreted as an infinitely long-lived dynamical bound state of a kink-antikink pair.
Visually, they take the form of either a spatially localized field profile that oscillates in time (for the tightly bound case); or a kink-antikink 
pair repeatedly colliding and moving away from each other, before turning around then colliding again (for the weakly bound case).
In particular, in the tightly bound limit the breathers have the same basic structural properties as an oscillon.
However, unlike oscillons, breathers do not decay and have infinite lifetimes~\cite{PhysRevLett.30.1262}.
Given their structural similarity, it is natural to look for a connection between oscillons and breathers.
Even when this is not within the scope of this exploration, infinite lifetime of the breathers may descend from integrability properties of the sine-Gordon 
equations. Such a connection could potentially provide an explanation for the longevity of the oscillons in terms of a weak breaking of integrability.

The primary focus of this paper is to provide an explicit connection between spherically symmetric oscillons and one-dimensional sine-Gordon breathers.
To make the connection most transparent, we primarily focus on the sine-Gordon model in spatial dimensions $D>1$; however, we briefly extend our approach to oscillons in 
monodromy models to illustrate the generality of the approach.
As is well known, these higher-dimensional sine-Gordon theories are no longer integrable, and thus do not posess infinitely long-lived spatially localized soliton solutions.
\footnote{Of course, partially localized planar symmetric breather solutions continue to exist.}
However, at least in low-dimensions ($D=2$ and $D=3$), the sine-Gordon model supports oscillon solutions.
Further, these oscillons tend to be spherically symmetric.
A natural conjecture is that these oscillons descend from the breathers upon breaking the integrability of the one-dimensional sine-Gordon equations.
To make this connection explicit, here we explore spherically symmetric oscillon-like solutions to the sine-Gordon model in dimensions $D\neq 1$.
This implies modifying the one-dimensional sine-Gordon equation by adding a first-order derivative term proportional to $\varepsilon \equiv D-1$.
Moreover, it is important to notice that one-dimensional breathers and oscillons have the same boundary condition at the origin.
We can thus interpret the equation for the radial profile as a deformation of the one-dimensional sine-Gordon equation.
In order to smoothly connect to the one-dimensional breather solutions, we allow $\varepsilon$ (and thus the spatial dimension) to be a real, rather than integer, parameter.
When $\varepsilon \ll 1$, the deformation to the equations of motion is small, and we therefore expect breathers to be approximate solutions 
to the higher-dimensional sine-Gordon equation. 

Motivated by this, we evolve a family of initial radial breather profiles under the $D$-dimensional spherically symmetric sine-Gordon equations for a range of choices of $D$.
As expected, for $\varepsilon \ll 1$, the entire family of initial breather profiles evolve into oscillons (i.e.,\, long-lived spatially-localized oscillating structures).
This provides an explicit link between oscillons in the sine-Gordon model and breathers, as anticipated above.
In this limit, the oscillons are very long-lived and do not decay for the duration of our simulations ($\sim1000$ oscillations).
The use of standard perturbative methods is not sufficient to approximate the oscillons produced when $\varepsilon \gtrsim 1$. Is in this regime where the use 
of the numerical renormalization (NDRG) \cite{GalvezGhersi:2021sxs} may be useful to build renormalized oscillons using a breather-like parameterization. The derivation of 
a semi-analytical formula to predict oscillon lifetimes in $D\neq 1$ dimensions may be possible after combining results from the perturbative and nonperturbative 
regimes. We will leave these tasks for a future project.   

There are a myriad of measurable (and sensible) parameters to determine the dynamical state of the oscillons produced in this way, such as the curvature at the origin, emitted 
energy measured at the tails, average radius, width, damping rate, etc. In this study, we choose (i) the amplitude and (ii) oscillation frequency at the origin, as well as (iii) the 
energy of the solution as diagnostic parameters to provide a reduced description of the solution's state. With respect to the amplitude and oscillation frequency, we explicitly show 
that attractor behavior \cite{Hindmarsh:2006ur, Gleiser:2009ys, Salmi:2012ta} appears as breathers deform into oscillons. The oscillons produced by breather deformation 
have a range of energies and oscillation frequencies. Within that range, we find a relation coupling the oscillation frequency and the energy, which is consistent with what 
is known for the breathers' energy as a function of its frequency in the limit $\varepsilon \ll 1$. The same relation allows us to learn about (a) the existence of oscillons with 
maximum energy/minimum frequency; and (b) signs of a minimum energy/maximum frequency cutoff for $\varepsilon\sim\mathcal{O}(1)$. A continuum of oscillons, bound by 
maximum and minimum energy configurations, collapses to essentially form a single oscillon when $\varepsilon \sim 2$. The collapse of states shows how critical behavior manifests 
in oscillon formation.
 
The connection between breathers and oscillons reveals a preference to form oscillons from breathers with more potential than kinetic energy. As an experiment, we modified the equations of motion to show that 
such a preference is due to the instantaneity of the dimensional transition. The same language of continuous deformations can be used to modify the potential, and possibly extend these results to other 
types of deformations. Concretely, we deform the positive sinusoidal potential into the axion monodromy potential \cite{McAllister:2008hb,Flauger:2009ab,Dong:2010in,Jin:2020vbl}, which is known to support 
oscillons. 

The remainder of this manuscript is organized as follows. In \secref{sec:setup}, we quickly revise the sine-Gordon model, presenting the breather solution 
and its properties. We introduce the dimensionally deformed sine-Gordon model to produce spherically symmetric oscillons using breathers as initial conditions.
In \secref{sec:sols_and_params} we show the spatial structure and evolution of stable and decayed solutions of the dimensionally-deformed sine-Gordon equations. 
We outline our method to measure oscillation frequencies, which we treat as a control parameter to compare breathers with oscillons. We show explicitly the presence of 
attractors in parameter space. In \secref{sec:dim_def}, we sample both the oscillation frequencies and energies of the oscillons produced by a range of dimensional 
deformations. After sampling over a span of 2500 initial breather profiles, we find oscillons undergoing periodic phases of contraction and expansion of their cores. Oscillons 
are well-approximated by breathers in the limit $\varepsilon\ll 1$; but the connection between them is more subtle in $D \gtrsim 2$ spatial dimensions. From the results in \secref{sec:dim_def}, 
We show how the features of the oscillon attractor vary with the dimensionality. The collapse of minimal and maximal energy oscillons to yield a single state leads us to discuss the 
presence of critical behavior in \secref{sec:critical}. In \secref{sec:SG_td}, we show the results of an implementation considering time-dependent dimensional 
transitions. We investigate how different durations affect the oscillation frequency of oscillons, and validate the frequency extraction procedure presented in \secref{sec:sols_and_params}. 
Section~\ref{sec:SG_monodromy} extends our framework to potential deformations by introducing a tunable model to deform the sine-Gordon potential into the axion monodromy potential. 
As the model deforms, solutions accumulate to yield maximum energy/minimum frequency oscillons as in the case of potential deformations. We present in \appref{app:numerical}
the pseudospectral method used to produce stable numerical solutions and to process data.  Finally, in \secref{sec:discussion}, we discuss and conclude.

\section{Defining sine-Gordon breathers and oscillons}\label{sec:setup}

Our goal in this paper is to relate oscillons appearing in relativistic field theories to breathers, which are a special class of solutions of the one-dimensional sine-Gordon equations.
In this section, we present some important background material and describe the framework we will use to establish the connection. 
We first review some relevant properties of the one-dimensional sine-Gordon equation and breathers.
We then present the dynamical equation governing spherically symmetric solutions to the $D$-dimensional sine-Gordon model, including an interpretation of the equation as a deformation from the one-dimensional equation.
This motivates us to lift the one-dimensional breather profiles to $D$-dimensional radial profiles for use as initial conditions, with the expectation that they dynamically relax into an oscillon state.

\subsection{Breathers and the 1D Sine-Gordon Model}\label{subsec:breathers}
The sine-Gordon model is the theory of a relativistic scalar field evolving in a cosine potential
\begin{equation}\label{eqn:sg-potential}
  V_{\rm SG} = \mu^2\phiScl^2\left[1-\cos\left(\frac{\phi}{\phiScl}\right)\right] \, ,
\end{equation}
where $\phiScl$ and $\mu$ are parameters setting the characteristic field and mass scales of the potential.  
It is also convenient to introduce the typical energy density scale $V_0 \equiv \mu^2\phiScl^2$.

In one spatial dimension, the corresponding equations of motion are
\begin{subequations}\label{eqn:sg1d}
\begin{align}
  \frac{\ud\phi}{\ud t} &= \Pi \\
  \frac{\ud\Pi}{\ud t}  &= \frac{\partial^2\phi}{\partial x^2} - \mu^2\phiScl\sin\left(\frac{\phi}{\phiScl}\right) \, .
\end{align}     
\end{subequations}      
To fix terminology, we refer to these equations as either the one-dimensional sine-Gordon equations or the one-dimensional sine-Gordon model.
The one-dimensional sine-Gordon equation posesses a number of very special and closely related properties: integrability, the existence of an infinite hierarchy of conserved charges~\cite{Kasuya:2002zs,Ferreira:2010gh,Ferreira:2013nda,Blas:2016bvt}, and exact solutions via an inverse scattering transform.

For our purposes, the most important property is the existence of a family of spatially localized solutions with exact temporal periodicity---the breathers.
In particular, breathers have an infinite lifetime, which is intimately tied to the integrability of~\eqref{eqn:sg1d}.
It is convenient to parametrize a breather located at the origin by its frequency $\omegaB$ and initial phase $\theta_0$

\begin{subequations} \label{eqn:breather-profile}
\begin{align}
 &\mathcal{R}(x) = \frac{\sqrt{\mu^2-\omegaB^2}}{\omegaB}\mathrm{sech}\left(\sqrt{\mu^2-\omegaB^2}x\right) \\
 &\Psi(x,t) = \mathcal{R}(x)\cos\left(\omegaB t - \theta_0\right)\\
 &\frac{\phiB}{\phiScl} = 4\tan^{-1}\left(\Psi\right) \\
 &\frac{\piB}{\mu\phiScl} = -4\frac{\omegaB}{\mu}\frac{\mathcal{R}}{1+\Psi^2}\sin\left(\omegaB t - \theta_0\right) \, .
\end{align} 
\end{subequations}
We must have $\omegaB < \mu$.
This reflects the intuitive fact that the breather is a bound state of a kink-antikink (${\rm K\bar{K}}$) pair, and should have frequency less than that of a freely propagating wave.

\brprof

\Figref{fig:breather-profiles} illustrates the breather profiles for a few values of $\omegaB$.
For reasons that will become clear momentarily, we plot the profiles in terms of the one-dimensional radius $r = \left| x \right|$.
Since each breather solution has even symmetry about the origin, no information is lost in this change of coordinates.
There are two distinct asymptotic regimes for breathers.
When $\omegaB \sim 1$, the ${\rm K\bar{K}}$ pair are tightly bound, and the breather takes the form of a localized field configuration undergoing nearly harmonic oscillations.
Meanwhile, when $\omegaB \ll 1$, the breathers represent a very weakly bound ${\rm K\bar{K}}$ pair undergoing repeated collisions.
Between collisions, the kink and the antikink become well-separated from each other.
Since these weakly bound breathers do not resemble oscillons, they are not of direct interest to us here.

The oscillation frequency (i.e.,\,inverse period) also fixes other structural properties of the breather, including the peak amplitude of oscillation at the origin. Having the analytic 
solution given by Eq.~\eqref{eqn:breather-profile}, we can compute the peak amplitude at the origin $(\mathcal{A})$  
\begin{equation}\label{eqn:breather-amp}
  \frac{\mathcal{A}}{\phiScl} = 4\tan^{-1}\left(\frac{\mu}{\omegaB}\sqrt{1-\frac{\omegaB^2}{\mu^2}}\right) \, ,
\end{equation}
the energy
\begin{equation}\label{eqn:breather-energy}
  \frac{E_{\rm B}}{\mu\phiScl^2} = 16\sqrt{1-\frac{\omegaB^2}{\mu^2}} \, ,
\end{equation}
and the damping rate of the amplitude envelope
\begin{equation}\label{eqn:breather-large-r}
  \lim_{\mu r\to\infty}\frac{1}{\mu r}\ln\left(\frac{\mathcal{A}}{\phiScl}\right) \sim \sqrt{1-\frac{\omegaB^2}{\mu^2}} \, .
\end{equation}  

\subsection{Dimensional Deformations: The Radial Sine-Gordon Equation}
We now consider the sine-Gordon model in $D>1$ dimensions.
An interesting type of localized object, called an oscillon, has been observed in this model for $D=2$ and $D=3$~\cite{Fodor:2009kf,Fodor:2019ftc}.
In addition to the sine-Gordon model, they have been observed in a wide variety of nonlinear field theories.
As mentioned previously, oscillons are spatially localized structures that oscillate in time.
Oscillons thus share several key structural features with breathers.
However, unlike breathers, oscillons have a finite lifetime.
Although, there are cases where lifetimes can be so long that it is hard to be precise about the exact instant where these decay.
Most oscillons dynamically relax to a spherically symmetric state.
It is therefore sufficient to consider their radial profiles, which we will do here.

Restricting to spherically symmetric solutions, the radial profile in $D$ dimensions satisfies
\begin{subequations}\label{eqn:sg-deformed}
\begin{align}
  \frac{\ud\phi}{\ud t} &= \Pi \\
  \frac{\ud\Pi}{\ud t} &= \left[\frac{\partial^2}{\partial r^2} + \frac{\varepsilon}{r}\frac{\partial}{\partial r} \right]\phi - \mu^2\phiScl\sin\left(\frac{\phi}{\phiScl}\right) \, ,
\end{align}     
\end{subequations}
where we have introduced $\varepsilon \equiv D-1$. Even when the study of shape asymmetries in oscillons \cite{Adib:2002ff, Wang:2022rhk} is a valid extension of our work, we leave the 
perturbative treatment of eccentricity for future research. To distinguish it from the one-dimensional ($\varepsilon=0$) case, we will refer to~\eqref{eqn:sg-deformed} as either the 
dimensionally-deformed sine-Gordon equation, or the $D$-dimensional sine-Gordon model. Comparing to~\eqref{eqn:sg1d}, we interpret the term proportional to $\varepsilon$ as a perturbation to the 
one-dimensional sine-Gordon equation. Therefore, to smoothly connect to~\eqref{eqn:sg1d}, we will take $\varepsilon$ to be a positive real parameter, rather than restricting to integer dimensions.
This provides a tunable way to control the breaking of key properties of the one-dimensional sine-Gordon equation, such as integrability and the presence 
of an infinite tower of conserved charges.

With this view of~\eqref{eqn:sg-deformed} as a deformation away from the one-dimensional sine-Gordon equation,
we want to understand the fate of the breathers for $\varepsilon > 0$.
Motivated by this, we will consider initial conditions
\begin{subequations}
\begin{align}
  \phi(r,t=0) &= \phiB(r,t=0|\omegaB=\omegaini) \\
  \Pi(r,t=0) &= \piB(r,t=0|\omegaB=\omegaini) \, ,
\end{align}
\end{subequations}
where we defined $\omegaini$ to be value of the breather frequency $\omegaB$ used in the initial condition profile. 
Since the breather solutions~\eqref{eqn:breather-profile} have even symmetry about the origin, the corresponding $\varepsilon+1$-dimensional profiles do not have any singularities as $\mu r\to 0$.
With $\varepsilon = 0$, we obtain breathers as the solutions to the differential equation.
For $\varepsilon \ll 1$, we expect that dynamical evolution will result in a field configuration that is similar to a breather.
In particular, for $\omegaini\sim 1$, we expect to obtain spatially localized solutions that oscillate in time.
However, setting $\varepsilon \neq 0$ breaks the integrability of the original one-dimensional sine-Gordon model, and we expect that the resulting solutions will have a 
finite (although long) lifetime. In other words, we expect to obtain spherically symmetric oscillon solutions. Setting the initial oscillon frequency $\omegaini$ to be the 
breather frequency $\omegaB$ means a major simplification when studying the system, since it is well-known that this parameter is sufficient to fix all the properties of the 
initial profile. This also implies that the evolution of the oscillation frequency provides (at least partial) knowledge of the other features of the solutions. To set our conventions, 
we will refer to the stable, localized solutions of Eqns.~\eqref{eqn:sg-deformed} as spherically symmetric oscillons in $D\neq 1$ dimensions, obtained after the deformation 
of sine-Gordon breathers.   

Once the object of study has been defined, we can obtain some analytic insight into the deformation of the solutions at large radius.
Assume that the solution takes the form
\begin{equation}
 \phi \approx A(r)\cos(\omega t + \Theta) \, ,
\end{equation}
and consider the limit $r \to \infty$.
Since we are interested in localized solutions, we require $A \ll 1$ as $\mu r \to \infty$, so that
\begin{align}\label{eqn:deformed-large-r}
 A(r) &\sim r^{-\varepsilon/2}{\rm exp}\left(-r\sqrt{\mu^2-\omega^2}\right) \,\\
 &\sim {\rm exp}\left(-r\sqrt{\mu^2-\omega^2}\right)\left[1-\frac{\varepsilon}{2}{\rm ln} r+\mathcal{O}(\varepsilon^2)\right]\nonumber\, ,
\end{align}
in this limit. Assuming that a long-lived solution with this frequency $\omega$ exists, we see that the $\varepsilon$ deformation induces 
a corresponding deformation to the $r \to \infty$ asymptotic of the breather with the same oscillation frequency given in~\eqref{eqn:breather-large-r}.
Once again, we see that only states with $\omega < \mu$ describe localized solutions. 

Detailed understanding of the ultimate fate of the breather initial conditions under the dimensionally deformed sine-Gordon equation requires numerical solutions.
This includes determining the values of $\omega$ for which long-lived solutions exist, which is not captured by the asymptotic estimate above.
We make use of a 8th-order Gauss-Legendre method for the time-evolution.
Oscillons evolving for long time intervals require resolving propagating radiative modes towards large radii. This requires an enormous amount of 
grid points, which make the computational task infeasible. The addition of perfectly matched layers (PMLs) allows us to only require sufficient resolution 
inside the boundary layers. Details of the setup and appropriate dimensionless units are presented in \appref{app:numerical}.

\solsnfreq

In what remains of this paper, we will use the setup described to understand how breather solutions are modified as we deform away from the one-dimensional sine-Gordon equation.  We will primarily focus on the dimensional deformation outlined in this section.  While our main focus will be on time-independent deformation parameter $\varepsilon$, we briefly consider time-dependent $\varepsilon$ in~\secref{sec:SG_td}.  Finally, to show the generality of our results, we briefly extend our approach to potential deformations in~\secref{sec:SG_monodromy}.

\section{Anatomy of the deformed solutions and diagnostic parameters}
\label{sec:sols_and_params}
Ultimately, we want to understand the fate of initial breather profiles as the initial condition parameters $\omegaini$ and $\theta_0$ are varied.
Additionally, we want to understand this dependence as we adjust the deformation parameter $\varepsilon$.
Efficiently comparing solutions in these scans requires us to encode the properties of the resulting solutions (oscillons or otherwise) in a few key parameters.
This is analogous to the encoding of the breather properties in the single parameter $\omegaB$.
To set the stage for an initial condition scan, in this section we first look at the detailed evolution from a few fiducial choices of $\omegaini$, $\theta_0$, and $\varepsilon$.
As expected, we find oscillons that form from the breather initial conditions.
We also introduce a convenient set of reduced parameters which we will use to describe the resulting evolution.

\figparam

In the left two panels of \figref{fig:sols_n_freq}, we illustrate two prototypical field evolutions starting from breather initial conditions.
The left panel shows an initial condition that settles down into an oscillating long-lived spatially localized state---an oscillon.
A more detailed look at the spatiotemporal structure of the solution reveals small dissipative effects (i.e.,\, changes to the core and tails of the radial profile) associated with the emission of classical radation.
Meanwhile, in the center panel we see an example where the field quickly decays and no oscillon is formed.
In order to set nomenclature for the remainder of the paper, we will refer to these as oscillons and decayed solutions, respectively.

Since we want to connect the properties of the oscillons to properties of the breathers, it is convenient to parametrize the oscillons in terms of a few key structural properties.
There are a plethora of reasonable quantities we could choose, such as: the width of peak, the damping rate at infinity, oscillation frequencies, energy-weighted average radius, 
curvature at the center, etc. As we will be scanning over initial conditions, we want quantities that can be robustly measured using automated procedures.
With this in mind, we now discuss the reduced parameters we will use to describe the field solutions.
We focus on quantities that are useful to describe the oscillons, rather than decayed solutions.
This parametrization is not meant to be a ``complete'' description of the oscillon dynamics,
but rather a convenient reduction of the dimensionality of the configuration space.

One simplification is to consider the evolution at a single point, with a convenient choice being the origin $r=0$.
The top right panel of \figref{fig:sols_n_freq} shows the corresponding evolution at $r=0$ for the oscillon illustrated in the left panel.
We see that the evolution is characterized by a damped oscillation
\begin{equation}
  \phi(r=0,t) \approx \mathcal{A}(t)\sin(\omegaosc t+\varphi_0) \,,
\end{equation}
where $\varphi_0$ is an arbitrary initial phase.
From the left panel of~\figref{fig:sols_n_freq}, we see that a similar decomposition with the same $\omegaosc$ holds for other radii near the core of the oscillon.
The existence of a single envelope function $\mathcal{A}$ (rather than separate functions describing the upper and lower envelopes) is consistent with the even 
symmetry of the potential.
This general behavior is quite common, although we will find interesting oscillon-like solutions where the profile $\mathcal{A}$ develops an additional low frequency 
modulation $\omega_{\rm mod} < \omega_{\rm osc}$.
Therefore, rather than consider the full time-stream, we further compress the information into a (time-dependent) amplitude $\mathcal{A}$ and oscillation frequency $\omega_{\rm osc}$.
A nice benefit of this approach is that we are directly using the oscillation frequency of the breather as a parameter in our initial conditions.
Finally, we empirically observe a slowing of the parameter flow once the solutions reach the attractor, such as the logarithmic dependence on time shown in~\figref{fig:param_flow}.
This will motivate an approximate treatment of the parameters as constant in future sections.

We now outline our method to extract $\mathcal{A}$ and $\omegaosc$ from simulation data.
The peak amplitude $\mathcal{A}$ is extracted directly from the time-stream.
In order to have reasonable resolution of the temporal peak locations, we choose the output timestep $dt_{\rm out}$ to sample around $20$ points per oscillation.
The peaks in the sampled timestream are then tagged using the \texttt{find\_peaks} function of \texttt{scipy.signal}~\cite{2020SciPy-NMeth}.
In most cases, we then use a cubic spline fit using \texttt{UnivariateSpline} in \texttt{Scipy}.
The exception to this is when the initial transient phase there is a rapid amplitude change.
In this case, we instead use a 10th order polynomial fit based on \texttt{polyfit} in \texttt{NumPy}, which provides a better global fit.
The late time evolution of the amplitude is insensitive to these two choices.
As a test of robustness, we repeated the above procedure retaining only a subset of the peaks and found the amplitude flow was insensitive to the details of this subsetting procedure, 
as long as the peaks continued to sample the full timestream. An example amplitude fit is shown in the top right panel of \figref{fig:sols_n_freq}.

For the oscillation frequency $\omegaosc$, it is more convenient to work in Fourier space
\begin{equation}
   \tilde{\phi}(\omega) = \sum_{t_i} e^{i\omega t_i}\phi(r=0,t_i)
\end{equation}  
and then compute the power spectral density
\begin{equation}
  \mathcal{P}_\omega \equiv \left|\tilde\phi(\omega)\right|^2 + \left|\tilde\phi(-\omega)\right|^2 
\end{equation}
as a function of temporal frequency $\omega$.
We then identify $\omegaosc$ as the frequency for which $\mathcal{P}_\omega$ has maximal power
\begin{equation}\label{eq:omegaosc_def}
  \omegaosc = \underset{\omega\in[0,\mu]}{\rm argmax}\left(\mathcal{P}_\omega\right) \, ,
\end{equation}
here the constraint $\omega \leq \mu$ restricts us to consideration of oscillations associated with a bound state.
An illustration of $\mathcal{P}_\omega$ and the extracted $\omegaosc$ is shown in the bottom right panel of~\figref{fig:sols_n_freq}.
The low frequency power below $\omegaosc$ is primarily due to the non-periodicty of the signal and implicit windowing effects.
Higher-order spectral peaks may also appear but they tend to be subdominant
To study the time-dependence of $\omegaosc$, we instead compute the short time Fourier transform with \texttt{signal.stft} from \texttt{Scipy}, using the default smoothed Hann 
window to smooth. The window size is chosen to capture around $80$ oscillations of the field, yielding a frequency resolution of $\mathcal{O}(1\%)$.
We then determine $\omegaosc(t)$ separately for each of the windowed transforms.

\Figref{fig:param_flow} shows the evolution of $\mathcal{A}$ and $\omegaosc$ for four choices of $\omegaini$ with $\varepsilon = 0.5$.
From the left two panels, we see that both parameters rapidly settle onto an attractor solution.
Further, once they reach the attractor the parameters evolve very slowly.
These observations will be important in the next section.
The attractor is further illustrated in the right panel, where we show the parameter flow in the $(\mathcal{A},\omegaosc)$ plane.
For comparison, we also include the corresponding relationship for the breather solutions ($\varepsilon = 0$)
\begin{equation}
 \frac{\omegaosc}{\mu} = \cos\left(\frac{\mathcal{A}}{4\phiScl}\right) \, .
\end{equation}
We note that (at least for these parameters) the oscillon parameter flow is approximately aligned with the breather relationship, although there is of course no flow of these 
parameters when $\varepsilon = 0$. The existence of oscillon attractors is consistent with existing intuition in the literature~\cite{Hindmarsh:2006ur, Gleiser:2009ys}. Although it 
is not shown here, we observe that not all of the solutions converge towards the oscillon attractors: as the modulation frequency $\omega_{\rm mod}$ of amplitude modulated 
solutions starts to reduce, parameter flows branch off the attractors. 

\ctfreqphase

While studying the time-dependence of the solution at the origin is extremely useful, there is additional information stored in the full radial profile.
There are again many parameters that one could extract, but here we will focus on the energy of the field configuration.
Ideally, we would separate the bound oscillon component of the field from the propagating radiation.
Unfortunately, since the oscillon solutions tend to continuously radiate energy, this separation is somewhat poorly defined.
However, while the oscillon profile remains localized near the origin, the radiation propagates to large radii, where it is damped away by our absorbing layer.
Since the oscillons are slowly radiating, we therefore take the energy in our simulation volume as a proxy for the energy of the oscillon.
Given the (pointwise) energy density
\begin{align}\label{eq:rho}
  \rho(r,t) &= \frac{1}{2}\left(\frac{\pd\phi}{\pd t}\right)^2+\frac{1}{2}\left(\frac{\pd\phi}{\pd r}\right)^2 \notag \\
  &+ \mu^2\phiScl^2\left(1-\cos\left(\frac{\phi}{\phiScl}\right)\right) \, , 
\end{align}
we can compute the total energy of our $D=\varepsilon + 1$ dimensional solutions
\begin{equation}\label{eq:energy_def}
  E = \frac{2\pi^{\frac{\varepsilon+1}{2}}}{\Gamma\left[\frac{\varepsilon+1}{2}\right]}
\int_0^{R_{\mathrm{max}}} \ud r\, r^{\varepsilon} \rho(r,t)\, .
\end{equation}
As explained in~\appref{app:numerical}, we compute this integral using numerical quadrature, and since our basis functions live on the semi-infinite interval we have 
$R_{\rm max} = \infty$.

\section{Generation of oscillons from dimensional deformations}
\label{sec:dim_def}

\diagnmod

The previous sections showed that oscillons can form from breather initial conditions in the dimensionally-deformed sine-Gordon model, while also  demonstrating the existence of an attractor solution in field configuration space.
In this section, we will explore how oscillon properties change as we scan over the parameters (encoded in $\omegaini$ and $\theta_0$) of the initial breather profiles.
This provides an explicit connection between the breathers of the one-dimensional sine-Gordon model, and the oscillons of the higher dimensional sine-Gordon model.
For $\varepsilon \ll 1$, the dimensionally deformed sine-Gordon equation represents a small perturbation on the one-dimensional version, and we expect that the resulting 
oscillons properties will closely resemble the breathers. Of course, as we increase $\varepsilon$, we expect that the oscillons (if they form) may deviate significantly from 
the initial breather solutions. With this in mind, we divide our results into the $\varepsilon \lesssim 1$ and $\varepsilon \gtrsim 1$ cases, which we refer to as the perturbative 
and nonperturbative regimes, respectively. Since the gradual increase in the dimensionality is important in our discussions, there will be instances 
(in our figures) where we combine results from both regimes.\footnote{It is not simple to define a ``clean cut'' between the perturbative and the nonperturbative regimes. 
Our approach is, instead, to show the distinctive features of the parameter flows in each case.}

\subsection{Case $\varepsilon\lesssim 1$: Oscillons from perturbative deformations}\label{subsec:pert_dim_def}

We now make an explicit connection between breather solutions of the one-dimensional sine-Gordon model~\eqref{eqn:sg1d} and oscillon solutions of the dimensionally 
deformed sine-Gordon model~\eqref{eqn:sg-deformed}.
In this subsection, we focus on the case of small deformation parameter $\varepsilon \lesssim 1$ and explore the impact of progressive growth in the dimensionality.
In particular, we investigate the oscillon frequency $\omegaosc$ and energy $E$, which were introduced in \Secref{sec:sols_and_params}. Before proceeding, let us remark
a consequence of the results shown in the previous section: strictly speaking, both the energy and the oscillon frequency flow with time. However, \figref{fig:param_flow} shows 
that once the solution reaches the attractor line, the reduced parameters $\mathcal{A}$ and $\omegaosc$ evolve slowly. Therefore, for the purposes of comparing a broad range 
of initial conditions, it is reasonable to approximate $\omegaosc$ and $E$ as time-independent, which we will do throughout this section.

First we study the oscillation frequency $\omegaosc$ as we vary the parameters of the initial breather profile.
We uniformly sample $\log_{\rm 10}\omegaini/\mu \in [-1;-0.02]$ and $\theta_0 \in [0,\pi]$.
The lower bound of $\omegaini$ ensures that the initial profiles have a localized peak at the origin, as illustrated in~\figref{fig:breather-profiles}.
Meanwhile, the upper bound is driven by numerical considerations, since solutions with slowly damping profiles are difficult to resolve numerically.
Throughout this subsection, we use a total integration time of $\mu T = 10^{4}$, which allows for a few thousand oscillations of the field at the origin in cases where an oscillon forms.
The corresponding frequency resolution is $\Delta\omegaosc / \omegaosc \sim N_{\rm osc}^{-1} \sim 10^{-3}$, where $N_{\rm osc}$ is number of field oscillations during the integration.

The resulting oscillation frequencies $\omegaosc$ are illustrated in \figref{fig:ct_eps_freq} for four choices of $\varepsilon$.
The color pallete represents the oscillon frequency $(\omegaosc)$ span, ranging from the lowest oscillation frequency (visible as wide plateaux in the maps) in ivory, while its variations 
colored up to red brick correspond to higher frequencies. Precise values of what is meant by lower and higher frequencies depend on the specific value of $\varepsilon$.
Regions yielding unstable solutions are colored in gray and labeled with the caption ``unstable'' and have $\omegaosc=\mu$.
In all cases, we see the emergence of a large ``plateau'' of oscillons (shown in ivory) with nearly identical frequencies $\omegaosc$.
We further illustrate this plateau in the inset figure of the top right panel.
This is consistent with the existence of an attractor line, as seen in~\figref{fig:param_flow}.
Further, it suggests that the attractor has an ``origin point'' that acts as a quasi-fixed point where many initial conditions rapidly accumulate during a transient phase, followed 
by a subsequent slow evolution along the remainder of the attractor line.
For this choice of initial condition parameters, the plateau boundary has nontrivial structure, which also extends to the $\omegaosc$ isocontours more generally.
We will discuss the physical origin of this structure below.

For the three panels with $\varepsilon \leq 0.5$, all of our breather initial conditions settle into long-lived solutions.
This coincides with the (perturbative) intuition outlined above that oscillon and breather solutions should be closely related for the case $\varepsilon \ll 1$.
This is consistent with perturbative (in the amplitude of the oscillations at the origin) treatments of oscillon dynamics~\cite{Fodor:2008es,Fodor:2019ftc, Levkov:2022egq}, 
which find a continuum of solutions with arbitrarily small amplitude and corresponding oscillation frequencies arbitrarily close to $\mu$.
However, our use of an upper bound on $\omegaini$ means we cannot fully verify this claim,
due to numerical difficulties in evolving very broad solutions. We leave to future work the interesting question of whether oscillon solutions of arbitrarily small amplitude exist in the $\varepsilon \ll 1$ regime.

For $\varepsilon = 0.75$ (as seen in the bottom right panel), we see the emergence of a new feature---some of our breather initial conditions fail to form an oscillon 
but instead rapidly decay, indicated by the gray region in the figure. We can view this as the breakdown of our perturbative intuition for the case of small amplitude 
solutions. Another indication of failure from the perturbative picture is the existence of a minimum frequency oscillon. 

Examining the gray region, we see some preference to form oscillons when the initial conditions have more potential than kinetic energy.
As with the examples in the previous section, as the oscillons evolve their frequencies increase and they approach the end of their life.
Consequently, if we were to consider longer timescales we expect the size of the decayed region to expand.\footnote{In addition to these scenarios, 
we will also find solutions with insufficient energy to form oscillons at any time.} 
We bin the oscillation frequencies $\omegaosc$ and plot color coded contours in \figref{fig:ct_eps_freq} for different breather-like initial conditions, which are labeled by their 
frequencies $\omegaini$ and phase $\theta_0$.

We now take a more detailed look at solutions in the transition region between oscillon forming and decaying initial conditions.
Since the frequency of the oscillons slowly increases with time, we expect solutions in this transition regime to be be closely related to the final oscillon decay process and 
solutions that are slightly displaced from the oscillon attractor.
In the left panel of \figref{fig:diag_n_mod} we show the evolution at the origin for a few solutions in this transition regime (indicated by the blue rectangle in the lower right 
panel of \figref{fig:ct_eps_freq}) for $\varepsilon=0.75$.

A distinguishing feature of the solutions is the existence of amplitude modulation and the corresponding emergence of a second timescale (dubbed from now on as $t_{\rm mod}$).
Within the transition zone, as we consider solutions with larger $\omegaosc$ (corresponding to increasing $\omegaini$ at fixed $\theta_0$), we find both the magnitude and 
timescale of the amplitude modulations increases.
This continues until we hit the regime of decayed solutions and no oscillon forms.
Alternatively, as we decrease the value of $\omegaosc$ the amplitude of the modulation decreases as does its characteristic timescale.
For sufficiently small $\omegaosc$ the modulation becomes imperceptible and we obtain an effectively single timescale object.
Although not explicitly illustrated here, we also (a) found amplitude modulated solutions for $\varepsilon=0.125$, $0.25$, and $0.5$ within the regions 
indicated by red and brown contours in \figref{fig:ct_eps_freq}; and (b) confirmed in parameter space that amplitude modulated solutions deviate off the 
oscillon attractor as soon as $\omega_{\rm mod}\neq \omegaosc$.  

\likelihood

The dynamical origin of the amplitude modulation in $\phi(r=0,t)$ can be better understood using the full spacetime structure of the solutions.
From the left panel of \figref{fig:diag_n_mod}, we see that (at least at the origin) the modulated solutions involve two hierarchically separated timescales
\begin{enumerate}
\item a fast timescale $t_{\rm fast} \sim \omegaosc^{-1}$, and
\item a much slower timescale $t_{\rm mod}\sim \omega_{\rm mod}^{-1}$ associated with the modulation of the amplitude.
\end{enumerate}
In order to study the modulation itself, we want to separate out the slow modulated dynamics from the much shorter timescale dynamics encoded in $\omegaosc$.
After rasterizing the image, we noticed that most of the high-frequency features of the image are suppressed.
The right two panels of \figref{fig:diag_n_mod} illustrate the evolution of the slow component for an example modulated solution.
In the middle panel, we show the evolution of $\left|\phi(r,t)/\phiScl\right|$.
From this spacetime picture, we see that the amplitude modulation at the origin is a manifestation of a slow contraction and expansion of the oscillon core.
In the right panel, we plot the radial structure of the evolving energy density for the same oscillon shown in the middle panel.
As the profile of the core expands and contracts, we see correlated bursts of classical radiation produced that then propagate away from the oscillon core at (approximately) the 
speed of light. Analogous solutions showing periodic phases of contraction and expansion also appear in two (and more) spatial dimensions, and they will be presented in 
the appendix, in \secref{app:2d_modul}, as the outcome of a different numerical setup. Previous efforts have presented amplitude modulation in oscillons (see 
\cite{Hindmarsh:2006ur,PhysRevD.74.124003}, for example) from initial Gaussian profiles and other potentials. Our contribution is not only to explicitly illustrate the spatial structure of 
such solutions; it is also to show that these describe the dynamics in the stability limit.

Note that these modulated solutions are not captured by the commonly assumed quasibreather prescription
\begin{align}
\frac{\phi(r,t)}{\phi_{\star}} = \sum_{n\in \mathbb{N}}R_n(r,\omega)\sin(n\omega t +\delta_n)\,,
\label{eq:qb}
\end{align}
which expresses the solution in multiples of the ``fundamental frequency'' $\omega$~\cite{Vakhitov1973,Cyncynates:2021rtf,Zhang:2020bec}, and corresponds to $\omegaosc$ 
in agreement with the nomenclature we used in this paper. 
Therefore, conclusions about oscillon properties based on this ansatz are not directly applicable to the modulated solutions in the transition regime.
Despite this, the spatial structure visible in the middle and right panels of \figref{fig:diag_n_mod} reveals that the amplitude modulated solutions remain spatially localized, 
and therefore fall under the broad definition of oscillon used here. We suspect that these solutions are related to the emission of staccato radiation in oscillons 
\cite{Dorey:2019uap, Nagy:2021plv}. 

\perten 

From \figref{fig:ct_eps_freq}, it is clear that oscillon formation is fairly robust to changes in the form of the initial breather profile, at least for $\varepsilon \ll 1$.
\Figref{fig:lk_eps} provides an alternative empirical representation of this robustness.
Using our ensemble of initial conditions uniformly sampled in $\log_{10}(\omegaini/\mu)$ and $\theta_0$, we construct the empirical distributions of $\omegaosc$ as 
$\varepsilon$ is varied.
These distributions are illustrated in~\figref{fig:lk_eps}.
We see the distributions deform form a two-component mixture (when $\varepsilon = 0.125$ and $0.25$) to a three-component mixture (when $\varepsilon = 0.75$), with 
$\varepsilon=0.5$ serving as a transition state between the two.
For the smaller values of $\varepsilon$, the distribution is well modelled as a two-component system: the first component is an approximate $\delta$-function of frequencies with 
$\omegaosc = \omega_{\rm min,\varepsilon}$, while the second component is a continuum of frequencies. These correspond to an attractor point in solution space and points along 
the attractor line, respectively. As for the first component, it indicates an important point of our discussions: there is a minimum frequency for oscillons to form. 
Examining both the color codes in \figref{fig:ct_eps_freq} and the lower bounds of the histograms in \Figref{fig:lk_eps}, we observe that $\omega_{\rm min,\varepsilon}$ grows with 
the dimensionality. As for the second component, the continuum of solutions is consistent with the presence of small amplitude 
oscillons~\cite{Fodor:2008es,Fodor:2019ftc, Levkov:2022egq}. As we will show below, such solutions are well-represented by breather perturbations. 
Within the initial frequency prior, we do not observe any decayed solutions for values of $\varepsilon<0.75$. 
For $\varepsilon=0.75$ we see the emergence of a third $\delta$-function like component with $\omegaosc=\mu$, corresponding to the decayed solutions.

This allows us to observe how the distributions (i.e., the histograms colored for different values of $\varepsilon$) deform progressively from being unimodal ($\varepsilon\ll 1$)
to be bimodal ($\varepsilon\sim 1$), and the range of oscillation frequencies contracts and shifts toward larger frequencies as $\varepsilon$ grows.
The interval shift is also visible from the displacement of the ensemble's mean, this is depicted by the white dot of each distribution. Extending these statistical results to other 
dynamical variables (such as the energy, for example) is not recommendable. The prior parameter distribution is determinant to its final shape, and its effects are hard to 
dissociate without denser parameter sampling. 

\ennonpert
\simbrosc

The emergence of smooth isocontours of $\omegaosc$ as we scan over breather initial conditions indicates that many initial breather profiles can collapse into an 
oscillon with nearly the same frequency. This degeneracy suggests a further reduction of the initial parameter space, where we consider constant phase curves 
(with $\theta_0=0$ fixed) as a proxy for the isofrequency surfaces in \figref{fig:ct_eps_freq}.
We verified for several cases that the energy/frequency flow lines do not depend on the choice of initial phases.
Our objective with this is to visualize how the relationship between the oscillon energy and frequency depend on $\varepsilon$. 
\Figref{fig:pert_en} shows $\omegaosc$ as a function of the initial energy and the oscillon energy at $\mu t = 10^4$. 
From this figure, we identify two important features: 
\begin{itemize}
\item{The collapse of different initial states to yield an oscillon with minimal frequency $(\omegamin)$ and maximal energy $(\Emax)$.  Both the maximum energy and minimum frequency grow with the dimensionality of the solution.  These solutions correspond to the plateau region in~\figref{fig:ct_eps_freq}.}  
\item{A continuum of states with frequencies greater than $\omegamin$ and energies smaller than $\Emax$.  The range of frequencies decreases with increasing dimension.}
\end{itemize}
The continuum of states (also visible in the smallest bars of the first three histograms of \Figref{fig:lk_eps}) is consistent with the perturbative expectation that for $\varepsilon\neq 0$ 
each breather profile will undergo a small deformation into an oscillon.  The emergence of a maximal oscillon energy $\Emax$ is a nonperturbative effect in the sense that the resulting 
oscillon has properties very different from the corresponding breather for many of the initial conditions.
Our intention is to represent the dynamical state of the solution by introducing a (non-invertible) map between breather energies and frequencies to oscillon parameters 
measured at $\mu t = 10^4$. Therefore, features from the initial parameter distributions are mapped to the flow lines in the $(\omegaosc, E)$ plane. As an example of this, 
we observe that for $\varepsilon \ll 1$, 
the maximal value of $\omegaosc$ (and corresponding minimal energy $E_{\rm min}$) is just an artifact of our initial condition sampling. Such a bound results from mapping the 
initial frequencies upper bound to the oscillon frequencies $\omegaosc$. Existing work on small amplitude oscillons has argued that there are a continuum of oscillon solutions 
with frequencies arbitrarily close to $\omegaosc = \mu$ and arbitrarily small energies \cite{Fodor:2008es,Fodor:2019ftc,Levkov:2022egq}, from which we can infer that it is 
reasonable to set $\omegamax = \mu$ and $\Emin=0$ in the $\varepsilon \ll 1$ limit. Unfortunately, these solutions are very wide, generating a large dimensionless hierarchy 
between the width of the oscillon and the typical wavelength of emitted radiation. This makes numerical investigation of this regime difficult, and we leave the phenomenology of 
solutions ``in the gap'' to future work.

\ecritnonpert
\perttonpert

As $\varepsilon$ grows, decayed solutions start to appear. Thus, given a sufficiently fine grid of initial configurations, it may be possible to compute the minimal 
energy/maximum frequency of an oscillon for $\varepsilon\gtrsim 0.75$. The left panel of \figref{fig:diag_n_mod} shows that such a solution may show periodic amplitude modulation.
Figure \ref{fig:pert_en} is also useful to show how energy and frequency curves (plotted in dots) approach to their initial values (in dashed lines) 
as $\varepsilon$ gets smaller. This is also an indication that the breather and oscillon profiles look similar in this regime.
The same figure also shows that the connection between oscillons and breathers is more subtle as 
the dimensionality increases. Providing, therefore, a hint on how this feature may be used to provide a self-consistent definition of the perturbative regime. The validity of this and other 
definitions will be explored in a future project.     

\subsection{Case $\varepsilon\gtrsim 1$: Beyond the perturbative regime}\label{subsec:non_pert}

Thus far we have explored oscillons in the regime with $\varepsilon \lesssim 1$, corresponding to spatial dimensions $D \lesssim 2$. 
Since $\varepsilon$ acts as a control parameter for the deformation away from the one-dimensional sine-Gordon model, this roughly corresponds 
to the regime where we expect oscillons and breathers to be related perturbatively in $\varepsilon$.
In particular, we expect that the differences in energy, frequency and shape parameters of the oscillons and breathers will be perturbative in $\varepsilon$.
We now consider the regime $\varepsilon \gtrsim 1$, where this assumption about the closeness of oscillon and breather profiles is somewhat dubious.
Some evidence of this can already be seen in the bottom right panel of \figref{fig:ct_eps_freq}, corresponding to the case $\varepsilon=0.75$, where some of the initial breather profiles (with nonzero initial kinetic energy) decay rather than forming oscillons. Moreover, these results also raise interesting questions about existence of a frequency gap and/or energy gap.

From the right panel of \figref{fig:en_non_pert}, we see that as we continue to increase $\varepsilon$, a larger fraction of the parameter space for the initial breather profiles decay rather than settle into an oscillon solution.
This includes some of the profiles with $\theta_0 = 0$ (\ie zero kinetic energy).
In the left panel of \figref{fig:en_non_pert}, we illustrate how the energy in the initial radial profile depends on the initial condition parameters $\omegaini$ and $\theta_0$.
Comparing the initial energy isosurfaces in the left panel to boundary of the region of decayed solutions suggests that for $\varepsilon \gtrsim 1$ there is a minimum energy 
oscillon configuration, and the initial conditions that fail to form an oscillon fall below it. The formation of stable oscillons is subtle since energy dissipation is a necessary part 
of the process. Thus, in $D\geq 2$ there are initial configurations radiating away most of its energy before forming an oscillon. 
More generally, the similarity between the constant energy isosurfaces and constant $\omegaosc$ isosurfaces indicates that the frequencies of oscillons that from the breather 
initial conditions are largely determined by the initial energy available in the simulation volume.
We confirmed that similar agreement occurs for other choices of $\varepsilon$.

An important motivation of this project is the similarity between oscillons' and breathers' radial profiles. Hence, we will proceed with our discussions on oscillon's parameterization 
using breathers from a slightly different perspective than the we one followed so far, \ie by comparing the radial profiles of oscillons and breathers. Let us consider the evolving oscillons 
of the dimensionally-deformed sine-Gordon equations in \eqref{eqn:sg-deformed}, where we use breather profiles as initial conditions. After a few hundred oscillations, we measure the 
height of the oscillon peak $(\mathcal{A}_{\rm osc})$ at $r=0$ (as in \Secref{sec:sols_and_params}) at some instant where the oscillon has 
reached its maximum amplitude. We compute its ``instantaneous'' breather frequency $\omega_{\rm inst}$ from the oscillon amplitude 
by using
\begin{align}
\frac{\omega_{\rm inst}}{\mu}=\cos\left(\frac{\mathcal{A}_{\rm osc}}{4\phiScl}\right)\, ,
\label{eq:omega_b}
\end{align}
and build a radial breather profile $\phiB(r,t=0|\omegaB=\omega_{\rm inst})$ from \eqref{eqn:breather-profile}.  In \figref{fig:sim_br_osc}, we show the evolution of the oscillon 
profile and compare its shape (within a limited timespan) with the breather built in this way, for $\varepsilon=2$ and after $\mu t \sim 4\times10^2$. 
From our results, we infer that it is possible to find a breather radial profile that approximates the shape of an oscillon at a fixed instant of time. 
Moreover, we are able to replicate this procedure at different times regardless of the oscillon's dimensionality, as long as this is stable. Differences in radial profiles 
appear in the tails, and grow as $\varepsilon$ increases. Furthermore, when $\varepsilon\ll 1$, oscillon dynamics is well-represented by time-dependent breathers; 
and as the dimensionality grows, oscillons tend to dephase quicker. We have not tested if this also happens for different initial conditions; but we find it holds for the 
one-dimensional potential deformations presented in \secref{sec:SG_monodromy}. In addition to the existence of amplitude modulated solutions, the possibility of parameterizing 
oscillons using breathers gives us another reason to consider a nonlinear mode mixing formula instead of the quasibreather ansatz suggested in~\cite{Cyncynates:2021rtf,Zhang:2020bec}. 
This result also motivates us to extend this similarity through the entire oscillon evolution (if possible). This extension demands time-dependent frequencies instead of fixed values 
(\ie $\omegaB\rightarrow \omegaB(t)$). In principle, such a change in the parameterization may be sufficient to capture the evolution of the oscillon profile and its oscillation phase. 
Numerical renormalization \cite{GalvezGhersi:2021sxs} suggests a reasonable procedure to build semi-analytical oscillons. We will explore its applications in a future project. 

We study the reduced space of $\omegaosc$ and $E$ as in the perturbative regime (in Subsec.~\ref{subsec:pert_dim_def}). As a consistency check, we found that 
the only effect of choosing $\theta_0\neq 0$ is to shift the states towards lower energies/higher frequencies as phases increase, leaving the flow lines invariant. 
We followed the same procedure used to find our results in \figref{fig:pert_en}, \ie by measuring the oscillation frequency as a function of the initial energy, as well as 
the energy at later times now in the case $\varepsilon\gtrsim 1$. The upper and lower panels to the left of \figref{fig:e_crit_eps} show that a fraction of the solutions 
have decayed in a similar way to what we observed in the case $\varepsilon=0.75$ in our discussions of the perturbative regime. The upper panel (corresponding to 
2D solutions) shows an intermediate state labeled as $(\Emin, \omegamax)$, since it is the oscillon with the highest frequency and the lowest energy in our sample. 
This is an amplitude modulated solution; it is located in between decayed and non-decayed solutions in the same way as we observed in the case $\varepsilon=0.75$ 
in the left panel of \figref{fig:diag_n_mod}. Finding the exact location of the maximum frequency states depends on the sampling of initial conditions. Before decaying, 
energies and frequencies of the amplitude modulated oscillons do not vary significantly with time.

\largeampb

As for the panel at the bottom, corresponding to $\varepsilon=1.8$, we do not find any intermediate states. The spread of energies to the left and right of 
$E_{\rm max}$ can be interpreted as a signal of the maximum and minimum energy states (and the whole line of states in between, representing a continuum 
of oscillons in the small epsilon limit) collapsed to a point in parameter space. As we will show shortly, such a collapse occurred at some smaller value of $\varepsilon$.

In all of the cases, it is clear that there is (approx.) an oscillon with maximum energy, which can be produced by a large family of breathers with initial energies larger than a threshold, 
where such a threshold is represented by an isosurface of constant energy, in an analog way to what is depicted in \figref{fig:en_non_pert}. 
Oscillons maximal energy ($\Emax$) can be determined more robustly than minimum (or intermediate) energies or frequencies. We can confirm this by examining the blobs in 
semitransparent black from the two left panels of \Figref{fig:e_crit_eps}, which (for $\omegaosc<\mu$) concentrate to form a solid black region around a narrow energy band. 
We estimate the maximum energies for a few values of $\varepsilon>1$ from the mean energy of the states within the darker regions. Estimation errors correspond to the 
standard deviations measured around the mean energy. With all of this information, in the right panel of \figref{fig:e_crit_eps} we fit $\Emax$ as a power law in $\varepsilon$ of the form  
\begin{align}
\frac{\mu^{\varepsilon+1}}{V_0}(E_{\mathrm{max}}-E_0^{>}) =\left(\frac{\varepsilon}{\varepsilon^{>}_{\star}}\right)^{p^{>}}\,,
\label{eq:en_fit} 
\end{align}
where the fitting parameters are $E^>_0=50.26\pm 2.44$, $p^>=5.15\pm 0.15$ and $\varepsilon_{\star}=0.947\pm 0.001$. Energies and frequencies vary faster 
in time as the dimensionality increases, due to the reduction of oscillon lifetimes. Thus, error bars enlarge since it is harder to measure fixed values of $\Emax$ and $\omegamin$.  
The points with error bars shaded in blue represent the effect of time evolution in the fit: points with higher values of $\varepsilon$ are the first to escape 
from the power law, since the corresponding oscillons decay faster as the dimensionality increases. Expressions fitted such as Eq.~\eqref{eq:en_fit} have no 
dependence on the initial breather parameters $\omegaini$ and $\theta_0$, since oscillons clustered in the $\omegamin$ blobs are the approximately 
the same for all of the initial frequency and phase choices (as long as $\theta_0\neq \pi/2$). 

The collapse of the minimal and maximal energy configurations into a small fuzzy region in parameter space (treated approximately as a point) is a characteristic feature of 
the nonperturbative regime of dimensional deformations. A sufficient amount of states located in the transition region between oscillons and decayed solutions is required to study 
the collapse. One way to increase the number of configurations in this region is to include solutions evolved from breathers with different initial phases. Spanning over 
phases does not vary our estimations of $\Emax$ and $\omega_{\mathrm{min},\varepsilon}$. Hence, in addition to the initial frequency span, in \figref{fig:pert_npert} initial phases are 
also mapped in the interval $\theta_0\in[0;\pi/2)$ to represent over 500 configurations in each constant $\varepsilon$ flow line. Even when discontinuities in the maximum 
oscillon frequency (as a function of $\varepsilon$) reveal that the sampling is still too sparse to resolve $(\omega_{\rm max},\Emin)$ accurately. However, it is dense enough to 
show drastic changes in the number of oscillons found in a certain range of frequencies and energies. Such changes manifest as gaps in \figref{fig:pert_npert}, and suggest the 
existence of a minimum energy/maximum frequency state when $\varepsilon\gtrsim 1$.  Additionally, our results confirm the collapse of a continuum of states (limited by maximal 
and minimal energy oscillons) close to the transition from the perturbative to the nonperturbative regime. At $\varepsilon=1.125$, we find a localized range of frequencies where 
oscillons can be found, here the maximum and minimum frequencies are significantly closer than in cases with smaller epsilon. The marginalized empirical distributions can be found 
in the inset plotted in the upper corner of the figure. In this inset, the oscillation frequency interval is restricted to $\omegaosc/\mu\in[0.88; 0.95]$ to show deformations in the distributions, 
which agree with the collapse of the continuum of states to a single point when $1.25\leq\varepsilon< 1.375$. It is possible (but not very likely) that such a point is actually a very narrow 
line for $\varepsilon\approx 2$. As seen in the lower left panel of \figref{fig:e_crit_eps}, states spread diffusely around the maximal energy/minimum frequency band, which starts to appear 
at $\varepsilon=1.31$. With our simulations, we were not able to clearly distinguish more than one state in that small region. The simulations dubbed as $(\omega_{\rm max},\Emin)$ in red 
circles for the cases $\varepsilon=1$ and $\varepsilon=1.0625$ (as well as some of their nearest neighbors) correspond to amplitude modulated oscillons undergoing periodic phases of contraction 
and expansion in their cores. These are located in the intermediate region between oscillons and decayed solutions, which is consistent with our results in 
subsection~\ref{subsec:pert_dim_def} for the case $\varepsilon=0.75$ depicted in \Figref{fig:diag_n_mod}.

Employing alternative parameterizations leads to many opportunities and possible explotations; in particular, it is reasonable to evaluate how an increase in the energy affects 
oscillon stability. If we consider 
\begin{align}
\frac{\phi(r,t)}{\phi_{\star}} = \alpha\phiB(r,t|\omegaB=\omega_{\rm ini})\,,
\label{eq:sg_new_param}
\end{align} 
as an initial condition injecting kinetic and gradient energy for $\alpha>1$. Potential energy can also grow for small amplitude states oscillating around $\phi=0$. 
If we consider this initial condition for $\alpha=1$, $\theta_0=0$ and $\omegaini/\mu=0.9$, and evolve it for $\varepsilon=2$ (i.e., in three spatial dimensions), the 
solution does not form an oscillon. As an experiment, we increased the amplitude of the same configuration by a factor of $\alpha=20$ to see the effects of an 
arbitrarily large amplitude boost in the solution. In \figref{fig:large_alpha}, our results show that the solution corresponds to a series of spherically symmetric 
energy shells, field configurations oscillate around more than one minima of the sine-Gordon potential. Certainly, the solution does not have any similarity with 
the oscillons discussed throughout this paper, and oscillon lifetimes are not boosted by the extra initial energy injected. Bursts of classical radiation escape from 
the solution throughout its evolution. It is clear that energy is no longer a localized quantity, and the frequency of the solution at the origin may not be a relevant 
parameter anymore. Therefore, increasing the initial breather’s amplitude by using the parameterization in Eq.~\eqref{eq:sg_new_param} does not necessarily 
support the formation of long-lived oscillons in higher dimensions. This agrees with many preceding results using enlarged Gaussian profiles as initial states.

\rcondensed

\section{Critical behavior}\label{sec:critical}
  
In the preceding sections, we found evidence of an oscillon attractor in the space of spherically symmetric
solutions to the dimensionally deformed sine-Gordon model. Further, this attractor is dynamically accessible from a wide
range of radial breather initial conditions. In addition to this, we found that amplitude modulated solutions, which are intermediate 
states between oscillons and decayed solutions, progressively deviate away from the osciilon attractor.  
In this section, we quantify the properties of this oscillon attractor as $\varepsilon\equiv D-1$ is varied. After collecting our results from the 
perturbative and nonperturbative regimes, the features of the oscillon attractor are consistent with the presence of critical behavior.  Thus far, 
most of our discussions focus on the oscillation frequencies $\omegaosc$ and energies $E$ as diagnostic parameters describing oscillon dynamics. 
These two parameters are clearly well-motivated physically, and also directly illustrate the similarity between the oscillon solutions and corresponding 
breathers when $\varepsilon\ll 1$, as shown in \Figref{fig:pert_en}.

\tabfitparams

First, let us summarize the key properties of the oscillon attractor that we found in \Secref{sec:dim_def}.
\begin{enumerate}
\item{For all values of $\varepsilon> 0$, we found a maximum oscillon energy $\Emax$, corresponding to a minimum oscillation frequency $\omegamin$. 
Breather initial conditions with $E_{\rm init} > \Emax$ tended to rapidly evolve towards this oscillon configuration.}
\item{As we increased $\varepsilon$, we eventually found that some of the breather initial conditions rapidly decayed
instead of forming an oscillon. The separation between decayed and oscillon solutions closely matched the energy of the initial breather configurations,
suggesting the existence of a minimum energy oscillon for sufficiently large values of $\varepsilon$. Our results in \Figref{fig:pert_npert}, where we found 
energy/frequency gaps with no oscillons, provide further evidence of this. The span of initial conditions in the region $|\omegaosc-\omegaini|\ll 1$ 
is too sparse to determine if this feature appears at a finite value of $\varepsilon$ or not. A technical reason to not sample this regime is that oscillons are very wide, 
and therefore, appropriate numerical implementations are computationally expensive.}
\item{As a result of these two properties, there are a continuum of oscillon solutions for $\varepsilon \ll 1$, labelled
alternatively by their energy $E_{\rm osc}$ or oscillation frequency $\omegaosc$.}
\item{As we continue to increase $\varepsilon$, the states $(\omega_{\rm max}, E_{\rm min})$, $(\omega_{\rm min}, E_{\rm max})$, and all the states in between 
approach each other, and the attractor line collapses down to a point.}
\end{enumerate}

\critical

It is of interest to understand how these key features of the oscillon attractor evolve with $\varepsilon$. Specifically, the maximum energy oscillon acts as a critical 
solution of sorts, since it forms the beginning of the oscillon attractor line. As a result, our breather initial configurations with $E_{\rm ini}>\Emax$ tend to cluster around 
this point as they dynamically evolve. In \Figref{fig:res_cond}, we gather the results of $\Emax$ and $\omegamin$ from the perturbative and nonperturbative regimes, 
knowing that these read from the accumulation of states around specific points of the $(\omegaosc, E)$ plane for every value of $\varepsilon$. Our results indicate that the 
$E_{\rm max}$ dependence of $\varepsilon$ cannot be well approximated by a single power law. We need two separate curves to do such a fit, one for $\varepsilon \lesssim 1$
\begin{equation}
\label{eqn:eps_en_pert}
  \frac{\mu^{\varepsilon+1}}{V_0}\left(E_{\rm max}-E_0^<\right) = \left(\frac{\varepsilon}{\varepsilon_\star^<}\right)^{p_<}\,,
\end{equation}  
and another one for $\varepsilon \gtrsim 1$
\begin{equation}
\label{eqn:eps_en_non_pert}
  \frac{\mu^{\varepsilon+1}}{V_0}\left(E_{\rm max}-E_0^>\right) = \left(\frac{\varepsilon}{\varepsilon_\star^>}\right)^{p_>} \, .
\end{equation}
The corresponding fit parameters are summarized in Table~\ref{tab:fit_p_np}. Naively, this suggests the presence of a phase 
transition of order higher than zero. However, the bottom panel of \Figref{fig:res_cond} shows that the minimum frequency $\omega_{\rm min}$ is well fit by only one  
power law
\begin{equation}
\label{eqn:eps_w}
  \frac{\omega_{\rm min}}{\mu} = \left(\frac{\varepsilon}{\varepsilon_\star^{\omega}}\right)^{p_{\omega}} \, ,
\end{equation}
with $\varepsilon_\star^{\omega} = 2.262\pm 0.078$ and {$p_\omega = 0.125\pm 0.003 \approx \frac{1}{8}$}. This implies that the discontinuity in powers seen in 
the maximum energy is either not universal, or with respect to the frequency it is represented by a higher order phase transition. For larger values of $\varepsilon$, 
the power law fit becomes poor, and the minimum frequency actually appears to decrease slightly before reaching a plateau. 
Expressions fitted as Eqns.~(\ref{eqn:eps_en_pert}--\ref{eqn:eps_w}) have no dependence on the initial breather parameters $\omega_{\mathrm{ini}}$ and $\theta_0$, since the stable 
oscillons are (approx.) the same for all of the initial frequency and phase choices in the ivory regions (as long as $\theta_0\neq \pi/2$). Thus, the measured values of 
$\Emax$ and $\omega_{\mathrm{min}}$ are insensitive to the initial breathers shape.

The collapse of a continuum of states, bound by states with minimum and maximum energy/frequencies, is one of the main results of our discussions in the nonperturbative 
regime in subsection \ref{subsec:non_pert}. To illustrate the collapse towards the minimum frequency line in the lower panel of \Figref{fig:pert_npert}: the maximum 
frequency states included come from (a) setting $\omegamax=\mu$ as an educated guess when $\varepsilon<0.75$, and (b) empirical maximum frequency values 
found from our simulations in the nonperturbative regime. Figure \ref{fig:critical} resembles a phase diagram, depicting the collapse of minimum and maximum frequency 
states to a single point. Such a collapse allows us to identify a ``triple point'' in the phase diagram, which can be located in interval $\varepsilon\in[1.25;1.375)$. The black dots denote the 
upper limit in oscillon frequencies spanned by our simulations. We have found oscillons in the regions hatched in blue, while the regions in red correspond to wide oscillons. Such 
solutions have not been explored due to the resolution limits of our numerical setup.  Even when we considered initial phases when sampling initial configurations to probe lower energies in 
the nonperturbative regime, maximum frequencies (within the purple rectangle) are still prone to large error bars. The collapse of all intermediate states to a single point, and the 
gaps between stable and unstable solutions are still visible in spite of this. Amplitude modulated solutions were found throughout the entire region hatched in blue as intermediate 
states between minimum and maximum frequency configurations; although these are easier to distinguish around $\varepsilon\gtrsim 1$.  

In what remains of this paper, we will explore alternative ways to deform the sine-Gordon model. In \Secref{sec:SG_td}, we suggest an implementation to consider 
dynamical transitions in the spacetime dimensionality and evaluate some of their effects. We build a tunable model deforming the sine-Gordon to the axion monodromy potential 
to extend our previous results in \Secref{sec:SG_monodromy}. Further in the text, the reader can learn about our numerical implementation in Appendix 
\ref{app:numerical}, and find the discussions in \Secref{sec:discussion}. It would be interesting to extend this treatment to consider other localized structures, 
such as solitons and strings produced by topological defects \cite{Hindmarsh:2007jb, Blanco-Pillado:2020smt, Blanco-Pillado:2022rad}, finding their connections 
(if any) with other integrable models. 

\breatherfit

\section{Time-dependent deformations}\label{sec:SG_td}
The concept of dynamical spacetime dimensionality has been suggested in a wide variety of scenarios \cite{Atick:1988si, tHooft:1993dmi, Carroll:2009dn, Ito:2017rcr}, 
and its effects in nonlinear field theories deserve attention. On the other hand, thus far all the spherically symmetric oscillons were produced by instantaneous dimensional deformations of sine-Gordon 
breathers. Therefore, in this section we explore dimensional modifications of the SG model having a finite duration, since it is valid to ask how the connections presented in \secref{sec:dim_def}
due to time-dependent dimensional deformations. To introduce dynamical dimensional transitions, let us consider the following action
\begin{align}
\label{eq:act_vard}
S_{\varepsilon_t}&=\int (\ell_{\rm T} r)^{\varepsilon_t}\bigg\{\frac{1}{2}\left(\frac{\pd\phi}{\pd t}\right)^2-\frac{1}{2}\left(\frac{\pd\phi}{\pd r}\right)^2\\
&-\mu^2\phiScl^2\left[1-\cos\left(\frac{\phi}{\phiScl}\right)\right]\bigg\}~dr dt\,.
\nonumber
\end{align}
For simplicity, we assume that the dimensional inverse length scale $\ell_{\rm T}$ is the same as $\mu$, which may not hold in a general setup. Sensitivity of our results with other choices will be explored 
in a future project. This action yields the dimensionally deformed equations of motion in \eqref{eqn:sg-deformed}. In this section, we modify the action by converting $\varepsilon$ into a time-dependent function 
denoted as $\varepsilon_t$, which is a straightforward deformation of the Minkowskian scalar field action in spherical symmetry. Introducing a such a dependence on real 
(instead of integer) values in the action analog to the dimensional regularization procedure applied in quantum field theory \cite{tHooft:1972tcz, Peskin:1995ev, 
Weinberg:1995mt}. After this redefinition, equations of motion follow from the functional derivative of \eqref{eq:act_vard}:
\begin{subequations}
\begin{align}
&\frac{d\phi}{dt}\equiv\Pi_{\varepsilon},\label{eq:dphi_dt_vard}\\
&\frac{d\Pi_{\varepsilon}}{dt}=-\Pi_{\varepsilon}\dot{\varepsilon}_t\ln (\ell_{\rm T} r) +\left[\frac{\varepsilon_t}{r}+\frac{\pd}{\pd r}\right]\frac{\pd \phi}{\pd r}-
\mu^2\phiScl\sin\left(\frac{\phi}{\phiScl}\right)\,,
\label{eq:dpi_dt_vard}
\end{align}
\end{subequations}
It is clear that in the case $\dot{\varepsilon}_t=0$ the equations reduce to spherically symmetric in $(1+\varepsilon_t)$ spatial dimensions. The term 
proportional to $\dot{\varepsilon}_t$ has a logarithmic singularity at $r=0$, but this is not a reason of concern since (a) the singularity is less severe than $r^{-1}$ 
and (b) it is only switched on during the transition. 

\tabduration
\varfreqphase

As for the functional form of $\varepsilon_t$, we continuously connect constant values of $\varepsilon$ by 
using cosine tapered functions \cite{bloomfield2004fourier}. Thus, we can write $\varepsilon_t$ as
\begin{align}
\varepsilon_t =
\begin{cases}
\varepsilon_{\mathrm{ini}} + \Delta D\left[\sin\left(\frac{\pi t}{2\sigma_t}\right)\right]^2 \,, & 
{0\leq t<\sigma_t,} \vspace{0.5em}\\
\varepsilon_{\mathrm{ini}} + \Delta D \,, & {t\geq \sigma_t} \,,
\end{cases}
\label{eq:eps_of_t}
\end{align}
which is a $C^1$ piecewise function continuous at $t=\sigma_t$. This function is very similar to a continuous step function, except that 
the input and output are exact instead of asymptotic, which allows us to be precise about the initial and/or final state of the dynamical 
system. $\sigma_t$ is the duration of the transition from $D=\varepsilon_{\mathrm{ini}}+1$ to $D=\varepsilon_{\mathrm{ini}}+\Delta D+1$ 
spatial dimensions, and it determines the speed of the dimensional deformation. $\varepsilon_{\mathrm{ini}}$ is the initial value of 
$\varepsilon_t$ and $\Delta D$ is the change in the number of spatial dimensions we want to achieve. The positive/negative sign of 
$\Delta D$ is used to denote if the transition is an increase/decrease in $\varepsilon_t$. $\dot{\varepsilon}_t$ is a single-peaked function 
of time, which becomes a ``delta kick'' in the limit $\sigma_t\rightarrow 0$. In the single-particle reduction of our system, such a spike 
can lead us to fractional kinetic energy gain or loss, similar to the scenario of an inelastic collision. 

As a proof of concept for the deformed field equations, we evaluate the transition from a breather-like spherically symmetric oscillon 
in three spatial dimensions to a one-dimensional breather. In this case, the initial condition has the same shape of the breather in 
Eq.~\eqref{eqn:breather-profile} with $\omegaB=0.1\mu$ and initial phase $\theta_0=0$. To represent the dimensional transition, we use 
$\varepsilon_{\mathrm{ini}} = 2$, $\Delta D=-2$ and $\sigma_t=0.1\mu^{-1}$ in 
Eq.~\eqref{eq:eps_of_t}, which is approximately instantaneous. Fig.~\ref{fig:3D_to_1D} shows that the solution evaluated at 
constant radius $r_{\mathrm{c}}=0$ can be written as 
\begin{align}
\frac{\phi(r,t)}{\phiScl}= \phiB(r=0,t|\omegaB=\omegaosc)\,,
\label{eq:sg_fit}
\end{align} 
with $\theta_{\mathrm{B}}\approx-4\pi/21$ and $\omegaosc\approx 0.381\mu$. The value of $\omegaosc$ was extracted
from the evolving field following the procedure described in Sec.~\ref{sec:sols_and_params} (as seen in the right panel of 
Fig.~\ref{fig:sols_n_freq}): by finding the dominant frequency of the solution evaluated at the origin. 
As plotted in the figure, this result is fully consistent with a well-known fact of the sine-Gordon model \cite{PhysRevLett.30.1262}: its 
solutions can only be combinations of breathers, solitons and non-linear waves. Simultaneously, we evaluated the consistency 
of the deformed field equation solutions with SG breathers, and validated the frequency extraction procedure explained in preceding sections.

We evaluate the sensitivity of the oscillation frequency $(\omegaosc)$ with the dynamical dimensional transition suggested in Eqns.~\eqref{eq:dphi_dt_vard} and 
\eqref{eq:dpi_dt_vard} in coherence with our work in the previous sections. None of the breathers has been deformed to compensate for 
the lack of energy in the one-dimensional initial conditions. Considering $\varepsilon_{\mathrm{ini}}=0$ and $\Delta D=+1$, we simulate 
the dynamical deformation of 1D breathers into 2D oscillons for the four different durations reported in Table \ref{tab:duration}. As we can notice, 
the first two cases $\sigma^{(1)}_t$ and $\sigma^{(2)}_t$ correspond to transitions happening in less than one oscillation. 
Cases $\sigma^{(3)}_t$ and $\sigma^{(4)}_t$ last more than a full oscillation period, observing that the duration of the extremal scenarios
is different by two orders of magnitude. We generate oscillation frequency maps in Fig.~\ref{fig:ct_eps_freq} in the same range of initial 
frequencies and phases used in the perturbative regime. In the four panels of Fig.~\ref{fig:var_eps_freq}, we present the oscillation 
frequency maps corresponding to the transition durations in the table. Observing that the symmetry of the cusp centered at $\theta_0=\pi/2$ is restored in the abrupt 
transition limit (in the lower right and left panels labeled as $\sigma^{(1)}_t$), used throughout the perturbative and nonperturbative regimes discussed 
in this manuscript.  However, we notice from our results that, essentially, the oscillation frequency range is (approximately) independent of the dimensional 
transition duration for the span of initial breathers used throughout the paper. 

The right column of Table \ref{tab:duration} reveals that the number of rapidly decaying solutions (within the gray contours) varies in less than 10\% for a two 
orders of magnitude change in the transition duration, which implies that the amount of oscillons is also approximately independent of the transition speed. However, 
it would not rigorous to extend these conclusions to different choices (and ways of sampling) of initial conditions. Similar deformations to the high duration maps 
$\sigma^{(3)}_t$ and $\sigma^{(4)}_t$ in Fig.~\ref{fig:var_eps_freq} can be reproduced if we change the initial frequency binning of the $\sigma^{(1)}_t$ panel, 
by mixing some fraction of the amplitude evolution from adjacent initial frequencies. For larger time intervals, such as in the panel labeled as $\sigma^{(4)}_t$, 
the cusps become less sharp, connecting smoothly the regions of initial parameter space where oscillons and rapidly decaying solutions exist. It is clear that the initial dependence tends 
to disappear as the transitions becomes slower. As shown in Table \ref{tab:duration}, the slowest transition has a relatively mild effect in changing the number of oscillons. Nonetheless, the 
same cannot be said about the amount of intermediate frequency states. In the same panel, we notice that the frequency gradient becomes smoother, and consequently, the number of amplitude 
modulated solutions increases with respect to the other cases. 

\deltapotential
\dvstatomega

\section{Oscillons in Other Models: Potential Deformations}\label{sec:SG_monodromy}
Thus far we have studied oscillons for a relativistic scalar field with canonical kinetic terms evolving in a cosine potential (i.e.,\, the sine-Gordon model).
By considering spherical solutions in non-integer dimensions, we were able smoothly connect oscillon solutions in these models to the breathers of the one-dimensional sine-Gordon model.
However, oscillons exist in a plethora of other relativistic field theories, and we would like to understand if sine-Gordon breathers can be related to these oscillons as well.
In this section, we extend the framework introduced above to the case of oscillons in theories other than the sine-Gordon model.
For concreteness, we will apply these methods to the axion monodromy model, which is well known to support oscillons~\cite{Olle:2020qqy,Zhang:2020bec,Cyncynates:2021rtf,Levkov:2023ncb}.

The potential for axion monodromy is given by
\begin{equation}\label{eqn:mono-potential}
  V_{\rm M} = \mu^2_{\rm M}\phi_{\rm M}^2\left[\sqrt{1 + \frac{\phi^2}{\phi_{\rm M}^2}} - 1\right] \, ,
\end{equation}  
and is illustrated in~\figref{fig:mono-potential}.
From a global perspective, the monodromy potential $V_{\rm M}$ is radically different from the sine-Gordon potential $V_{\rm SG}$.  For example, $V_{\rm M}$ has a single global minimum and no local maxima, while the sine-Gordon potential has a (countably) infinite number of degenerate potential minima and maxima.  However, a typical oscillon only probes a finite region away from the local potential minimum around which it oscillates.  As a result the deformation to the part of the potential actually probed by a given oscillon solution can be small.

Analogously to passing between spatial dimensions, we want a tunable parameter to that will allow us to smoothly deform our theory between the sine-Gordon potential and monodromy model.
While there are many ways such a parameter can be introduced, we adopt the following straightforward approach.
First, we need to match the characteristic time and field scales of the two potentials.
We match characteristic time scales by setting $\mu_{\rm M} = \mu_{\rm SG}$ so that the potential curvatures at the origin are equal.
To ensure that nonlinear corrections to both potentials appear at similar field excursions, we also set $\phi_{\rm M} = \phiScl$.
We then introduce the difference between the monodromy potential and the sine-Gordon
\begin{align}
 \Delta V &\equiv V_{\rm M} - V_{\rm SG}  \notag \\
          &= \mu^2\phiScl^2 \left[\sqrt{1+\frac{\phi^2}{\phiScl^2}} +\cos\left(\frac{\phi}{\phiScl}\right) - 2\right] \, .
\end{align}
Finally, we introduce a (tunable) deformed potential
\begin{align}\label{eqn:pot-deformed}
  V_{\varepsilon_{\rm V}} &\equiv V_{\rm SG}+\varepsilon_{\rm V}\Delta V\\
  &=\mu^2\phiScl^2\left[1-\cos\left(\frac{\phi}{\phiScl}\right)\right] + \varepsilon_{\rm V}\Delta V \, ,
  \nonumber
\end{align}
where the tunable parameter $\varepsilon_{\rm V} \in [0,1]$.
For $\varepsilon_{\rm V} = 0$ we recover the sine-Gordon potential, and for $\varepsilon_{\rm V} = 1$ we recover the monodromy potential.
\Figref{fig:mono-potential} illustrates this potential deformation procedure.
We see that within the local minimum at the origin (roughly for $-\pi \lesssim \phi/\phi_\star \lesssim \pi$), the deformed potentials (including the monodromy potential) are a small perturbation of the sine-Gordon potential. 
Although we only consider the axion monodromy potential here, it should be clear that the procedure is generally applicable.

For the purposes of this study, we will restrict ourselves to the one-dimensional case.
The corresponding equations of motion are
\begin{subequations}\label{eq:dpidt_1d}
\begin{align}
  &\frac{\ud\phi}{\ud t} = \Pi_{\phi},\label{eq:dphidt_1d}\\
  &\frac{\ud\Pi_{\phi}}{\ud t} = \frac{\pd^2\phi}{\pd x^2}-\mu^2\phiScl\sin\left(\frac{\phi}{\phi_{\star}}\right)-
\varepsilon_V\Delta V'(\phi) \, ,
\end{align}
\end{subequations}
where we now identify $\varepsilon_{\rm V}$ as the parameter controlling a deformation away from the one-dimensional sine-Gordon equation.
Although we will not explore this here, the potential deformations controlled by $\varepsilon_{\rm V}$ could be combined with dimensional deformations as in the preceding sections.

We now consider the evolution from breather initial conditions in the deformed potential~\eqref{eqn:pot-deformed} as the parameter $\varepsilon_{\rm V}$ is varied.
To ensure that the solution only probes regions where the deformed potential closely matches the sine-Gordon potential, we take {$\omegaini / \mu \in [10^{-0.1},10^{-0.015}]$}.
The lower bound ensures that the oscillating solutions are confined to a single potential well centered at $\phi=0$, where the sine-Gordon and monodromy potentials are similar to each other.
Meanwhile, the upper bound arises from numerical difficulties in evolving very broad oscillon profiles.
Empirically, we find oscillon solutions emerge from these initial conditions, but that the relaxation onto the oscillon attractor is somewhat slower than for the dimensional deformations studied above.
To capture the evolution along the attractor, we evolve our simulations for time $\mu T_{\max} = 2\times 10^4$, which is twice as long as the $\varepsilon \ll 1$ cases considered above.
We also find that the properties of the oscillon are approximately invariant to the initial phase,
which is consistent with the fact that the initial energy of the configuration in independent of $\theta_0$ for $D=1$.
Therefore, in what follows we fix the initial phase of the breather profiles $\theta_0 = 0$.

\Figref{fig:dv_omega_stat} summarizes the properties of oscillons that emerge from these scans over initial breather frequencies as we vary the potential deformation parameter $\varepsilon_{\rm V}$.
As in the previous sections, we focus on the oscillation frequency $\omegaosc$ and energy $E$ of the resulting oscillon.
Details of how we extract these quantities from simulation data are provided in \secref{sec:sols_and_params}.

For $\varepsilon_{\rm V}=0$ we are in the one-dimensional sine-Gordon limit, and the breathers are exact solutions to the equations of motion.
In this case, we see that the frequency distribution is unchanged, providing a basic sanity check on our results.
As we increase $\varepsilon_{\rm V}$, we observe the density of oscillation frequencies increasing at lower frequencies.
From our work in previous sections, we understand that it is possible to build an oscillation frequency map from the initial breather 
frequencies (\ie $\log_{10}\omegaosc(\log_{10}\omegaini)$). To shorten the notation, we denote $\mathcal{W}_{\rm osc}\equiv \log_{10}(\omegaosc/\mu)$
and $\mathcal{W}_{\rm ini}\equiv \log_{10}(\omegaini/\mu)$. Moreover, conservation of probabilities implies that the initial frequency distribution 
$P_{\omegaini}$, and the oscillation frequency distribution $Q_{\omegaosc}$ are related via
\begin{align}
P_{\omegaini}\ud\mathcal{W}_{\rm ini}=Q_{\omegaosc}\ud\mathcal{W}_{\rm osc}\,.
\end{align}
We assumed that $P_{\omegaini}$ is a discrete uniform distribution in $\mathcal{W}_{\rm ini}$, hence the distribution $Q_{\omegaosc}$ is the Jacobian
\begin{align}
Q_{\omegaosc}= \frac{1}{N}\left|\frac{\ud\mathcal{W}_{\rm osc}}{\ud\mathcal{W}_{\rm ini}}\right|^{-1}\,,
\end{align} 
where $N=50$ is the number of points sampled in the interval $\log_{10}(\omegaini/\mu) \in [-0.1;-0.015]$. The integral of $Q_{\omegaosc}$ along $\mathcal{W}_{\rm osc}$ (\ie its 
cumulative distribution) represents the map $\mathcal{W}_{\rm ini}(\mathcal{W}_{\rm osc})$, which is essentially the inverse map of $\omegaosc(\omegaini)$.
The left panel of \figref{fig:dv_omega_stat} shows the evolution of the Jacobian and the map $\mathcal{W}^{-1}_{\rm osc}(\mathcal{W}_{\rm ini})$ as the potential deforms. The accumulation 
of lower frequency states in the Jacobian suggests that the formation of a minimum frequency oscillon is a generic dynamical feature, and does not depend on the initial frequency prior. 
Although the Jacobian tends to become narrower in the upper end of the frequency span, the evidence may not be sufficient to prove the existence of a maximum frequency oscillon. 

The right panel of \figref{fig:dv_omega_stat} instead shows the relationship between the energy and oscillation frequency of the oscillons, which is the analogue of \figref{fig:pert_en} and \figref{fig:e_crit_eps}.
As with the perturbative dimensional deformations in \figref{fig:pert_en}, we see a continuum of breather energies and frequencies.
Further, it is clear that the breathers of the one-dimensional sine-Gordon model (the $\varepsilon_{\rm V} = 0$ line), map smoothly into the oscillons of the axion monodromy model, at least for this range of initial breather frequencies $\omegaini$.
This is strong evidence that the breathers provide a reasonable approximation to the oscillons, especially in the limit $\varepsilon_{\rm V} \ll 1$.
Since we have restricted to relatively large values of $\omegaini/\mu$, the maximum energy oscillon we observe is dictated by our initial conditions, rather than a physical mechanism.
If we were to explore smaller values of $\omegaini/\mu$, we expect a maximum energy plateau would appear as in the case of dimensional deformations.
Comparing the frequencies of the left end of the curves (which all have $\omegaini^{\rm (max)} = 10^{-0.015}\mu$), we see that $\omegaosc(\omegaini^{\rm (max)})$ decreases with $\varepsilon_{\rm V}$.
This agrees with the behavior seen in the left panel.
We see no evidence of a frequency gap or minimal energy solution, although a more definitive investigation of this requires extending our numerical techniques to the case of very wide oscillons, which 
we leave to future work. Similarly, we leave a more detailed exploration of the oscillon phase diagram (similar to our results in \secref{sec:critical}) for potential deformations to future work.
To be consistent with our results for dimensional deformations in \Figref{fig:pert_en}, the inset plotted in the lower right corner of the figure includes the initial energy lines to show the convergence 
of parameter flows as $\varepsilon_{\rm V}\rightarrow 0$. Confirming that one-dimensional oscillons are well-represented by breathers when $\varepsilon_{\rm V}\ll 1$. We are not able to explain why 
the states with the lowest frequencies coincide for all the values of $\varepsilon_{\rm V}$. We leave further investigations of this for a future project.   

\section{Discussion}\label{sec:discussion}

In this paper, we provided an explicit connection between one-dimensional sine-Gordon breathers with spherically symmetric oscillons. To achieve this, we studied the oscillons produced by 
deforming the breather solutions of the sine-Gordon equation, and viewed the dimensional term $\varepsilon r^{-1} \pd\phi/\pd r$ (with $\varepsilon\equiv D-1$) as a perturbation to the one-dimensional 
sine-Gordon equation.

In \secref{sec:setup}, we quickly revised the breather solution and its features and presented it as the initial condition of the evolving solution. A key point of this section is to 
understand that the breather needs (essentially) only one parameter to be fully characterized: its oscillation frequency. Hence, examining the evolution of the oscillon frequency $(\omegaosc)$ is a viable 
way to assess the dynamical state of the deformation. In \secref{sec:sols_and_params}, we outlined a procedure to extract the post-transient oscillation frequency of an oscillon, as well 
as its amplitude and energy. We did not intend to provide a ``complete'' description of the oscillon dynamics with this parameter choice, but rather a convenient reduction of the dimensionality 
of the configuration space. We explicitly showed the formation of an oscillon attractor in parameter space. Finding that once the solutions have reached the oscillon attractor, it is safe to 
consider their parameters to be approximately constant.

The deceleration of the parameter flow allows us to build an approximately static map connecting one-dimensional breathers and spherically symmetric oscillons. 
In \secref{sec:dim_def}, we solved the dimensionally deformed sine-Gordon equation to scan over different initial breather profiles. Such profiles are parameterized by 
their initial frequencies and phases. We divide our results in two scenarios: the perturbative $(\varepsilon\lesssim 1)$ and the nonperturbative $(\varepsilon\gtrsim 1)$ 
regimes of dimensional deformations. By choosing the measured oscillon energy and frequency to reduce the space of parameters, we explicitly show this 
connection in subsection \ref{subsec:pert_dim_def} via a non-invertible map in the perturbative regime. When $\varepsilon\ll 1$, we notice that the resulting distribution 
of oscillon energies and frequencies can be modelled as a two-component system: the first is an approximate $\delta$-function, which determines a maximum energy/minimum frequency bound 
for oscillons. The second component is a continuum of states corresponding to points along the attractor line. Oscillons along the continuum are well-represented by perturbative 
corrections of breathers. Resolving the maximum frequency/minimum energy limit involves solving wide oscillons, which is a complicated task due to the generation of a large 
hierarchy between the oscillon width and the wavelength of the emitted radiation. 

In our simulations, decayed solutions start to emerge as $\varepsilon$ gets closer to one. As in the small deformation limit, many of the states accumulate around a maximal energy configuration. 
In between decayed solutions and maximum energy states, we found oscillons having nontrivial radial structures for $0.75\lesssim \varepsilon \lesssim 1.0625$, observing that their evolution and 
radial profiles are incompatible with the quasibreather prescription. Still, it is correct to call them oscillons since their oscillating profiles and energy densities are spatially localized. 
These solutions undergo periodic phases of contraction and expansion of their cores, and in consequence their amplitudes are modulated. As core profiles expand and contract periodically, we 
observed correlated bursts of classical radiation produced propagating away from the oscillon core at (approximately) the speed of light. Apart from the natural oscillation timescale (scaling as 
$\omegaosc^{-1}$), we see the emergence of a second, much slower, timescale related to the amplitude modulation.  The location of these solutions in the oscillation frequency map gives us a reason 
to suspect that the emergence of the second timescale is related to oscillon decay rate. We leave further explorations of this possible connection for future research. 

With further growth in $\varepsilon$, we studied the $\varepsilon\gtrsim 1$ regime of dimensionally deformed breathers in subsection \ref{subsec:non_pert}. The connection between breathers and 
oscillons in this regime is more subtle than in the perturbative case. In this regime, we found that it is possible to construct a breather having a similar profile to a spherically symmetric oscillon at a 
fixed instant of time, regardless of the oscillon dimensionality. We have not explored yet if this result holds for different initial profiles; nonetheless, it holds for the potential modifications attempted 
in \secref{sec:SG_monodromy}. Similarities persist dynamically only in the case of small dimensional deformations, and dephase quicker as $\varepsilon$ grows. This result suggests the possibility of 
building semianalytical solutions, capturing the frequency, amplitude and oscillation phase as time-dependent parameters. We leave the implementation of semianalytic oscillons for a future project. 

As for the explorations in parameter space started in the nonperturbative regime, we generated (a) an oscillon frequency map -- extending of our procedures from the $\varepsilon\lesssim 1$ case -- and 
(b) an initial energy map by scanning over the same initial breather parameters previously used in the subsection \ref{subsec:pert_dim_def}. We overlapped the two maps to find that there is a minimum 
energy threshold to form an oscillon. For states with energies below that threshold, we showed that an arbitrary initial energy boost does not necessarily translates in increasing the oscillon stability. 
Finding the corresponding minimum energy/maximum frequency oscillon is a complicated task requiring a denser scan of initial breather profiles; however, our findings show that our scan is sufficient to prove its 
existence. On the other end of the oscillon attractor, the $\delta$-shaped distributions of states defining the maximum energy bound for oscillons also appear when $\varepsilon\gtrsim 1$. We find that a 
power law proportional to $\varepsilon^{5.15\pm0.15}$ is a good fit for the maximum energy as a function of $\varepsilon\gtrsim 1$. Although we could not determine the minimum oscillon energies 
precisely, the span of initial breather profiles is sufficient to find one of the key results of our analysis: the gradual collapse of the continuum of states, bound by maximum and minimum energy oscillons, 
to (approximately) a single stable solution for $\varepsilon\in[1.25;1.375)$.

The objective of \secref{sec:critical} is to show how critical behavior manifests in oscillon formation. To make this possible, we summarized most of our results in subsections \ref{subsec:non_pert}
and \ref{subsec:pert_dim_def}. The evolution of the maximum energy oscillon with respect to $\varepsilon$ needs two power laws to be represented: one for oscillons in the perturbative 
regime, and another when $\varepsilon\gtrsim 1$. Naively, this indicates the presence of a phase transition. On the other hand, it is sufficient to fit a single power law proportional 
to $\varepsilon^{\frac{1}{8}}$ to describe the minimum oscillon frequency. This result implies one of two possibilities: (a) the phase transition seen for the maximum energy is not universal, 
or (b) the order of the phase transition is higher when is plotted in terms of the frequency. 

Due to the high computational cost of solving wide configurations, we could not resolve the minimum energy endpoints of the oscillon attractors, nor their dependence on the dimensionality. 
However, it is reasonable to consider $\omegaosc=\mu$ as an educated guess for the maximum frequency bound in the perturbative regime. From our simulations in the nonperturbative regime, the stable 
solution with the highest frequency was used to provide a crude estimate of the maximum frequency oscillon. Combining our maximum frequency estimates with the minimum frequency yields 
a plot similar to a phase diagram. From this plot, we confirmed the collapse of minimum and maximum frequency (including a continuum of intermediate states) configurations to a single point, 
which is the main result of our explorations in the nonperturbative regime. We also found that some of the states in the region limited by maximum and minimum frequencies correspond to amplitude 
modulated solutions. 

In sections \ref{sec:SG_td} and \ref{sec:SG_monodromy}, we tested the connections between breathers and spherically symmetric oscillons in different dynamical setups. 
In \secref{sec:SG_td}, we considered the effects of dynamical spacetime dimensionality in oscillon formation. Observing that the cusps in the oscillation frequency isocontours (centered at $\theta_0=\pi/2$) 
tend to dilute as the dimensional transitions have longer durations. Thus, there is no preference to form oscillons from breathers with more potential or kinetic energy.  Additionally, the framework 
implemented in this section allows us to validate the frequency extraction procedure presented in \secref{sec:sols_and_params}. In \secref{sec:SG_monodromy}, we built a tunable potential as an alternative 
way to produce one-dimensional oscillons from breathers. This potential transformed the periodic sine-Gordon potential into the monodromy potential in incremental steps. The evolution of the frequency Jacobian 
with the growth of the deformation parameter suggests that the accumulation of states to form a maximum frequency oscillon is generic. We found no evidence of a frequency gap or a minimum energy bound for 
oscillons.
 
\acknowledgments

The authors would like to thank J. Richard Bond, John Dubinski, Andr\'es Evangelio, Andrei Frolov, Daniel Horna, Lorena Luj\'an, Jorge Medina, Thomas Morrison, Levon Pogosian, 
Guillermo Quispe, Leo Stein and Diego Su\'arez for many fruitful conversations, technical and logistic support, and their valuable feedback in earlier versions of this paper. All of the computations were 
performed on the computing workstations at the Canadian Institute for Theoretical Astrophysics (CITA). The work of JG was partially supported by the 
Natural Sciences and Engineering Research Council of Canada (NSERC), funding reference \#CITA 490888-16, \#RGPIN-2019-07306. The work of JB 
was partially supported by the Natural Sciences and Engineering Research Council of Canada (NSERC) and by the Simons Modern Inflationary Cosmology 
program.

\appendix
\section{Numerical setup and convergence tests}\label{app:numerical}
In this appendix we provide details of our numerical setup, including the unit conventions used in the code, discretization scheme, and various tests of numerical convergence.

\subsection{Dimensionless Units}\label{subapp:dim_units}

Before detailing our numerical methods, we briefly review the units used in our code.
For notational consistency, we will denote dimensionless quantities by an overbar $\bar{\cdot}$.
We follow the convention $\hbar = c =1$ throughout, so that time and space have units of inverse mass.

For the sake of generality, suppose we have a potential
\begin{equation}
  V(\phi) = \mu^2\phiScl^2 \bar{V}\left(\frac{\phi}{\phiScl}\right) \, ,
\end{equation}
with $\mu^2$ fixed by the requirement
\begin{equation}
 V''(\phi_{\rm min}) = \mu^2 \, ,
\end{equation}
where $\phi_{\rm min}$ is the field value at the potential minimum we wish to expand around.
Given a potential of this form, we will measure the field in units of the characteristic scale $\phiScl$, and time and space in units of the inverse field mass $\mu^{-1}$.
To do this, we introduce the dimensionless field variable
\begin{equation}
 \bar{\phi} \equiv \frac{\phi}{\phiScl} \, ,
\end{equation}
and dimensionless time and space coordinates
\begin{equation}
 \bar{t} \equiv \mu t, \qquad \bar{x} \equiv \mu x \, .
\end{equation}
For consistency, the dimensionless field momentum is given by
\begin{equation}
 \bar{\Pi} \equiv \frac{\partial\bar{\phi}}{\partial\bar{t}} = \frac{\Pi}{\mu\phiScl}
\end{equation}
The dimensionless equations of motion for the radial profile in $D=\varepsilon+1$ dimensions in first-order form are then given by
\begin{subequations}
\begin{align}
 \frac{\ud\bar{\phi}}{\ud\bar{t}} &= \bar{\Pi}\\
 \frac{\ud\bar{\Pi}}{\ud\bar{t}}  &= \left[\frac{\partial}{\partial\bar{r}}+\frac{\varepsilon_t}{\bar{r}}\right]\frac{\partial\bar{\phi}}{\partial\bar{r}} - \frac{\ud\bar{V}(\bar{\phi})}{\ud\bar{\phi}} \, .
\end{align}
\end{subequations}

For the sine-Gordon potential that is the primary focus of this paper, we have
\begin{equation}
 \bar{V}(\bar{\phi}) = 1 - \cos\bar\phi\,.
\end{equation}

As for the case of time-dependent dimensional transitions developed in \Secref{sec:SG_td}, the dimensionless 
version of Eqns.~\eqref{eq:dphi_dt_vard} and \eqref{eq:dpi_dt_vard} yields

\begin{subequations}
\begin{align}
 \frac{\ud\bar{\phi}}{\ud\bar{t}} &= \bar{\Pi}\\
 \frac{\ud\bar{\Pi}}{\ud\bar{t}}  &= \left[\frac{\partial}{\partial\bar{r}}+\frac{\varepsilon_t}{\bar{r}}\right]\frac{\partial\bar{\phi}}{\partial\bar{r}} + \dot{\varepsilon}_t\bar{\Pi}\ln \bar{r} -\sin\bar{\phi} \, .
\end{align}
\end{subequations}

where we considered $\ell_{\rm T}=\mu$. For deformations involving the monodromy potential, as in~\secref{sec:SG_monodromy}, we have
\begin{equation}
 \bar{V}(\bar{\phi}) = \sqrt{1+\bar{\phi}^2} \, .
\end{equation}

In order to simply notation, throughout this appendix we work in the dimensionless units outlined above,
but will omit the overbars throughout.

\subsection{Spatial and Temporal Discretization}\label{subapp:num_set}
For our spatial discretization, we use a pseudospectral approach which we outline here.
Given a function $f(r)$ defined on the semi-infinite interval $[0,\infty)$, we expand it in a (truncated) basis of even Chebyshev rational functions on the doubly infinite interval 
\begin{equation}\label{eqn:spectral-expansion}
  f(r) = \sum_{n=0}^{N-1} c_n \cos\left(n\theta(r)\right) 
\end{equation}
where
\begin{subequations}
\begin{align}\label{eqn:angular-coord}
  \theta(r)  &= 2\cos^{-1}\left(\frac{r}{\sqrt{r^2+\lenPar^2}}\right) \, , \\
  r(\theta) &= \lenPar\cot\left(\frac{\theta}{2}\right) \, ,
\end{align}
\end{subequations}
with $r\in [0,\infty]$ and $\theta\in [0,2\pi]$.  Here $\lenPar$ is a tunable parameter that should be set to the typical ``size'' of the object in the radial grid.
As explicitly seen in~\eqref{eqn:spectral-expansion}, this expansion of the function is equivalent to an (even) cosine expansion in the mapped $\theta$ coordinate.
Alternatively, in the coordinate system
\begin{equation}
\label{eq:gl_grid}
 x = \cos\left(\frac{\theta(r)}{2}\right) = \frac{r}{\sqrt{r^2+\lenPar^2}} \, ,
\end{equation}
this is an expansion in the even Chebyshev polynomials.
For further details, see Boyd~\cite{Boyd1989ChebyshevAF}.
Although we will not explore them here, the cotangent mapping can be freely exchanged for other coordinate mappings adapted to a specific problem.

The expansion~\eqref{eqn:spectral-expansion} defines the field at all values $r\in [0,\infty]$ of the radial grid.
However, provided we have adequately resolved $f$ (\ie we have taken $N$ large enough) we can store all of the information contained in the $c_n$'s in $N$ spatial grid points.
For our purposes, it is convenient to choose the (mapped) Gauss-Chebyshev collocation points
\begin{subequations}
\begin{align}
  r_i &= \lenPar\cot\left(\frac{\theta_i}{2}\right)  \\
  \theta_i &= \left(N-i+\frac{1}{2}\right)\frac{\pi}{N} \qquad i=1,\dots,N \, .
\end{align}
\end{subequations}
Here is one of the key facts to understand our implementation: even symmetry prevents us from enforcing Neumann boundary conditions. 
In addition to this, it is not necessary to evaluate singular terms in the equations of motion. 
The cotangent mapping can be freely exchanged to other coordinate choices specific to the problem.
Therefore, a target function $f$ expandable in the even Chebyshev basis can be expressed as
\begin{align}
f(r_i) = \sum^{N-1}_{n=0}c_n\cos\left[n\theta(r_i)\right]\,,
\label{eq:exp_cos}
\end{align}
where $\theta(r_i)$ follows from \eqref{eq:gl_grid}
\begin{align}
\theta(r_i) = 2\cos^{-1}\left(\frac{r_i}{\sqrt{r^2_i+\ell^2}}\right) \, .
\label{eq:angle_tran}
\end{align}
As every spectral expansion, interpolation to points out of the collocation grid only needs from the expansion coefficients. 
For example, we can trace the value of $f$ at the origin by computing 
\begin{align}
f(r=0)=\sum^{N-1}_{n=0}(-1)^n c_n\,. 
\end{align}
The expansion also allows computing the derivative of the target function
\begin{align}
\frac{df(r_i)}{dr} = \sum^{N-1}_{n=0}\left[-n c_n\frac{d\theta(r_i)}{dr}\right]\sin\left[n\theta(r_i)\right] \, .
\label{eq:exp_sin}
\end{align}
We immediately identify the term in square brackets as the sine transform ($\mathcal{F}_{\mathrm{sin}}$) of the radial derivative.
Knowing that $d\theta(r_i)/dr=-2\ell(r_i^2+\ell^2)^{-1}$, the radial derivative is also equivalent to the following inverse sine transform
\begin{align}
\frac{df(r_i)}{dr} = \frac{2\ell}{\ell^2+r^2_{i}}\mathcal{F}^{-1}_{\mathrm{sin}}(nc_n)
\label{eq:der_sin}
\end{align} 
if we use \texttt{fftw3} \cite{Frigo:2005zln} to compute cosine (\texttt{FFTW\_REDFT10} --DCT type II) and inverse sine (\texttt{FFTW\_RODFT01}  -- DST type III) 
transforms, the elements of the $nc_n$ array need to be rearranged before applying an inverse sine transform.

As for the time evolution, we used an eighth-order Gauss-Legendre symplectic integrator \cite{10.2307/2003405}, which is the same 
used in \cite{Frolov:2017asg, GalvezGhersi:2021sxs}, where the time step for the evolution is limited by the Courant-Friedrichs-Lewy 
(CFL) condition:
\begin{align}
\Delta t_{\mathrm{CFL}}\approx\Delta x_{\mathrm{min}}\,,
\label{eq:CFL}
\end{align}
where $\Delta x_{\rm min}$ is the smallest spacing between grid points. This condition holds for semi-linear wave equations bounded potentials (and external) 
interactions in the equations of motion.

\enconsbr

\subsection{Perfectly matched layers (PMLs) and equations of motion in flux conservative form}\label{subapp:pml_flux}

Oscillons slowly dissipate energy during their evolution through the emission of outward traveling radiation, as shown in~\figref{fig:diag_n_mod} and~\figref{fig:large_alpha}, for example.
To maintain numerical accuracy, this radiation must be dealt with either by removing it from the simulation volume, or increasing the spatial resolution at large radii.
Since oscillons are long-lived, we want to integrate for extended periods of time.
Therefore, using the latter approach would require an inordinate number of grid points, resulting in a huge memory requirement and making the parameter scans computationally intractable.
Instead, we will follow the former approach and force the radiation to damp away at large radii through the use of perfectly matched layers (PMLs).
In this subsection, we will outline our numerical implementation of PMLs. We follow the procedure developed in Frolov et al~\cite{Frolov:2017asg}, which extends the PML approach presented 
in Johnson~\cite{Johnson-2108-05348}.
We begin with the equations of motion
\begin{subequations}
\begin{align}
  &\frac{\ud\phi}{\ud t} = \Pi\\
  &\frac{\ud\Pi}{\ud t}  = \left[\frac{\pd^2}{\pd r} + \frac{\varepsilon}{r}\frac{\pd}{\pd r} \right]\phi - \frac{\ud V}{\ud\phi}(\phi)
\end{align}
\end{subequations}

\hitpml

The procedure introduces two auxilliary fields, denoted here by $v\equiv r^{-1}\pd\phi/\pd r$ and 
\begin{align}
\frac{\ud w}{\ud t}\equiv (\varepsilon+1)v-\frac{\ud V}{\ud\phi}(\phi)\,,
\end{align} 
which absorbs the potential derivatives and the dimensional deformations. The definition of $v$ preserves the parity of the fields evolving in the lattice without introducing singular behavior. 
After the field redefinition $\Pi\rightarrow\Pi+w$, the equations written in flux conservative form now read as
\begin{subequations}
\begin{align}
  &\frac{\ud\phi}{\ud t} = \Pi-w\, \\
  &\frac{\ud\Pi}{\ud t} = r\frac{\pd v}{\pd r}\label{eq:dpidt_rad_ap_mod2}\, \\
  &\frac{\ud w}{\ud t} = (\varepsilon+1)v-\frac{\ud V}{\ud \phi}(\phi)\, \\
  &\frac{\ud v}{\ud t} = \frac{1}{r}\frac{\pd}{\pd r}(\Pi-w)\label{eq:cons_2_ap}\,,
\end{align}
\end{subequations}
where the last equation imposes the commutation of time and radial derivatives. The implementation of perfectly matched layers is based on the 
analytical continuation of the spacetime coordinates domain, resulting in the deformation of the radial derivative 
\begin{align}
\frac{\pd}{\pd r} \rightarrow \left(1+\frac{\gamma(r)}{\pd_t}\right)^{-1}\frac{\pd}{\pd r}
\end{align}
where $\gamma(r)$ is a function with compact domain, which is zero along the simulation length and behaves as a 
smooth incline in the last nodes of the grid, acting as an absorbing layer. 

\ampmodtwod

Once the derivative redefinition is applied in the Eqns.~\eqref{eq:dpidt_rad_ap_mod2} and \eqref{eq:cons_2_ap}, we find the set of equations to simulate
\begin{subequations}
\begin{align}
  &\frac{\ud\phi}{\ud t} = \Pi-w\,\label{eq:dphidt_rad_fin} \\
  &\frac{\ud\Pi}{\ud t} = r\frac{\pd v}{\pd r}-\gamma\Pi\, \\
  &\frac{\ud w}{\ud t} = (\varepsilon+1)v-\frac{\ud V}{\ud \phi}(\phi)\, \\
  &\frac{\ud v}{\ud t} = \frac{1}{r}\frac{\pd}{\pd r}(\Pi-w)-\gamma v\label{eq:cons_2_fin}\,,
\end{align}
\end{subequations}
where the cost is the introduction of two auxilliary variables, with two corresponding evolution equations. Writing flux conservative equations for the one-dimensional deformed 
system in Sec.~\ref{sec:SG_monodromy} and the time-dependent dimensional transitions in Sec.~\ref{sec:SG_td} is not substantially 
different from the procedure described above. In the latter case, we also need one more equation corresponding to an auxiliary variable 
$\psi$, defined to evolve as
\begin{align}
\frac{\ud\psi}{\ud t} = \dot{\varepsilon}_t\Pi\,,
\label{eq:aux_psi}
\end{align}
to absorb the inelastic collision term in \eqref{eq:dpi_dt_vard}. After redefining the time derivative by $\piDim+\psi\ln r-w\rightarrow 
\piDim$, the equations of motion with absorbing boundary layers can be written in their final form, 

\begin{align}
&\frac{\ud\phi}{\ud t} = \Pi+w-\psi\ln r,\label{eq:dphidt_rad_fin_t}\\
&\frac{\ud\Pi}{\ud t} = r\frac{\pd v}{\pd r}-\gamma\Pi\,,\label{eq:dpidt_rad_fin_t}\\
&\frac{\ud v}{\ud t} = \left[\frac{1}{r}\frac{\pd}{\pd r}\right](\Pi+w-\psi\ln r)-\gamma v\,,\label{eq:cons_1_fin_t}\\
&\frac{\ud w}{\ud t} = (\varepsilon_t+1)v-\frac{\ud V}{\ud \phi}(\phi)\,,\label{eq:cons_2_fin_t}\\
&\frac{\ud\psi}{\ud t} = \dot{\varepsilon}_t(\Pi+w-\psi\ln r)\,,\label{eq:cons_3_fin_t}
\end{align}
which is the extension of Eqns.~(\ref{eq:dphidt_rad_fin}--\ref{eq:cons_2_fin}) for the case of time-dependent dimensional transitions. 

For consistency, we evaluate energy conservation in the simulation length by considering the case $\varepsilon=0$ (i.e., during the 
oscillation of standing breathers). In Fig.~\ref{fig:en_cons}, we plot the energy conservation residuals $|\Delta E|\equiv |E(t)-E(t_0)|$ 
observing that conservation holds at the level of round-off errors in double precision and residuals do not grow in time. Additionally, it is important to show the effect 
of PMLs as filters of radiation escaping the simulation length. To do so, we allow the propagation of a free Gaussian wavepacket following 
the one-dimensional wave equation (with no potential) and compute its scalar flux 
\begin{align}
\mathcal{J}(t,r)\equiv \mu^2\frac{\pd\phi}{\pd r}\frac{\pd\phi}{\pd t}\,,
\label{eq:scalar_flux}
\end{align}
in the simulation domain. In Fig.~\ref{fig:hit_pml}, we observe the absorption of the Gaussian peak ``fired'' directly towards the PML. Reflected 
scalar flux is several orders of magnitude smaller compared to the ingoing flux, and becomes even smaller with subsequent reflections. A 
closer look at the red region shows that the solution decays progressively as it goes through the absorbing layer. Luckily, the cases of study 
do not involve (a) inhomogeneous media, or (b) angle-dependent absorption where PMLs are prone to fail. In the case of artificial numerical 
reflections, the safest way to proceed is to increase the resolution in the nodes where the absorbing layers are located (as suggested in 
\cite{Johnson-2108-05348}). This is beneficial to resolve the tail structure and its frequency peaks. 

\multires

\subsection{Intermittent expansion and contraction of the oscillon core in two-dimensional solutions}\label{app:2d_modul}

The presence of modulation in the oscillons amplitude is not an artifact of using radial equations of motion. 
Let us evaluate the two-dimensional equations of motion, given by  
\begin{align}
&\frac{\ud \phi}{\ud t} = \Pi\label{eq:phi_t_2d}\,,\\
&\frac{\ud \Pi}{\ud t} = \frac{\pd^2\phi}{\pd x^2}+
\frac{\pd^2\phi}{\pd y^2}-\mu^2\phiScl\sin\left(\frac{\phi}{\phiScl}\right)\label{eq:pi_t_2d}\,,
\end{align}
For these results, we use an independent piece of code working with 
periodic boundary conditions in a two-dimensional simulation box, being this sufficiently large to avoid interactions 
with classical radiation. In the right panel of Fig.~\ref{fig:osc_2d_amp_mod}, we show the evolution of a solution with the 
breather-like initial conditions in \eqref{eq:sg_new_param} projected in the $y=0$ plane, choosing $\alpha=1$ (i.e., without 
modifying the initial amplitude), $\theta_0=0$ and $\omegaini=0.63\mu$ to fix the initial 
breather's shape. The oscillon's core undergoes intermittent periods of contraction and expansion 
noticeable in distances comparable to the size of the object. Intermittent behavior does not seem to 
support energy equipartition, since during this phase radiation can be trapped and injected again before 
escaping the core. In the left panel of the same figure, we show amplitude modulation for the same initial 
breather parameters, which is also visible in the perturbative regime discussions in subsection \ref{subsec:pert_dim_def}. The 
existence of intermittent phases of contraction and expansion of the oscillon core (represented 
by amplitude modulations) has been tested in two independent numerical setups. Therefore, it is 
unlikely that these are consequences of some numerical artifact, or some long-time growing instability. 
Apart from this consistency check, we can also use the 2D solutions of Eqns.~\eqref{eq:phi_t_2d} 
and \eqref{eq:pi_t_2d} at constant initial phase $\theta_0$ to sample over different values of initial 
frequency $(\omegaini)$. We noticed that the modulation frequency of the amplitude envelope 
(similar to the red curve in right panel of Fig.~\ref{fig:sols_n_freq}) becomes larger as $\omegaini$ 
reduces, being this consistent with our observations in the perturbative regime. In principle, such 
a frequency can be treated as an additional diagnostic parameter, which is measurable and can 
be sampled over the span of initial parameters to be connected with the imaginary part of the frequency 
(if such a connection exists), which is the oscillon's decay rate. We will study its connections 
to the dynamical state of amplitude modulated solutions in a future project. 

\convrmax
\tabresolutions

\subsection{Convergence Tests}\label{subapp:conv_test}

In this section, we perform convergence tests for the amplitude modulated solution depicted in the middle and right panels of 
Fig.~\ref{fig:diag_n_mod} for $\varepsilon=0.75$. Considering that dimensional deformations stretch the breathers' length to form 
much wider oscillons, the numerical implementation needs at least $10^3$ nodes to resolve oscillons with sufficient dynamical range.  
We construct four radial semi-infinite domains following the Gauss-Chebyshev collocation grids for the length scale 
$\ell=10^2\mu^{-1}$, and considering the resolutions reported in Table \ref{tab:resolutions}, where the CFL time scale 
$\Delta t_{\mathrm{CFL}}$ follows from the condition in Eq.~\eqref{eq:CFL}, which follows from the dispersion relation for semi-linear wave equations with bounded potentials. 
The number of nodes in the lowest possible resolution is still considered to be ``high'' for spectrally accurate one-dimensional simulations.
Nevertheless, resolving radiation at large radii still requires enough resolution to be correctly attenuated by the PMLs.
In the left panel of Fig.~\ref{fig:multi_res}, we plot the spectral coefficients (found by computing the 
cosine transform of the solution) in terms of the number of nodes for all of the resolutions at fixed time $t=10^4\mu^{-1}$. We observe that 
keeping high frequency coefficients in the same magnitude as round-off errors requires a large number of collocation nodes. High frequency 
coefficients appear during the initial transient phase as scattering modes decay with fractional powers of the distance.  Spectral 
coefficients coincide for the first hundred nodes, which are sufficient to resolve the core, as we can observe in the right panel. Up to 
some extent, this justifies the invariance of the oscillation frequency maps in Fig.~\ref{fig:ct_eps_freq} with changes in the resolution. 
Solutions keeping all of the high-frequency terms with powers below machine precision are computationally expensive, needing at least 
8-10 times more k-modes to be fully resolved. In the right panel, we observe the field configuration as a function of the radial coordinate. 
The solutions at different resolutions (interpolated to the lowest resolution spatial grid) look almost identical: it is only at the origin where 
one-percent level errors can be assessed. 

An important feature we can extract from the right panel of Fig.~\ref{fig:multi_res} is that we can evaluate the convergence errors by 
considering the field values at the origin (or the closest point in the collocation lattice) at different resolutions. Considering the solution at the 
highest resolution as a reference, we can subtract the solutions from the other resolutions and evaluate the differences as functions of time. 
To compare the outcomes from different spatial resolutions, the time step $\Delta t=\Delta t^{\mathrm{mid-max}}_{\mathrm{CFL}}/8$ is 
kept as a constant in all the resolutions to avoid inaccuracies due to time interpolation. In Fig.~\ref{fig:conv_r_max}, we plot the 
difference between field configurations obtained at different spatial resolutions. Observing that numerical errors reduce as we use 
more modes to resolve the oscillating configurations, this figure is a piece of evidence indicating numerical convergence. Moreover, it 
is important to remark that reported errors do not grow in time for the highest resolutions.  As expected, for the lowest resolutions errors 
tend to increase when the core expands and contracts, which is the defining feature of amplitude modulated solutions.

\input{SG_dim_def.bbl}
\end{document}

%% file: figures.tex
\newcommand{\solsnfreq}{%
\begin{figure*}
\centering
\subfigure{
\includegraphics[width=.28\textwidth]{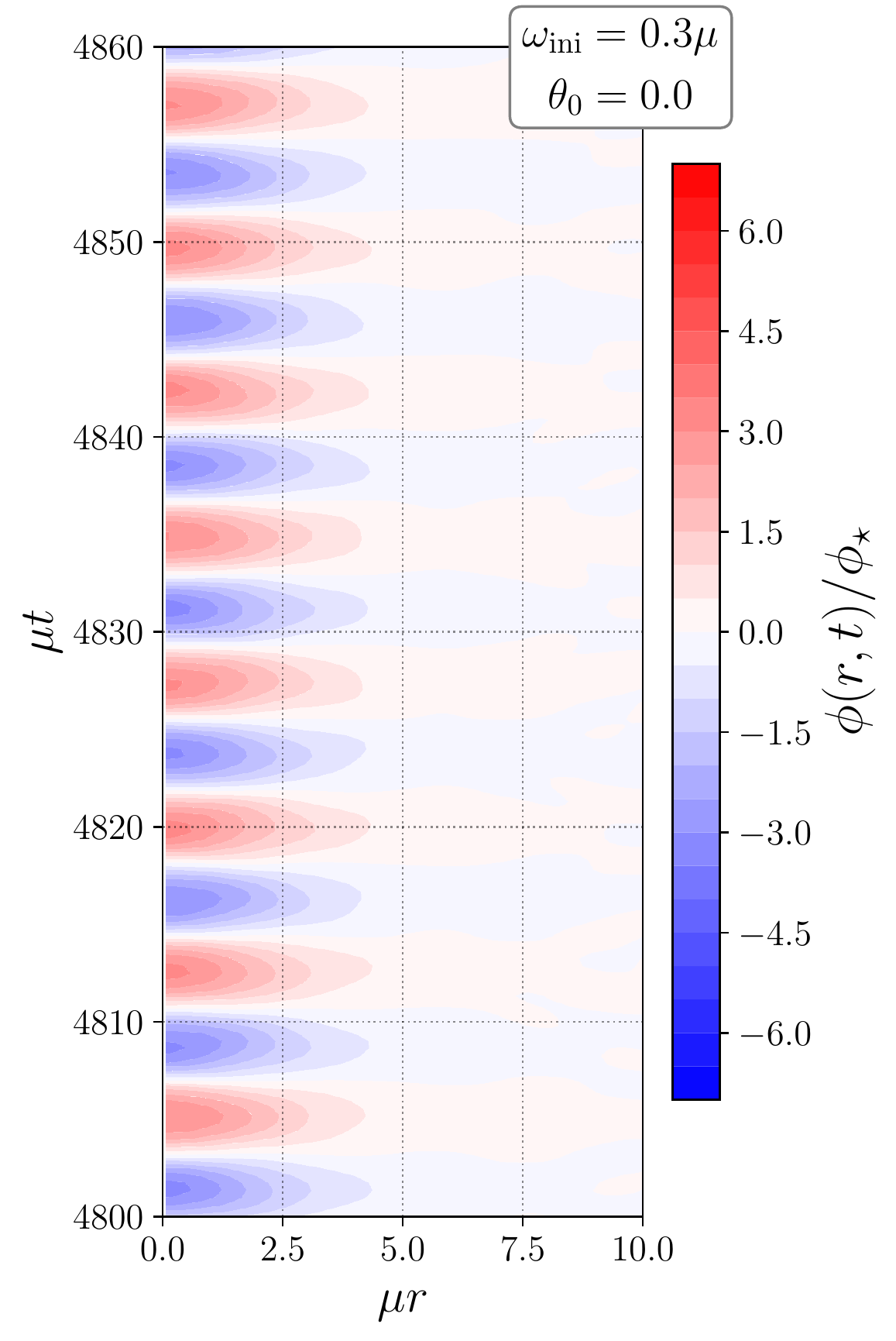}} \,
\subfigure{
\includegraphics[width=.28\textwidth]{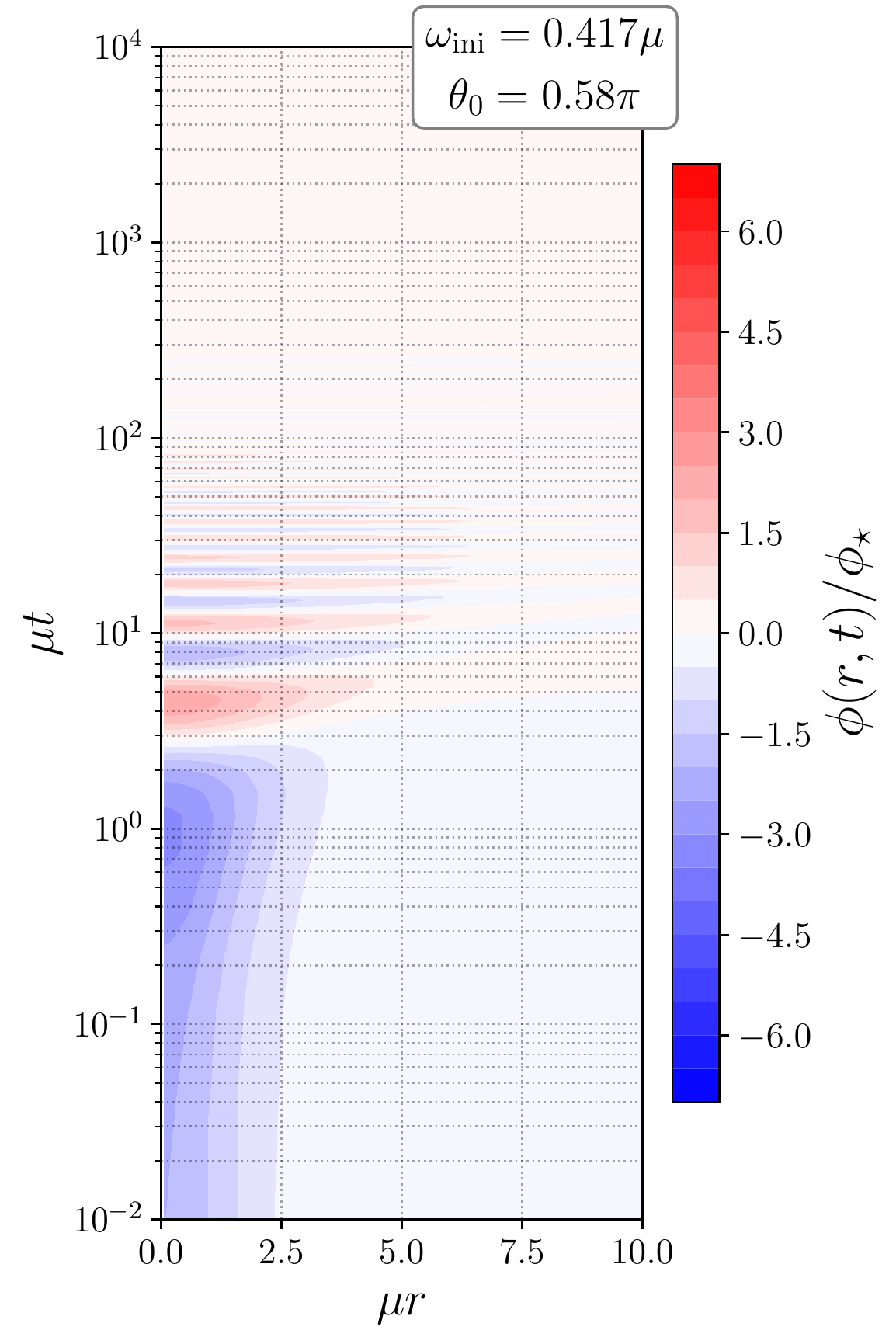}} \,
\subfigure{
\includegraphics[width=.268\textwidth]{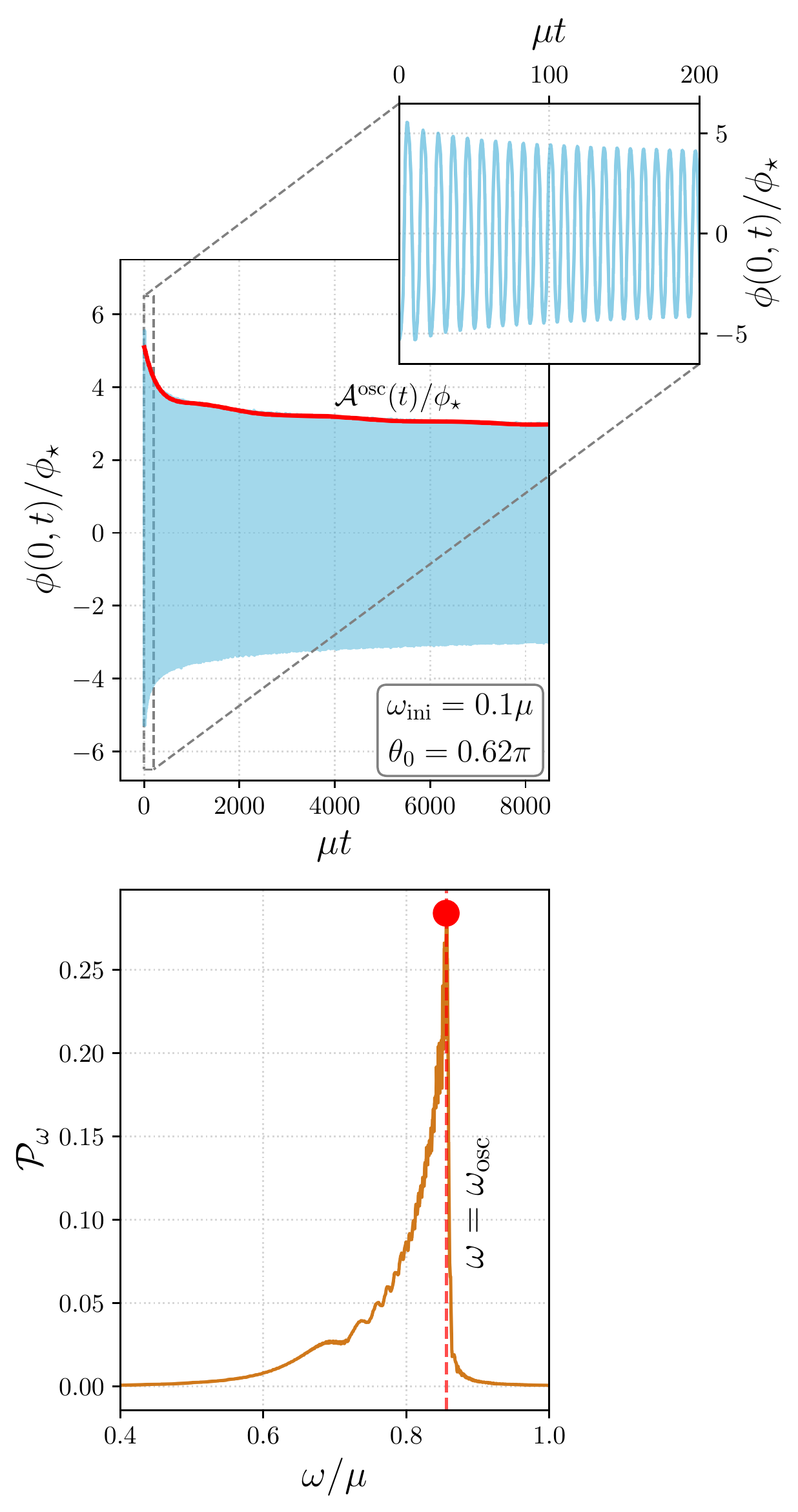}} 
\caption{\label{fig:sols_n_freq} Showing the spatial structure, time evolution of oscillon cores and 
determination of the oscillation frequency ($\omegaosc$) for $\varepsilon=0.75$. 
Left panel:  Typical evolution of the radial profile for a long-lived oscillon, deformed from a SG breather 
with initial frequency $0.3\mu$ and no phase. Middle panel: Radial profile of a decaying 
oscillon (observe the time axis in logarithmic scale). Right panel (top):  Time evolution of the oscillon 
evaluated at $r=0$ to determine the dominant frequency and the amplitude $\mathcal{A}$.
Right panel (bottom): From the temporal power spectrum $\mathcal{P}_{\omega}$ of the top panel, 
we determine $\omegaosc/\mu$ to be the angular frequency with the highest peak 
(marked by the red dot) in power.}
\end{figure*} 
}

\newcommand{\figparam}{%
\begin{figure*}[t!]
\centering
\includegraphics[width=1.0\textwidth]{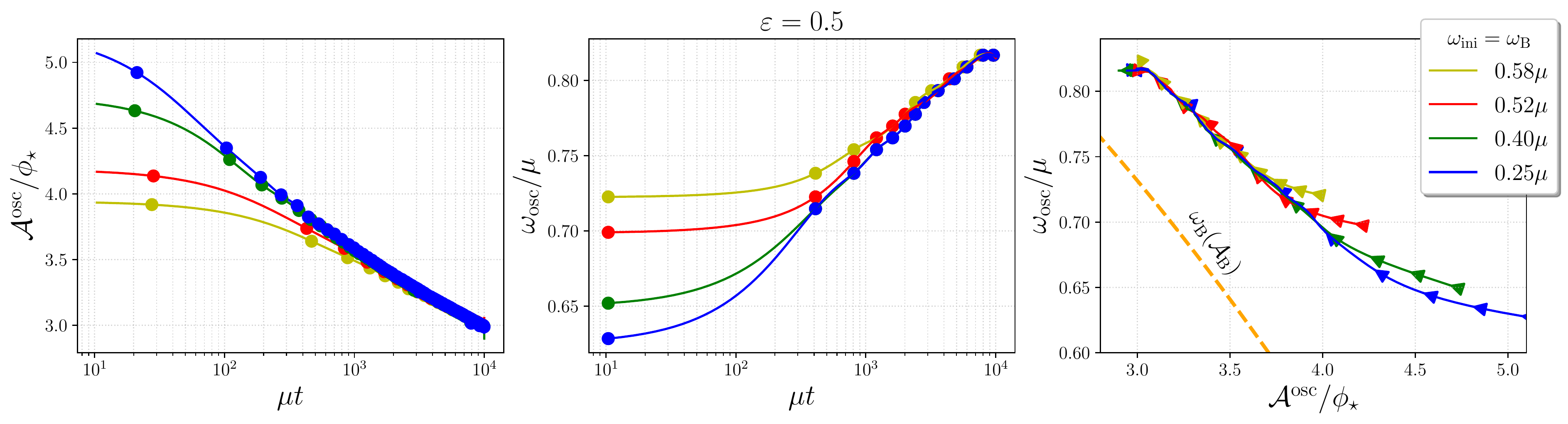}
\caption{\label{fig:param_flow} Parameter flow at $\varepsilon=0.5$ for four initial breather 
frequencies. In the first two panels, we show the evolution of $\mathcal{A}$ (i.e., the red envelope 
in Fig.~\ref{fig:sols_n_freq}) and the oscillon's frequency $\omegaosc$ as functions of time. 
We observe that both parameters evolve very quickly towards an attractor. Once the convergence 
occurs, the flows slow down but do not stop. In the last panel, we see that the flow in parameter 
space collapses into a line, aligned along the breather flow line (in orange, dubbed as $\varepsilon=0$) 
given by Eq.~\eqref{eqn:breather-amp}. The arrows only represent the direction of time, since the speed can 
be inferred from the first two panels of this figure.}
\end{figure*}
}

\newcommand{\simbrosc}{%
\begin{figure}[t!]
\centering
\includegraphics[width=0.5\textwidth]{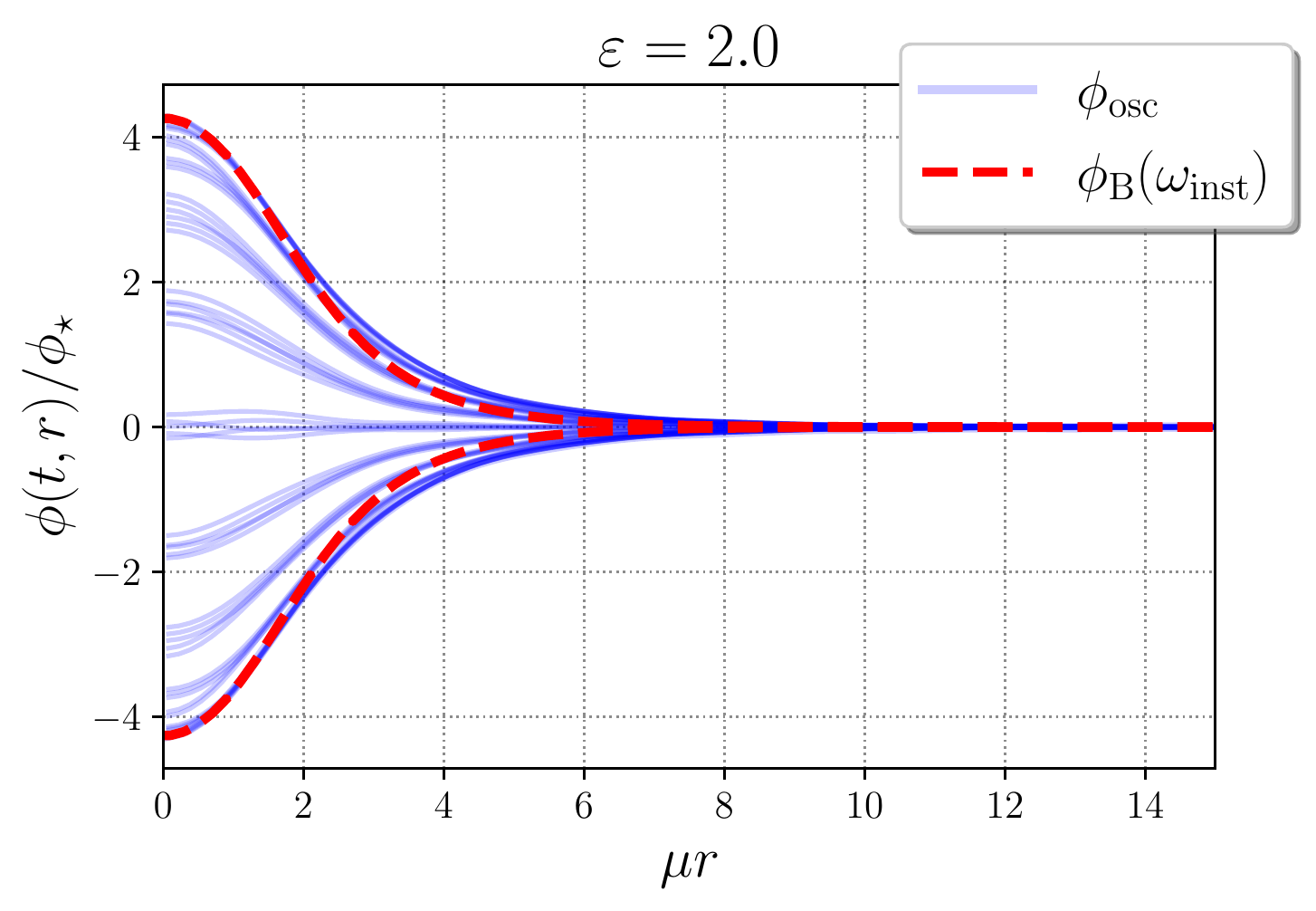}
\caption{\label{fig:sim_br_osc} Comparing the radial profile evolution of a 3D oscillon of the sine-Gordon model $(\phi_{\mathrm{osc}})$ 
(in solid blue lines), with a breather profile $(\phi_{\mathrm{B}})$ built at fixed time $\mu t\sim4\times10^2$ with frequency $\omega_{\rm inst}/\mu=0.482$ 
(in red dashed lines). The peaks formed by both solutions have very similar shapes.}
\end{figure}
}

\newcommand{\ctfreqphase}{%
\begin{figure*}
\centering
\subfigure{
\includegraphics[width=.35\textwidth]{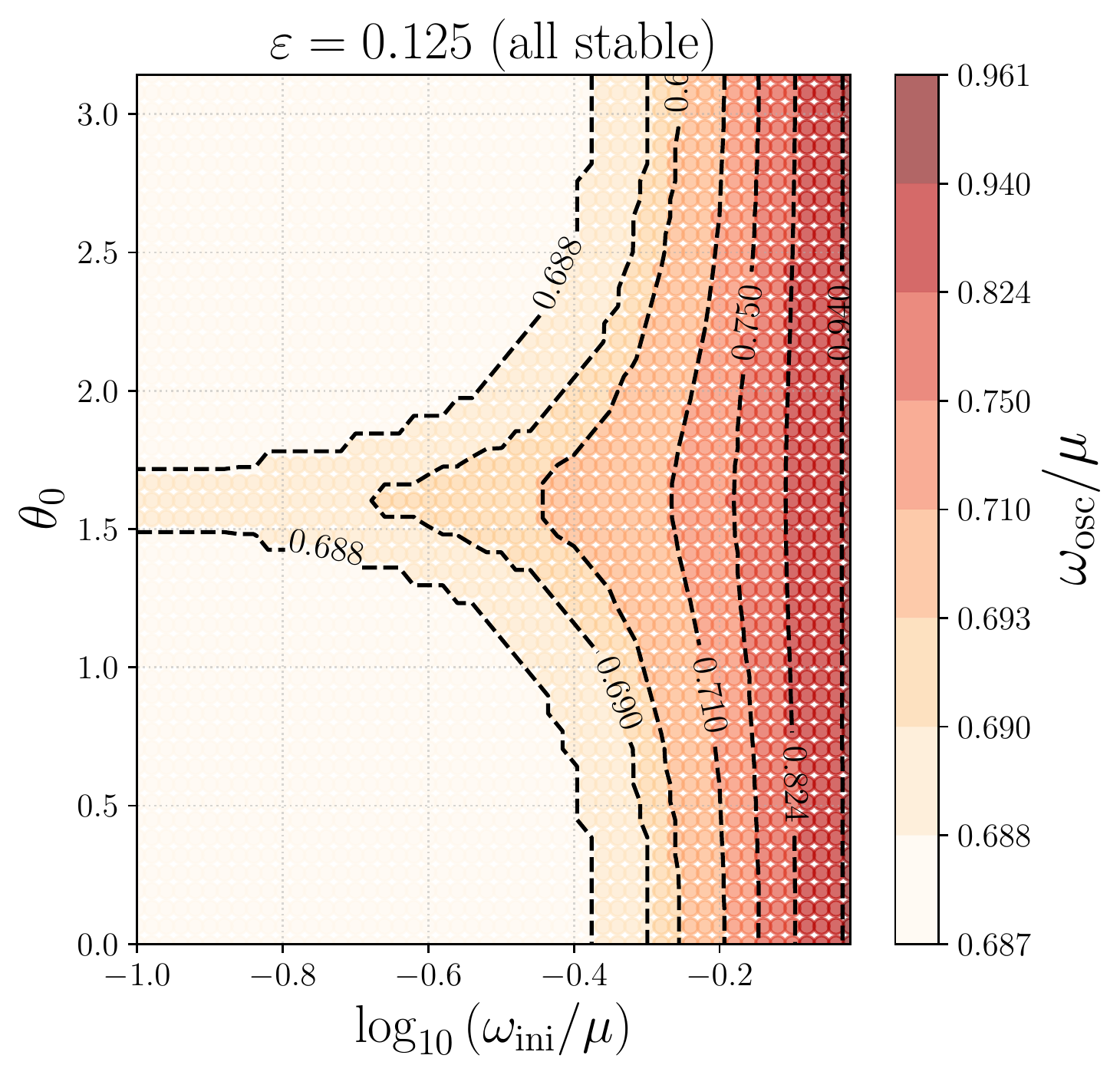}} \,
\subfigure{
\includegraphics[width=.475\textwidth]{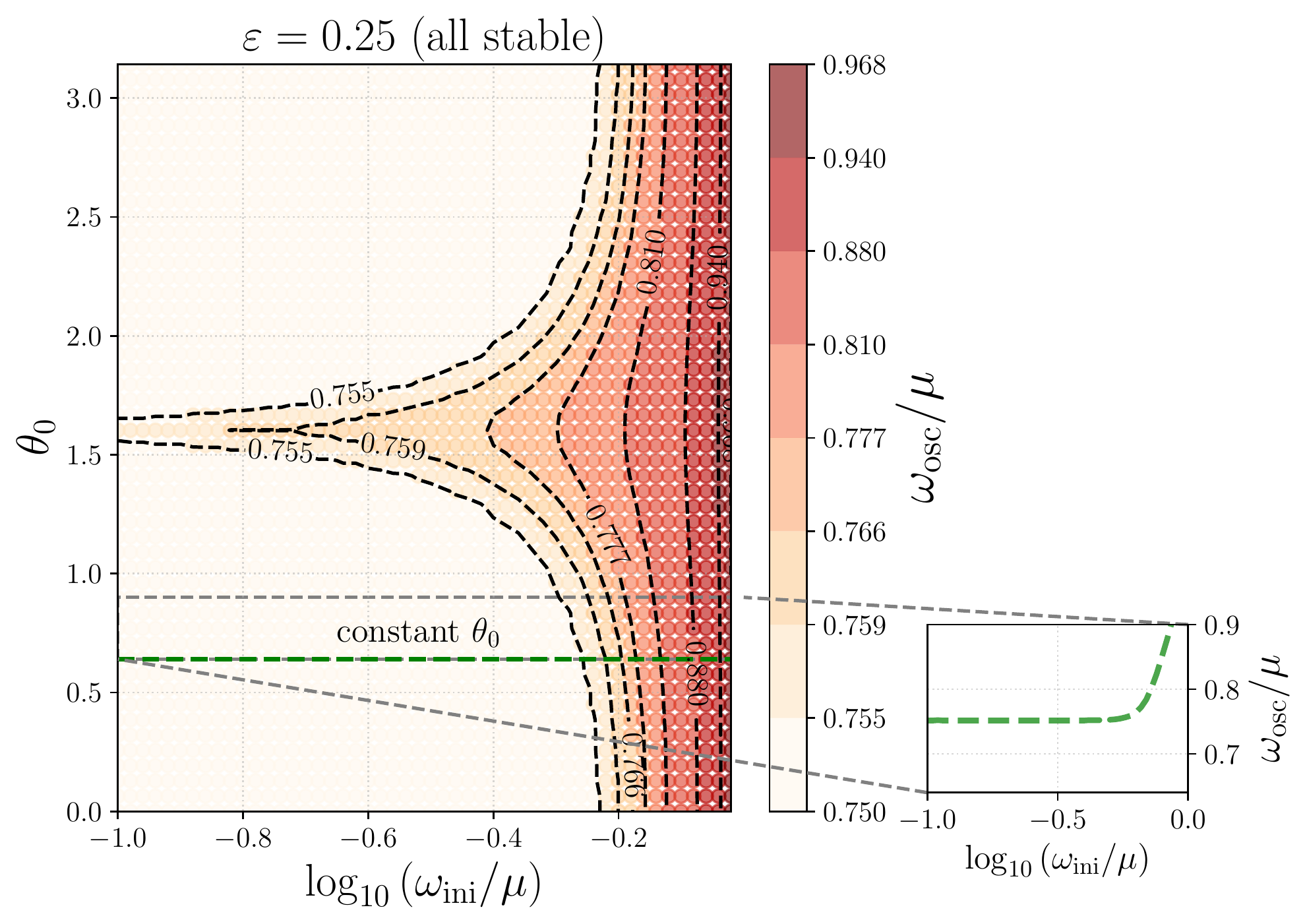}} \,
\subfigure{
\includegraphics[width=.35\textwidth]{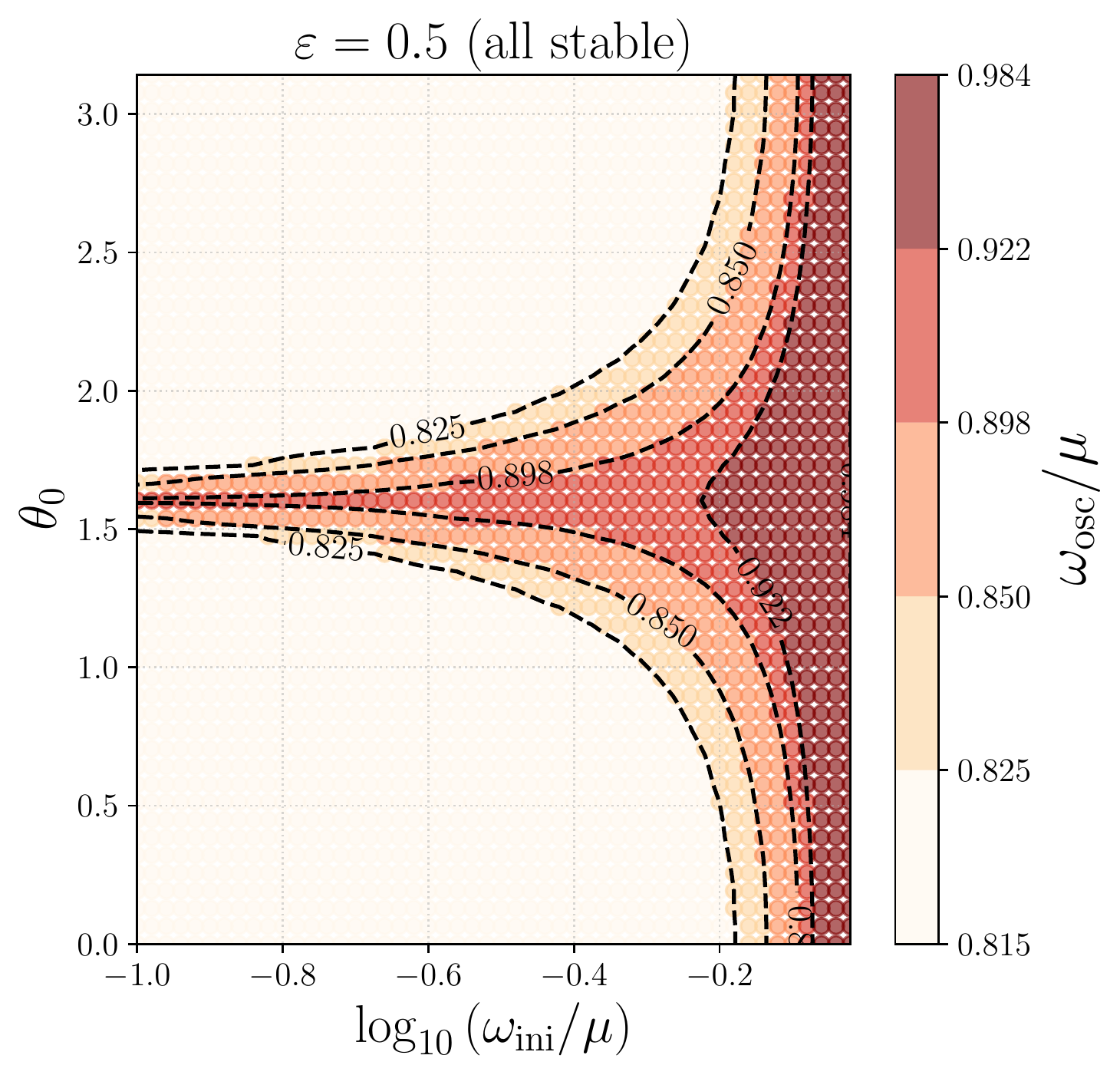}} \,
\subfigure{
\includegraphics[width=.35\textwidth]{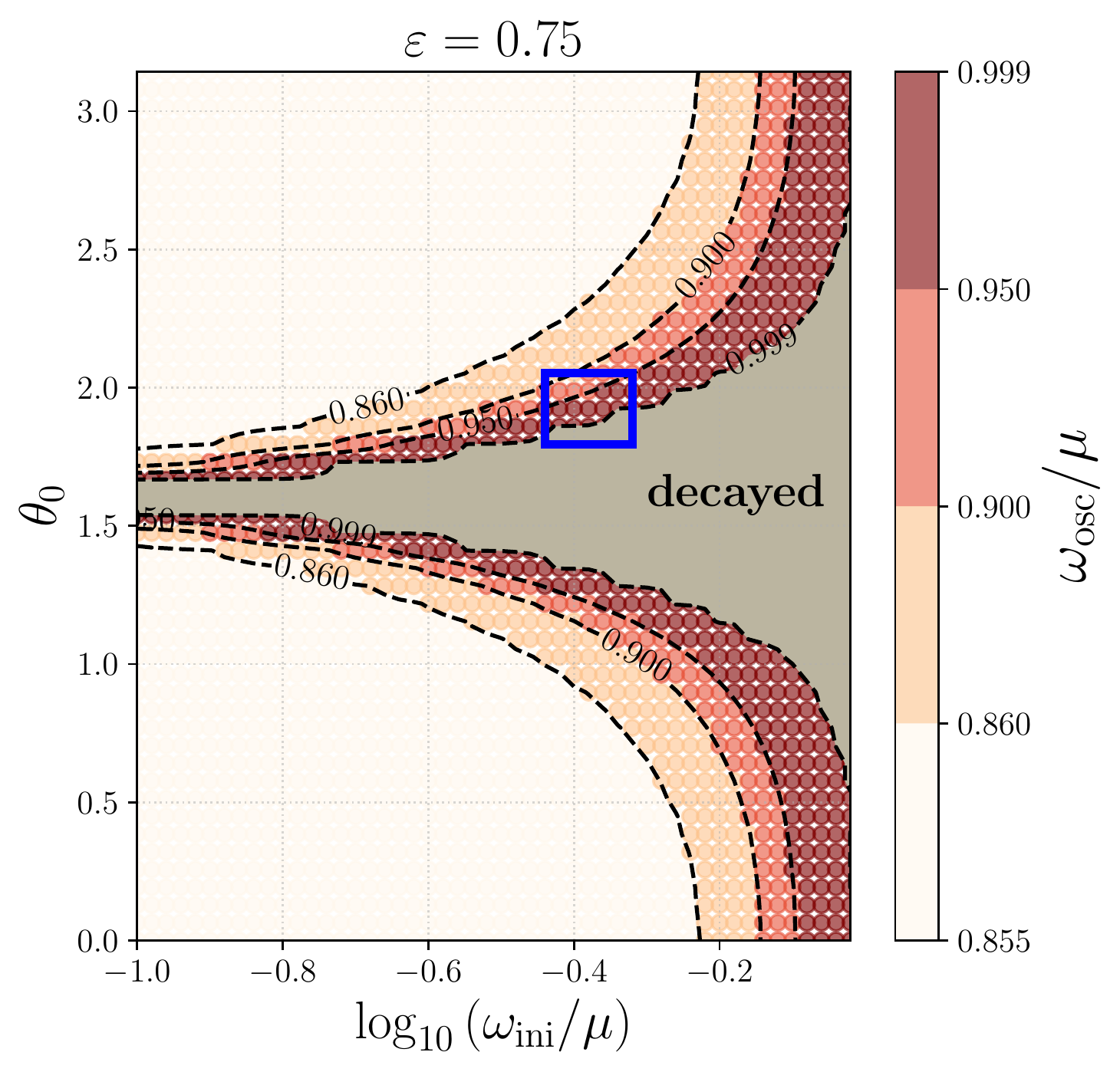}}
\caption{\label{fig:ct_eps_freq} Surfaces showing the oscillon frequency as a function of the phase and
frequency of the initial SG breathers for $\varepsilon = 0.125, 0.25, 0.5$ and $0.75$. Regions within isocontours of oscillation 
frequency are produced from a grid of $50\times 50$ initial configurations of frequencies $(\omegaini)$ 
and phases $(\theta_0)$, uniformly distributed in $\log_{10}(\omegaini/\mu)\in [-1,-0.02]$ and in $\theta_0\in [0,\pi)$. 
The frequency of stable oscillons (colored in ivory in all of the panels) formed by breathers with 
$\omegaini=\omegaB\lesssim 0.3\mu$ increases with $\varepsilon$. For the span of initial conditions considered here, the quantity 
of scenarios where oscillons (with $\omegaosc< \mu$) form tends to decrease as $\varepsilon$ increases. 
Solutions within the transition regions (in red) may have non-trivial modulation in their core during the transient. 
\Figref{fig:diag_n_mod} focuses in the cases enclosed in the blue rectangle ($\varepsilon=0.75$, in the right bottom panel), 
which exhibit time-dependent modulation in their amplitude. In the upper right panel (dubbed as $\varepsilon=0.25$), we plotted 
a green dashed curve as an inset (to the right) to show how the frequency changes for a frequency span at constant phase ($\theta_0=0.64$). 
Observing a constant frequency plateau which extends over the whole span of $\omegaini$ in the ivory region, and breaks as $\omegaini/\mu\rightarrow 1$. 
Throughout the remaining sections of this paper, the dependence of the oscillation frequency in $\omegaini$ transforms 
in various ways to represent the dynamical state of the oscillating field. The upper bound in $\omegaini$ is set to resolve oscillons 
within a simulation box of length $\ell=200\mu^{-1}$.}
\end{figure*} 
}

\newcommand{\diagnmod}{%
\begin{figure*}
\centering
\subfigure{
\includegraphics[width=.42\textwidth]{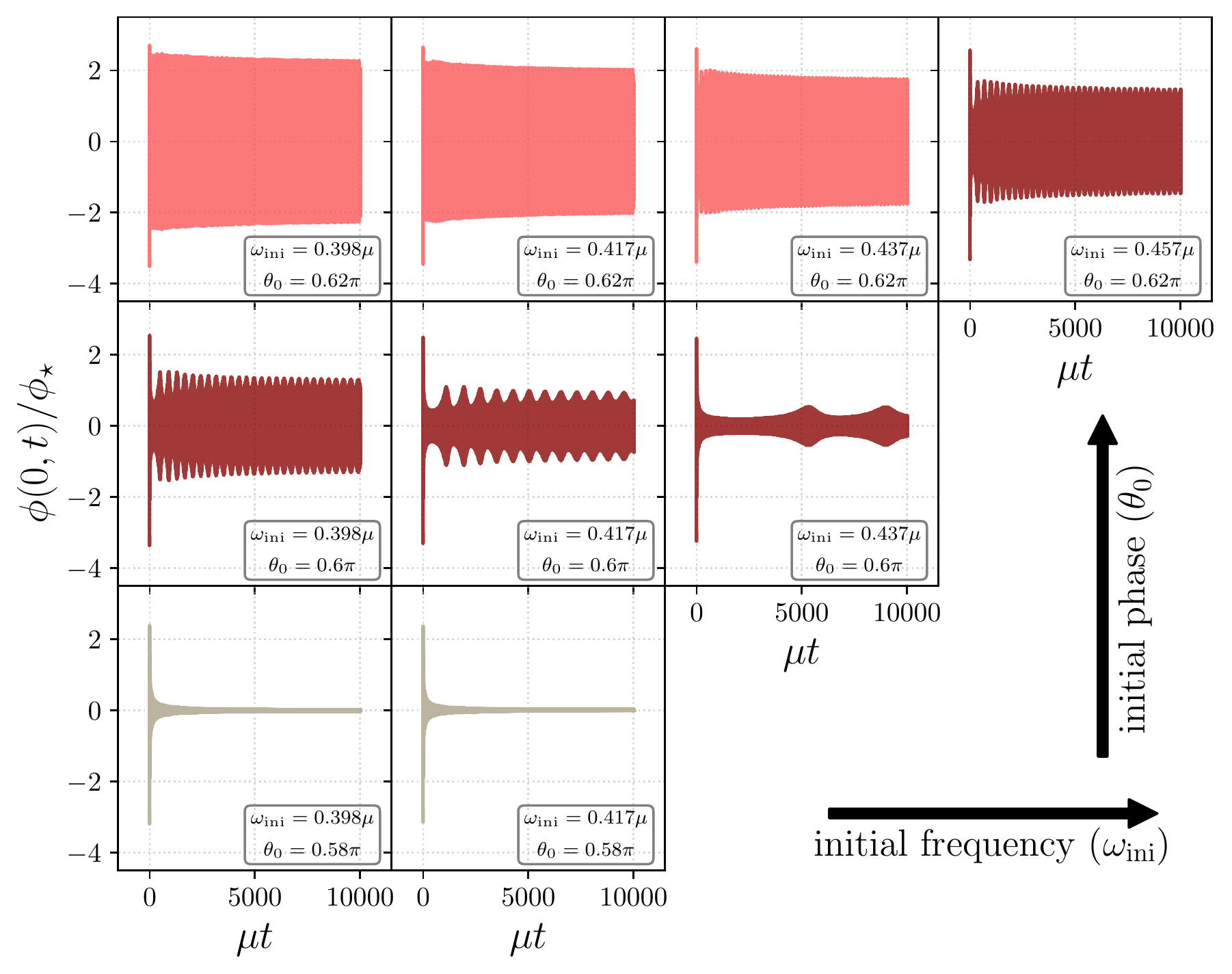}} \,
\subfigure{
\includegraphics[width=.245\textwidth]{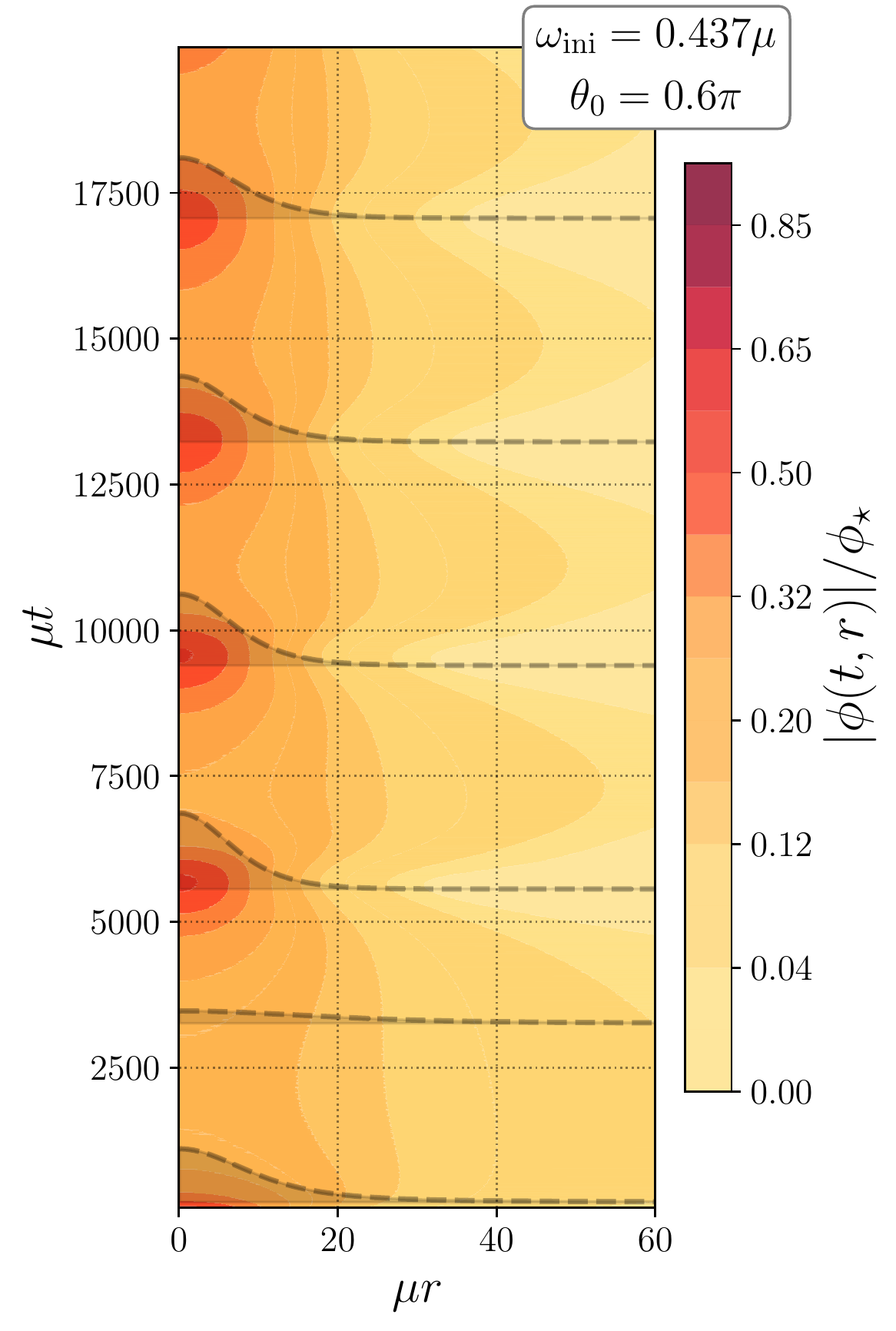}} \,
\subfigure{
\includegraphics[width=.272\textwidth]{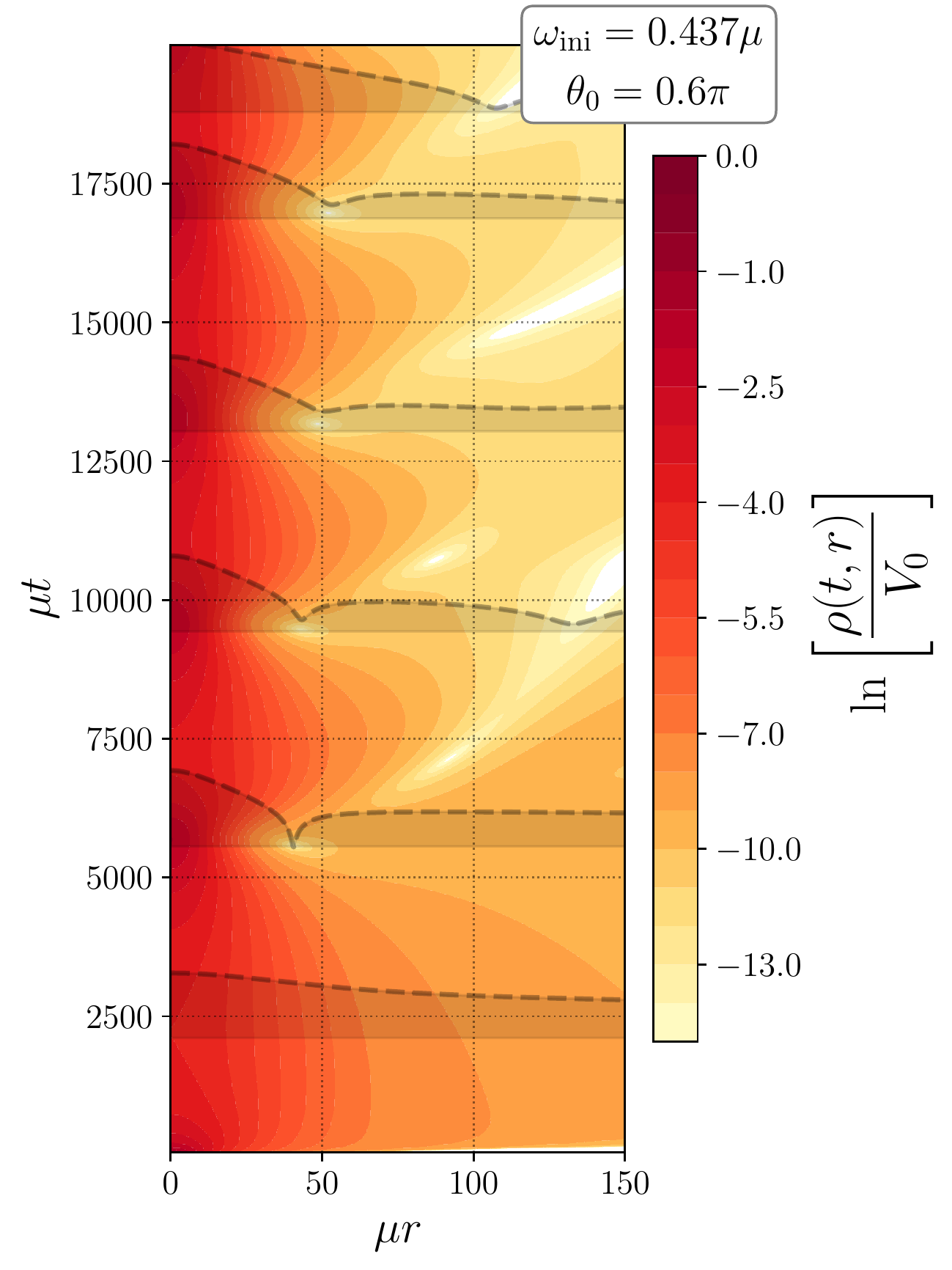}} 
\caption{\label{fig:diag_n_mod} Left panel: Time evolution of $\phi(0,t)/\phi_{\star}$ for some of 
the solutions in the blue rectangle (case $\varepsilon=0.75$ of Fig.~\ref{fig:ct_eps_freq}). 
We are using the same color code as in \figref{fig:ct_eps_freq}, showing that the solutions with modulated amplitude serve as ``transition solutions'' between the stable 
oscillons (in orange) and the fast decaying profiles (in gray). Interestingly, the frequency of the modulating envelope 
reduces as one approaches the unstable solutions.
Central panel: Spatial structure of an amplitude modulated solution for $\varepsilon=0.75$. Amplitude modulation is associated to periodic phases of contraction and expansion 
of the oscillon core. Right panel: Energy density as a function of radius and time for the solution in the middle panel.
Modulation occurs as energy leaves the core in a discrete number of bursts. In the middle and right panels, the black dashed lines correspond to constant-time snapshots of the 
field (central panel) and energy density (right panel), rescaled to fit in both panels. Rasterization suppresses most of the high-frequency structures in the evolution of field and 
energy density. Shaded areas below the dashed lines give a qualitative estimate of the field and energy density values. To show the peaks and throughs in the central and right 
panels, time slices in the middle and right panels do not match.}
\end{figure*} 
}

\newcommand{\likelihood}{%
\begin{figure}[t!]
\centering
\includegraphics[width=.48\textwidth]{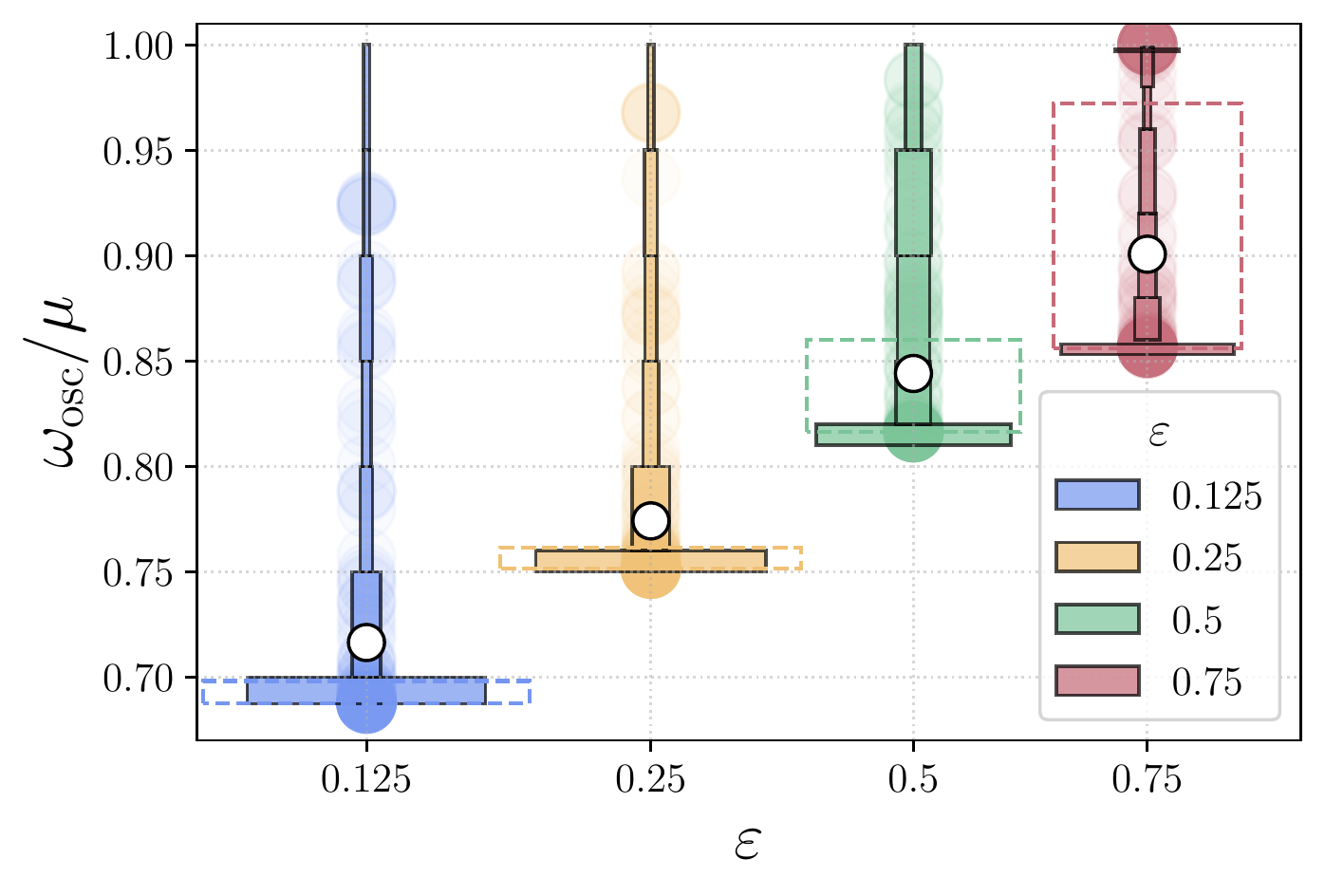}
\caption{\label{fig:lk_eps} After counting the number of solutions in each panel of 
Fig.~\ref{fig:ct_eps_freq}, we observe the deformation of the probability 
discrete probability distributions as a function of $\varepsilon$. Rectangles contain 75$\%$ of the solutions 
sampled for each value of $\varepsilon$. Although it is not shown here, up to 25$\%$ of the samples do not form stable oscillons 
at $\varepsilon=1$. The dots represent the mean frequency of the sample displacing upwards as $\varepsilon$ increases. Semi-transparent 
colored dots are also shown to represent the $\varepsilon$ dependence 
of the oscillation frequency distributions.  
}
\end{figure}
}

\newcommand{\perten}{%
\begin{figure}[t!]
\centering
\includegraphics[width=.48\textwidth]{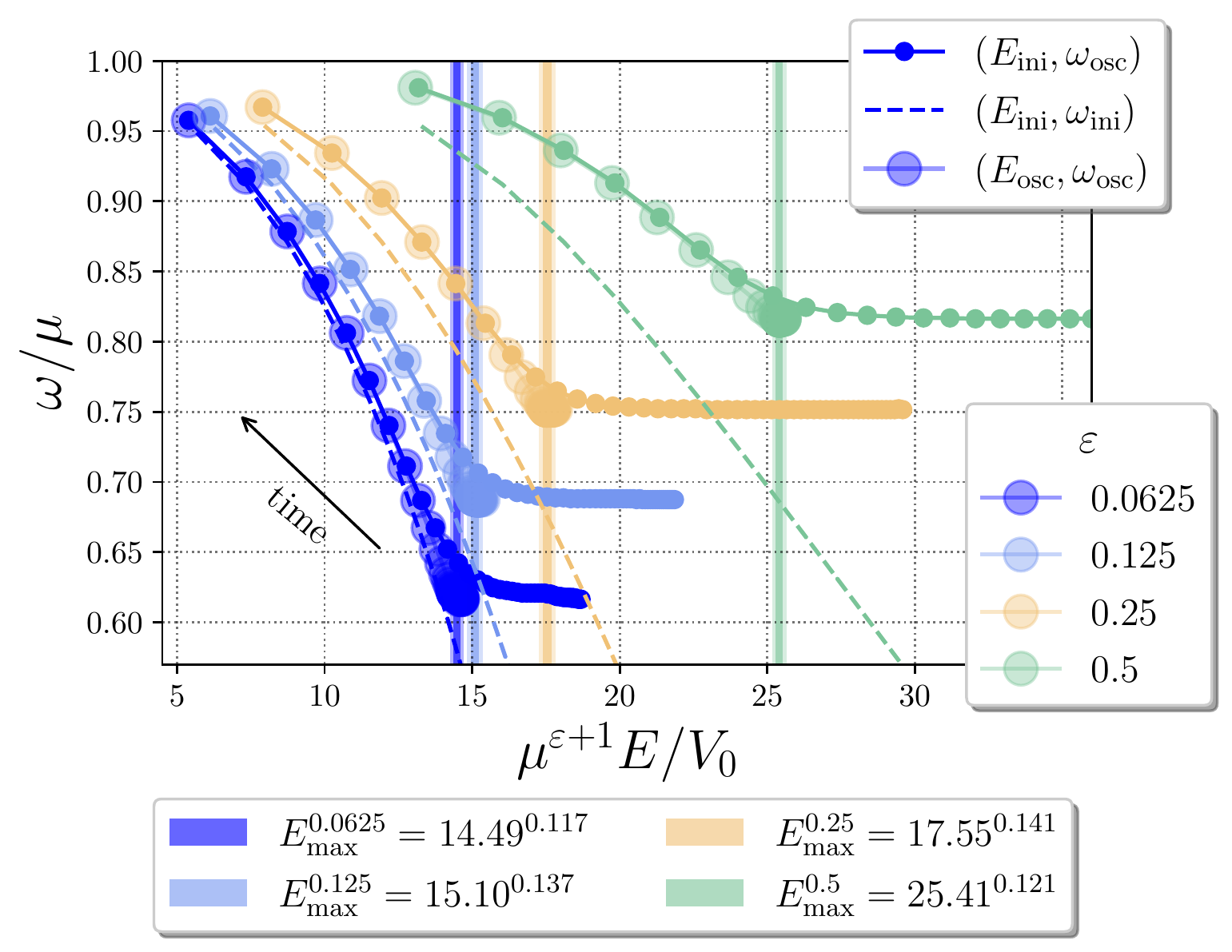}
\caption{\label{fig:pert_en} Convergence of energy and frequency curves (with oscillon energy and oscillation frequency measured at different times) is 
consistent with oscillons being well-represented by SG breathers when $\varepsilon\ll 1$. Oscillation frequency after a few hundred oscillations $(\omegaosc)$ 
is illustrated as a function of the initial energy ($E_{\rm ini}$) in solid dots, while oscillon energy $E_{\rm osc}$ and oscillation frequency $\omegaosc$ is plotted 
using semi-transparent dots. Dashed lines correspond to the initial frequency $\omegaini$ in terms of $E_{\rm ini}$. 
Vertical colored lines indicate the energy (denoted $E_{\mathrm{max}}$ in the legend), where solutions cluster to show the existence of a maximal energy for the oscillon. 
The values of $\varepsilon$ considered to produce the figure are shown in the color legend at the bottom right of the figure.}
\end{figure}
}

\newcommand{\ennonpert}{%
\begin{figure*}
\centering
\subfigure{
\includegraphics[width=.42\textwidth]{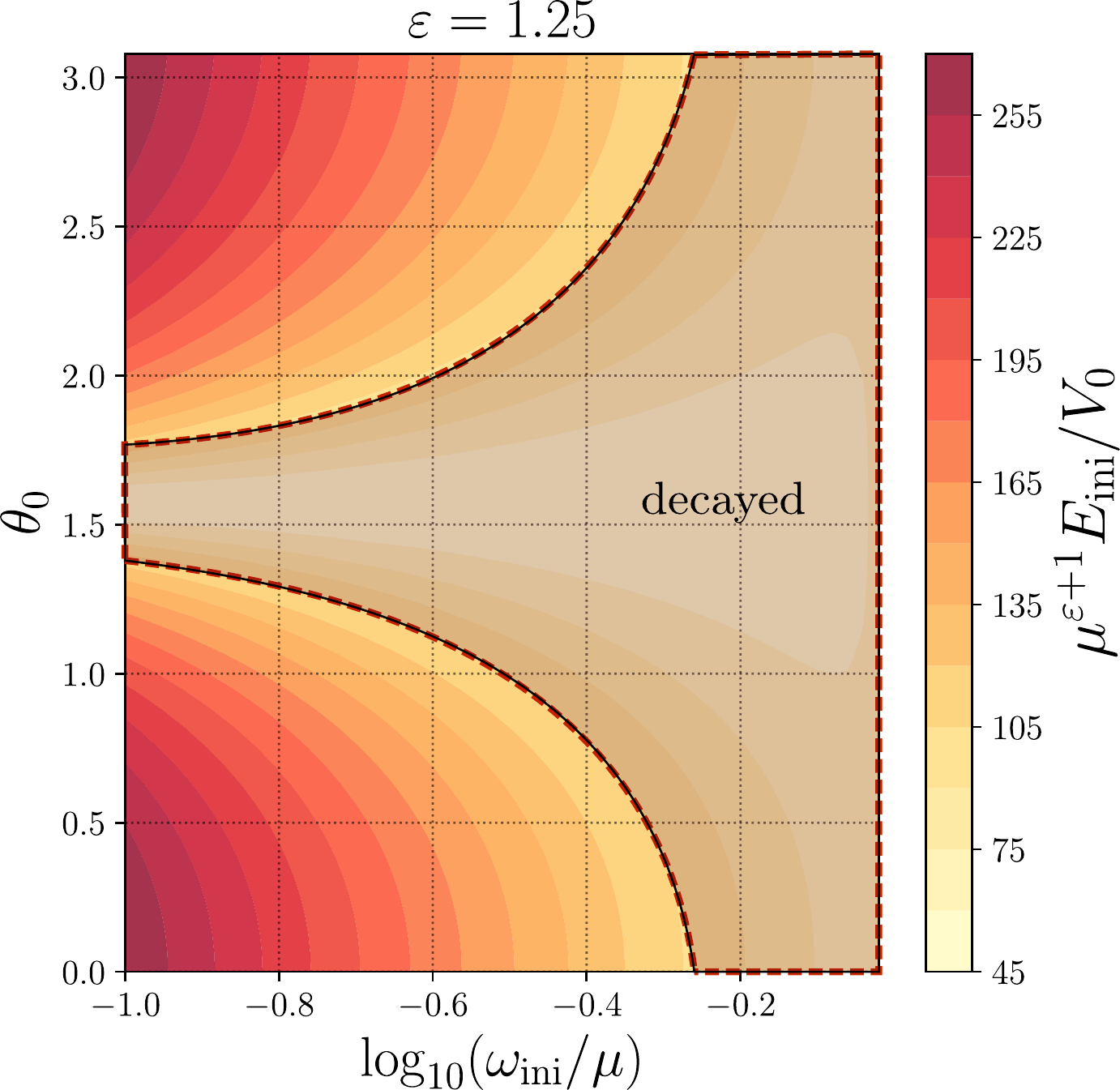}} \,
\subfigure{
\includegraphics[width=.43\textwidth]{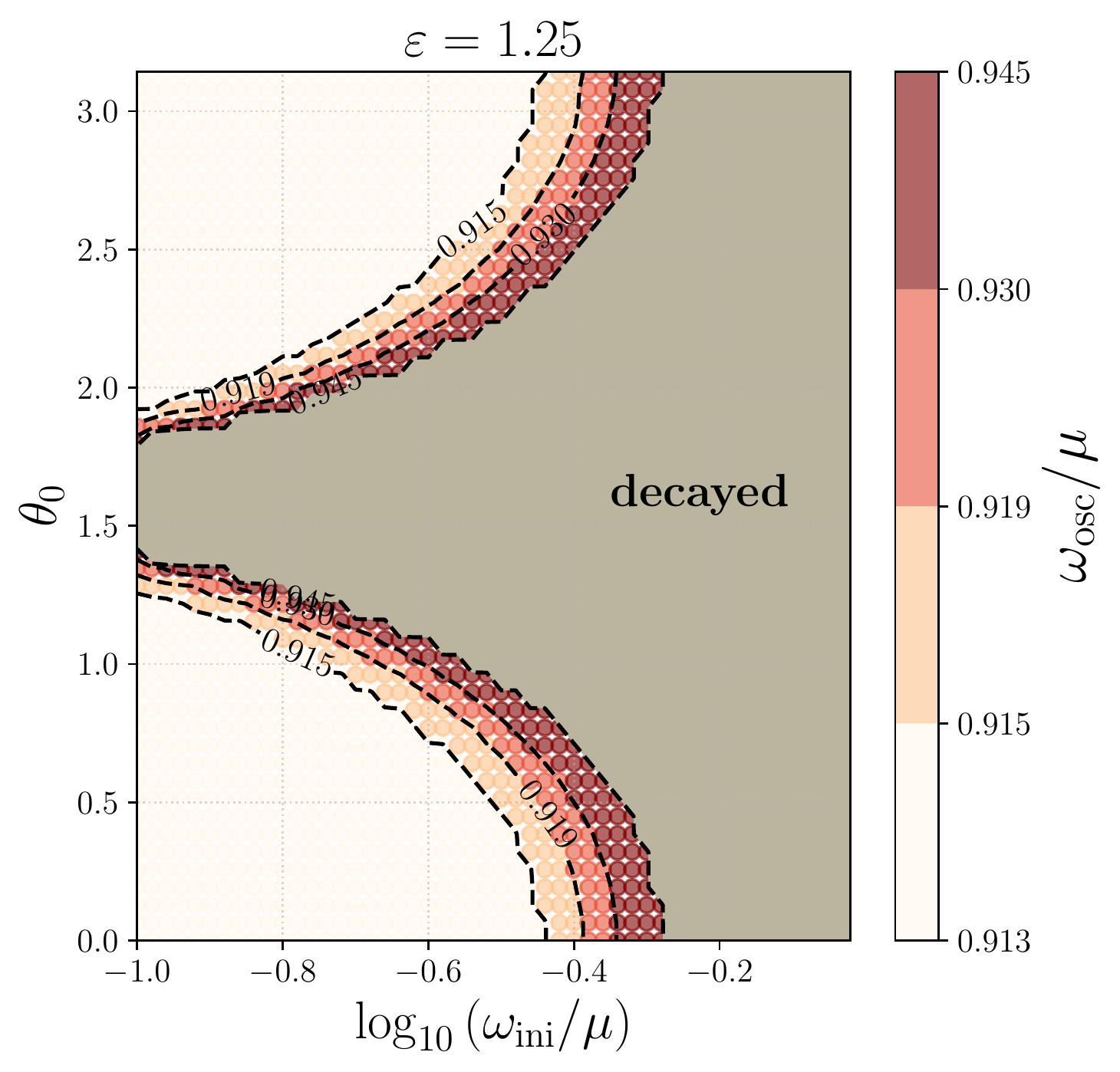}} 
\caption{\label{fig:en_non_pert} Initial energy (left panel) and oscillation frequency 
$\omegaosc$ (right panel) as functions of the initial breather parameters for $\varepsilon=1.25$. 
Parameterizing initial conditions with SG breathers discards a non-trivial fraction of the 
accesible states (inside the gray contours in the right panel, and in the shaded region 
in the left panel). Showing that in the nonperturbative limit, breathers may not have sufficient 
energy to produce stable solutions.}
\end{figure*} 
}

\newcommand{\ecritnonpert}{%
\begin{figure*}[t!]
\centering
\includegraphics[width=0.9\textwidth]{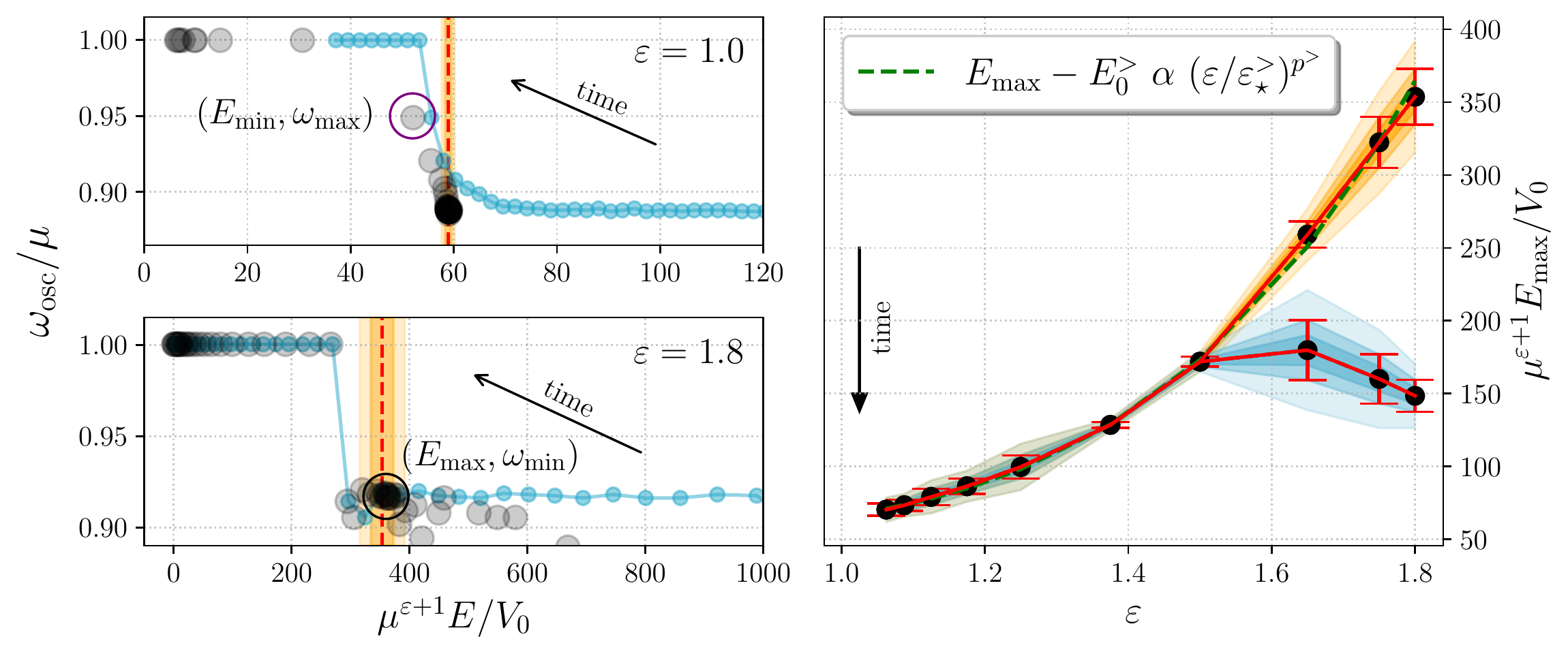}
\caption{\label{fig:e_crit_eps} Left panels: Oscillation frequency extracted at $\mu t=3\times 10^3$ as a function of the initial energy (small blue dots) and the energy after a few 
hundred oscillations (large semi-transparent black dots) for $\varepsilon=1.0$ (upper panel) and $\varepsilon=1.8$ 
(lower panel). Initial conditions correspond to 50 breathers with frequencies in the range $\log_{10} \omegaini/\mu\in [-1.0;0.02]$ (the same as in \Figref{fig:ct_eps_freq}), 
and $\theta_0=0$ as the initial phase. From the two figures, it is clear that the two clusters of black points denote states 
that at low energies (and $\omegaosc\sim\mu$) have already decayed. At higher energies, marked 
by the red dashed lines for frequencies $\omegaosc<\mu$, the clustered points correspond to the 
same stable oscillons. Right panel: Dots colored in solid black are the stable solution energies 
(indicated by red dashed lines in the left panels) where the stable states accumulate, rendering 
(approximately) a monotonically growing function of $\varepsilon$. Oscillons in $D\lesssim 3$ 
dimensions are the first to move away from the power law, as shown in the collection of points shaded 
in blue. As time progresses, oscillons slowly flow along the attractor (as shown in Fig.~\ref{fig:param_flow}) losing energy and increasing their oscillation frequencies.  
As a result, the corresponding dots (representing the states in the left panels) flow upwards following the black time arrow.
}
\end{figure*}
}

\newcommand{\tabfitparams}{%
\begin{table}[t!]
\centering
\scalebox{1.05}{
  \setlength{\tabcolsep}{0.3em}
  \begin{tabular}{|c|c|c|c|}
    \hline\hline \noalign{\smallskip}
  {case} & {$E_0[\mu^{\varepsilon+1}/V_0]$} & {$\varepsilon_{\star}$} & {$p$}\\
  \noalign{\smallskip} \hline \noalign{\smallskip}
    {$(<)$} & {$15.54\pm 0.66$} & {$0.21\pm 0.01$} & {$2.47\pm 0.07$}\\
    {$(>)$} & {$50.26\pm 2.44$} & {$0.947\pm 0.001$} & {$5.15\pm 0.15$}\\
    {$(\omega)$} & {--} & {$2.262\pm 0.078$} & {$0.125\pm 0.003$}\\
  \noalign{\smallskip} \hline\hline
\end{tabular}}
\caption{\label{tab:fit_p_np} Fitting coefficients and uncertainties for the perturbative and nonperturbative energy  and frequency fits presented in Figs.~\ref{fig:e_crit_eps} 
(green curve, right panel) and \ref{fig:res_cond}. $E_0^{<}$ is consistent with the energy of an infinitely separated ${\rm K\bar{K}}$ pair, while $E_0^{>}$ 
is the energy of the two-dimensional oscillon with minimal frequency. The last row contains the fitting coefficients of the minimum frequency $\omegamin$ as a function 
of $\varepsilon$, which is well-represented by the power law plotted in the lower panel of \Figref{fig:res_cond}.}
\end{table}
}

\newcommand{\rcondensed}{%
\begin{figure}[t!]
\centering
\includegraphics[width=.42\textwidth]{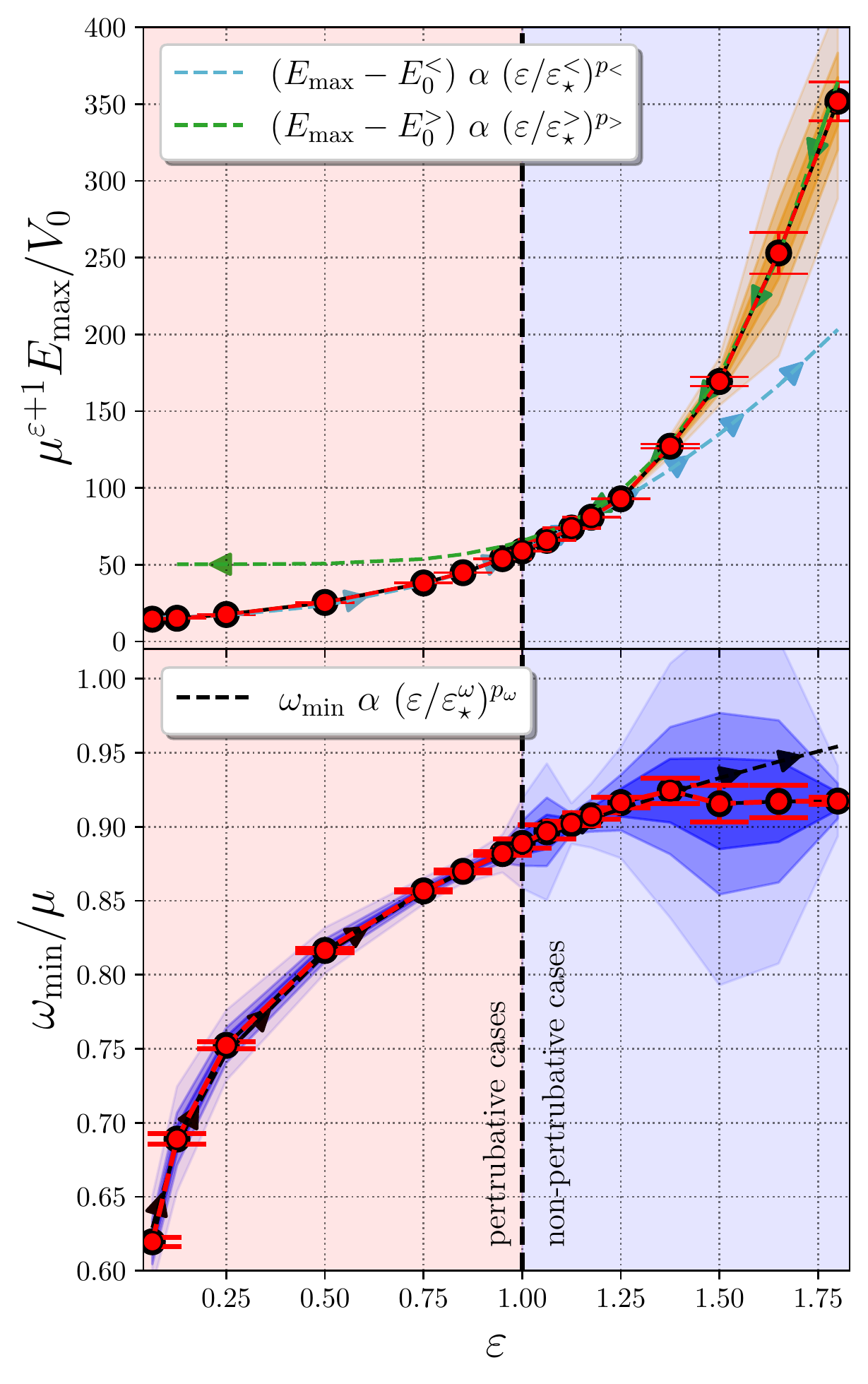}
\caption{\label{fig:res_cond} Maximum energy (upper panel) and minimum oscillation frequency (lower panel) of oscillons as a function of the number of spatial dimensions, 
after collecting the $\omegamin$ and $\Emax$ results in subsections \ref{subsec:pert_dim_def} and \ref{subsec:non_pert}. Dashed green and light blue lines in the top panel 
depict the $\varepsilon>1$ and $\varepsilon<1$ behavior of the maximum energy as a function of $\varepsilon$, respectively. In contrast to the maximum energy, the dashed 
black line shows that the minimum frequency fits using a single power law. Contours in blue and orange are the regions contained within a distance of one, two and five times 
the errors around the measured values. These contours should not be interpreted as confidence contours.}
\end{figure}
}

\newcommand{\critical}{%
\begin{figure}[t!]
\centering
\includegraphics[width=.48\textwidth]{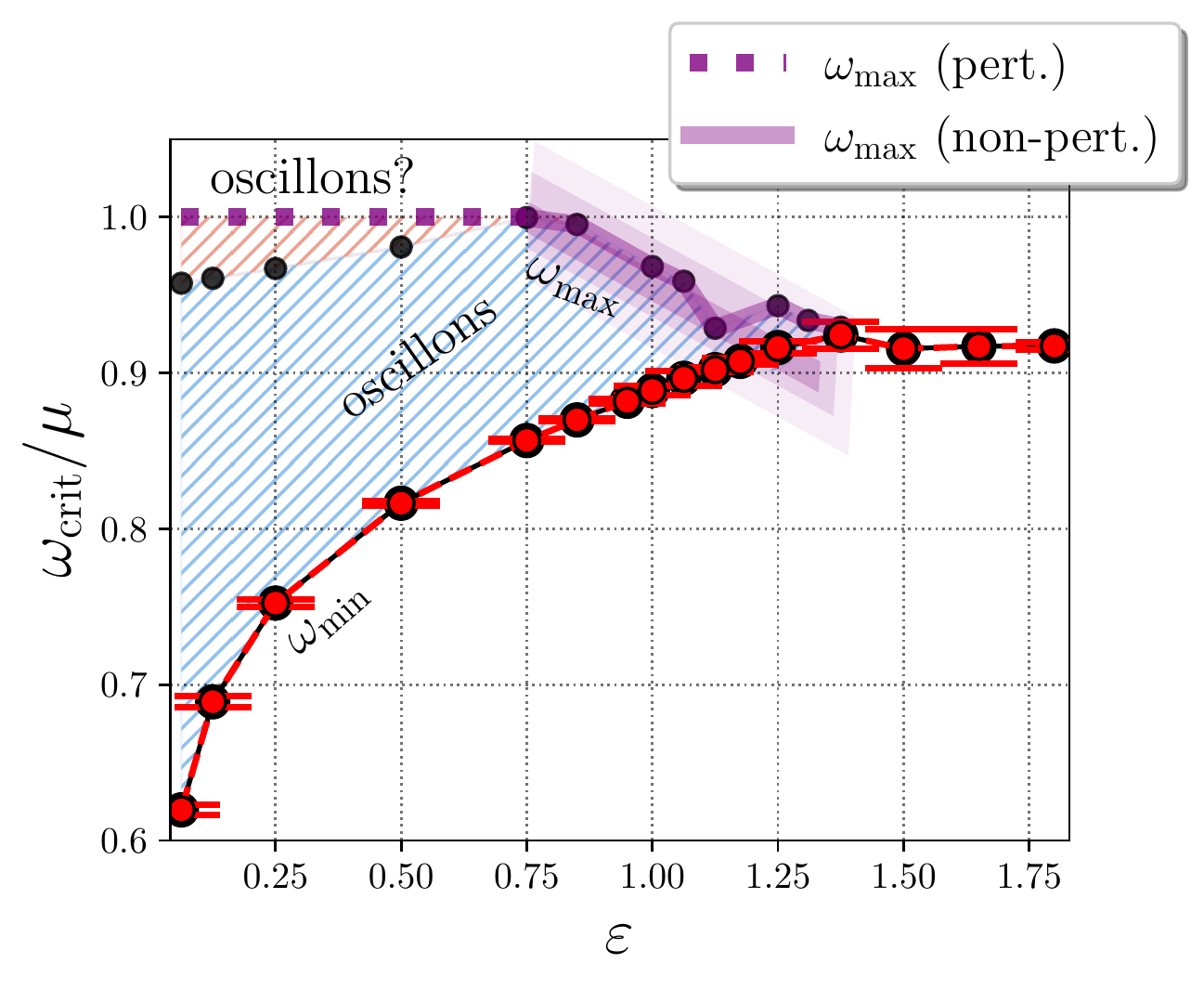}
\caption{\label{fig:critical} Phase diagram showing the collapse of states towards the minimum frequency curve in \Figref{fig:res_cond} as the dimensionality increases. The purple dotted line at 
$\omegamax=\mu$ corresponds to the maximum frequency estimates in the perturbative regime. Black dots represent the maximum oscillation frequencies measured 
in our simulations, which show for $\varepsilon\gtrsim 1$ the collapse of states in the nonperturbative limit. The area hatched in blue contains oscillons emerging from breathers within the range 
of initial frequencies and phases, and the red area is the region of the $(\varepsilon,\omegaosc)$ plane that has not been explored. The rectangle in purple denotes large 
error bars for the maximum frequencies for $\varepsilon>0.75$ due to sparse initial parameter sampling. The size of the rectangle does not intend to show the magnitude of 
the errors in $\omegamax$.  
}
\end{figure}
}

\newcommand{\perttonpert}{%
\begin{figure}[t!]
\centering
\includegraphics[width=.5\textwidth]{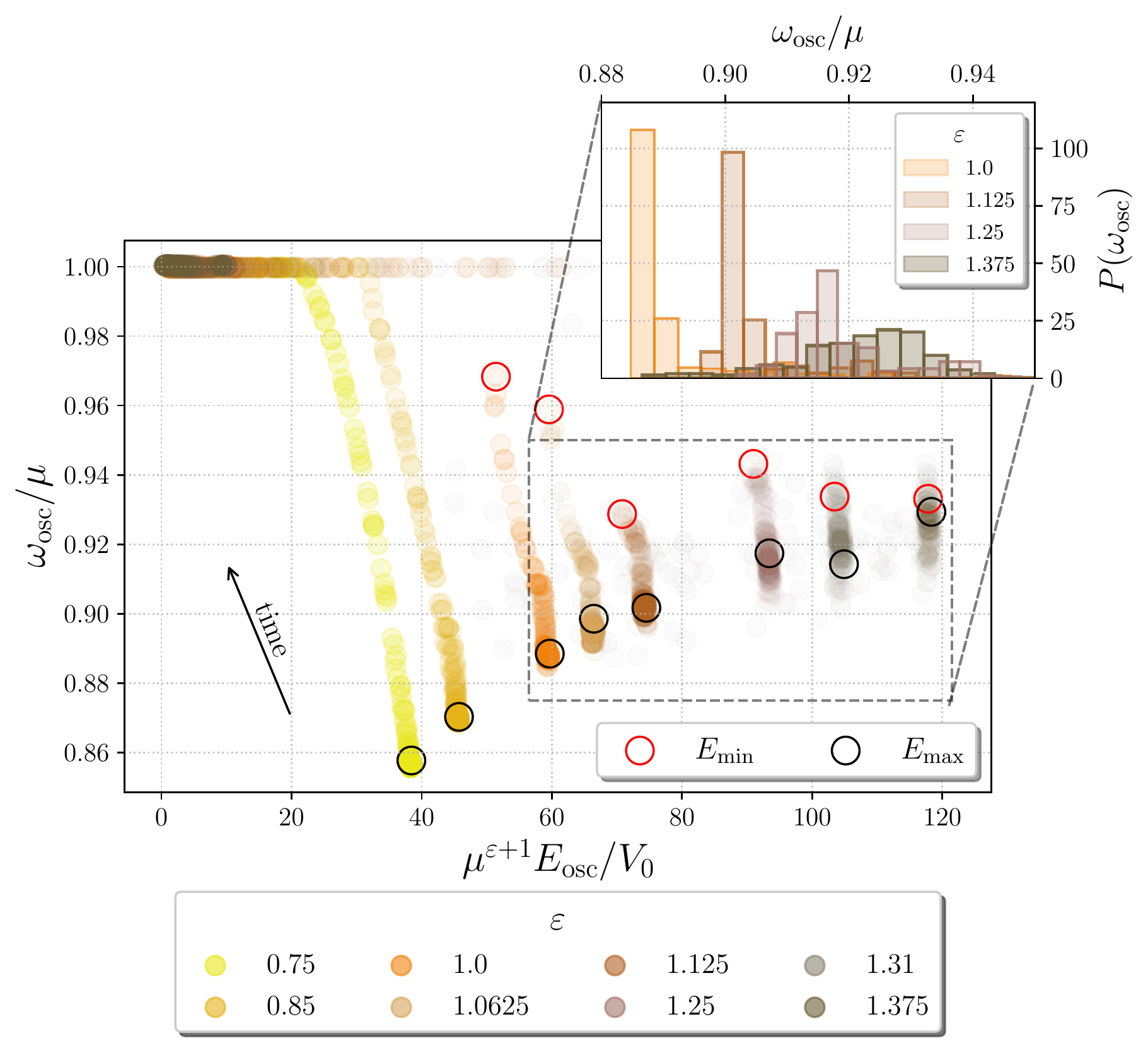}
\caption{\label{fig:pert_npert} Oscillon frequencies and energies in various dimensionalities. We chose fifty frequencies from a uniform span in the range 
$\log_{10}\omegaini/\mu\in [-1.0;0.0)$ and ten initial phases from the interval $\theta_0\in[0;\pi/2)$ for each value of $\varepsilon$. We notice the emergence of a 
frequency gap growing as $\varepsilon$ increases. The inset at the upper right corner shows the change in the empirical probability distributions for 
$\varepsilon\in[1.0;1.375]$. Histogram deformations explicitly show that the collapse of the oscillation frequency range to a point
(also shown in the lower left panel of \Figref{fig:e_crit_eps}) occurs for $\varepsilon\in[1.25;1.375)$. }
\end{figure}
}

\newcommand{\largeampb}{%
\begin{figure*}
\centering
\subfigure{
\includegraphics[width=.4\textwidth]{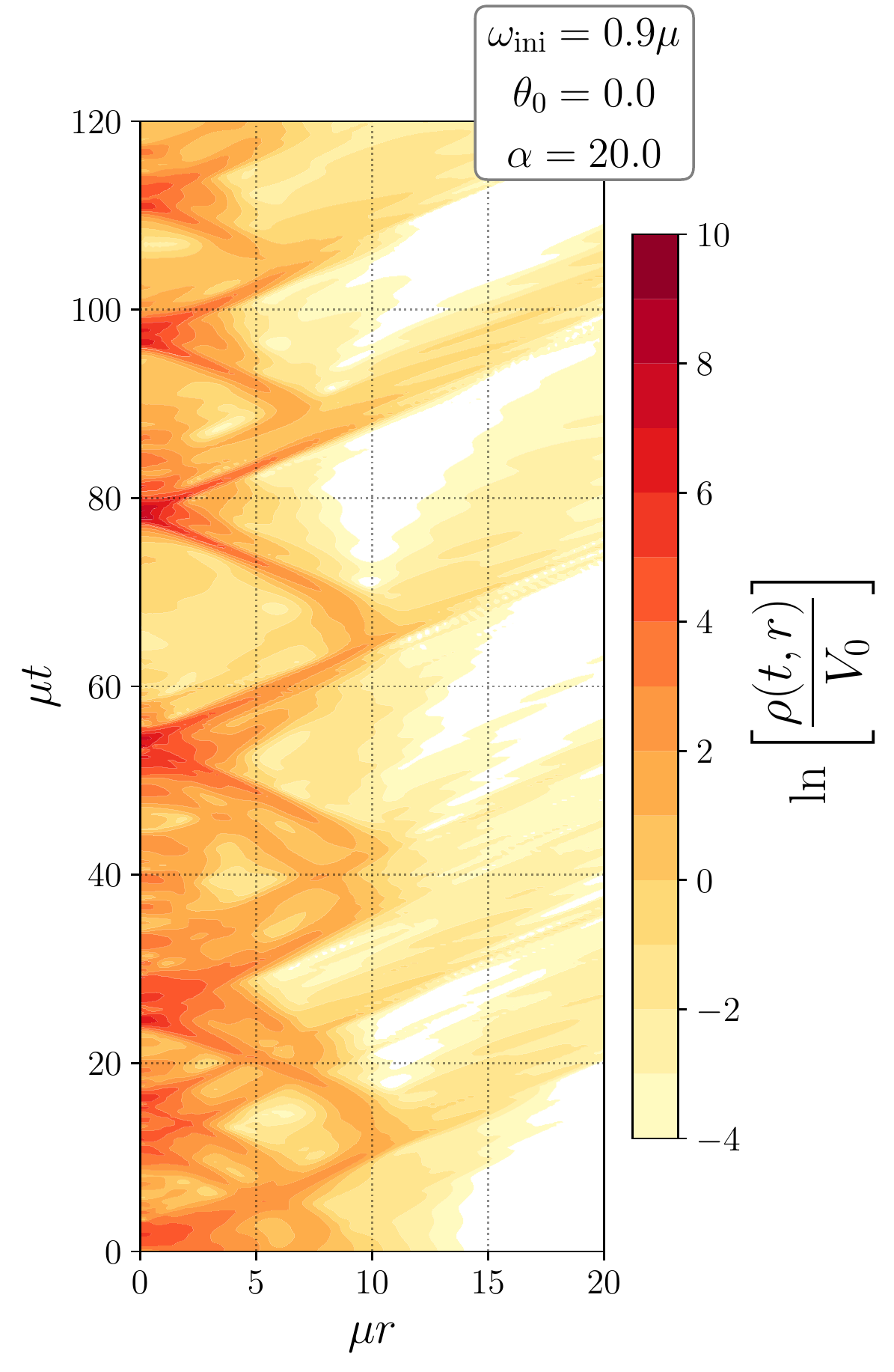}} \,
\subfigure{
\includegraphics[width=.4\textwidth]{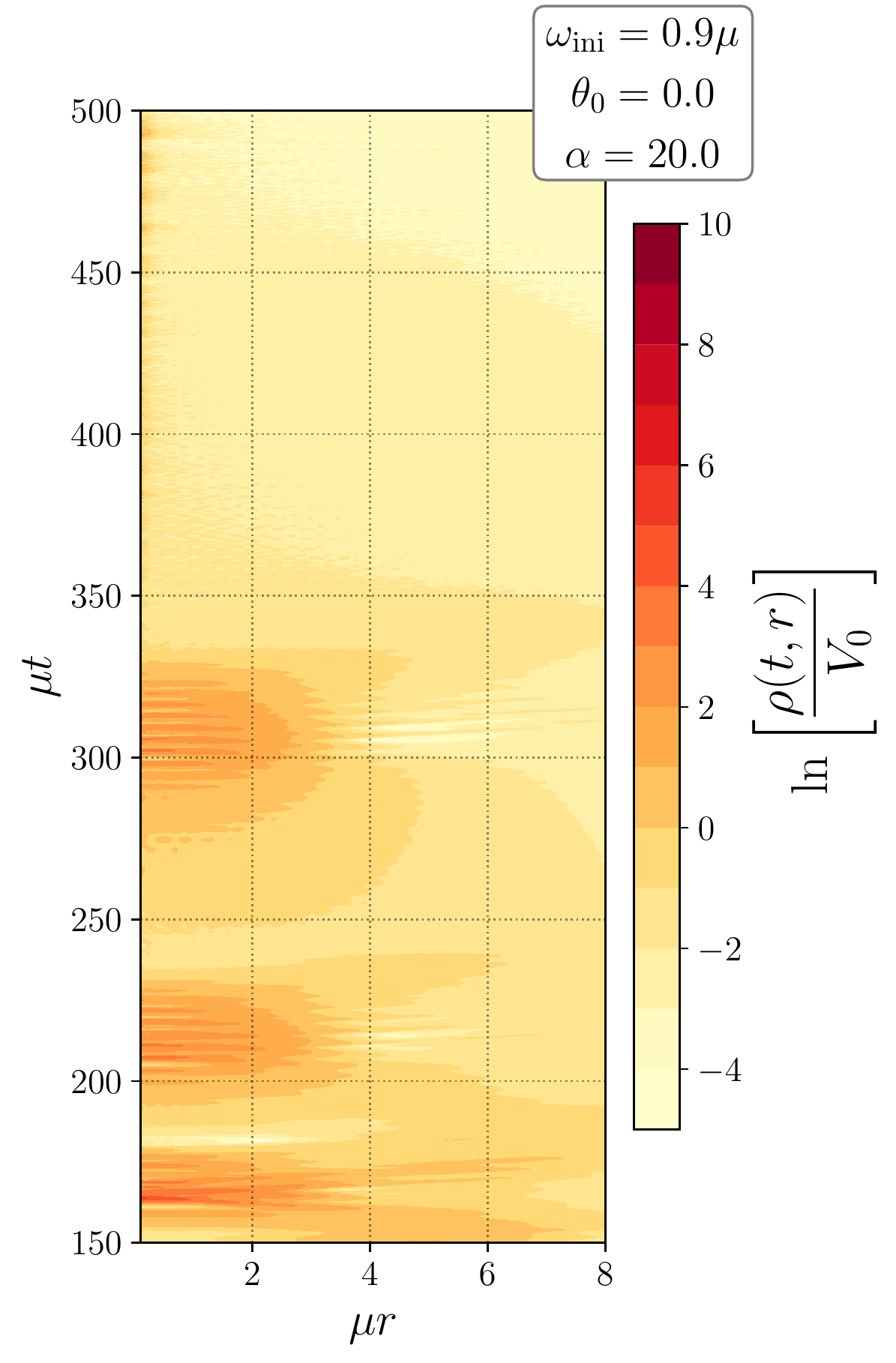}} 
\caption{\label{fig:large_alpha} Evolution of the energy density of a breather-like initial condition 
in the three-dimensional sine-Gordon model for $\omegaini=0.9\mu$ and $\alpha=20$. When 
$\alpha=1.0$, the frequency corresponds to a breather oscillating inside the well minimum centered 
at $\phi=0$. Left panel: Initial phase of the evolution corresponding to the collapse and expansion 
of spherical shells. Right panel: Dilution of the bound states, the solution disappears after two intermittent bursts.}
\end{figure*} 
}

\newcommand{\dvstatomega}{%
\begin{figure*}[t!]
\centering
\includegraphics[width=1.0\textwidth]{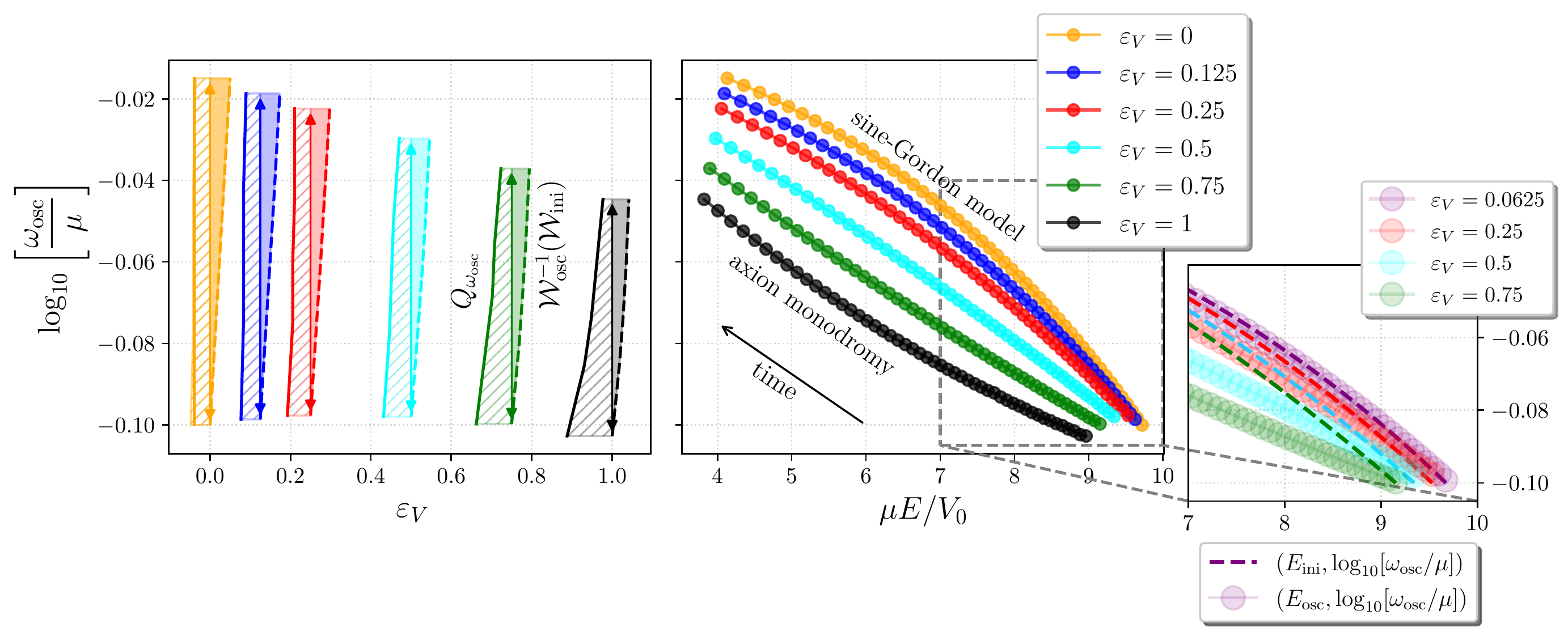}
\caption{\label{fig:dv_omega_stat} Left panel: Distribution of oscillon frequencies $\omegaosc$, assuming a log-uniform prior for 
$\omegaini$ with $-0.1 \leq \log_{10}\left(\omegaini/\mu\right) \leq -0.015$.
We sampled this prior using $50$ initial conditions with uniformly spaced $\log_{10}(\omegaini/\mu)$ in the indicated interval. 
The oscillation frequency interval deforms as a function of $\varepsilon_V$. The areas shaded with lines correspond to the Jacobian 
$Q_{\omegaosc}=N^{-1}|\ud \mathcal{W}_{\rm osc}/\ud \mathcal{W}_{\rm ini}|^{-1}$, while the areas in semi-transparent colors determine the 
inverse map $ \mathcal{W}^{-1}_{\rm osc}(\mathcal{W}_{\rm ini})$. As the frequency intervals contracts with the curvature growth of the sine-Gordon 
potential, a uniformly distributed sample in $\log_{10}(\omegaosc/\mu)$ deforms to have a higher number of 
available states at lower frequencies. Right panel: Continuous deformation of energy versus oscillation frequency curves for various intermediate stages of 
the potential deformation. The inset plotted to the right shows that flow lines concur in the limit $\varepsilon_{\rm V}$ , showing consistency with the small 
dimensional deformations limit in Fig.~\ref{fig:pert_en}.}
\end{figure*}
}

\newcommand{\breatherfit}{%
\begin{figure}[t!]
\centering
\includegraphics[width=0.5\textwidth]{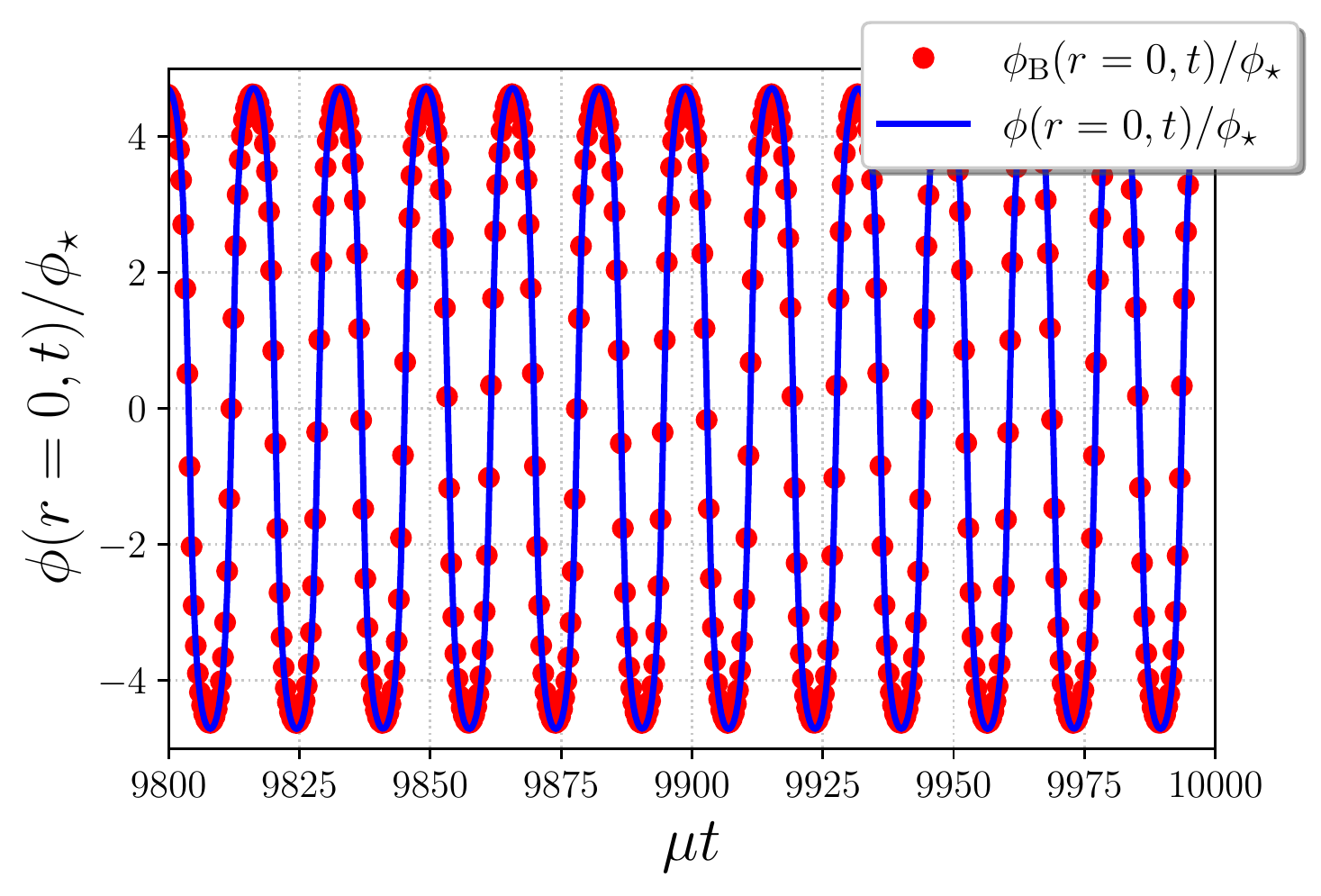}
\caption{\label{fig:3D_to_1D} Comparing the evolution of the solution of Eqns.~\eqref{eq:dphi_dt_vard} and 
\eqref{eq:dpi_dt_vard} (labeled as $\phi(r=0,t)/\phiScl$ and plotted in a solid blue line) transitioning from 3D to 
1D in a couple of oscillations with a SG breather evaluated at the origin. The initial sine-Gordon breather (labeled as 
$\phiB(r_{\mathrm{c}},t)/\phi_{\star}$ and plotted in red dots) is a good fit of the solution considering a phase of 
$\theta_{\mathrm{B}}\approx-4\pi/21$. Such a result validates the frequency extraction procedure introduced 
in \Secref{sec:sols_and_params}, and used throughout this manuscript.}
\end{figure}
}

\newcommand{\varfreqphase}{%
\begin{figure*}
\centering
\subfigure{
\includegraphics[width=.4\textwidth]{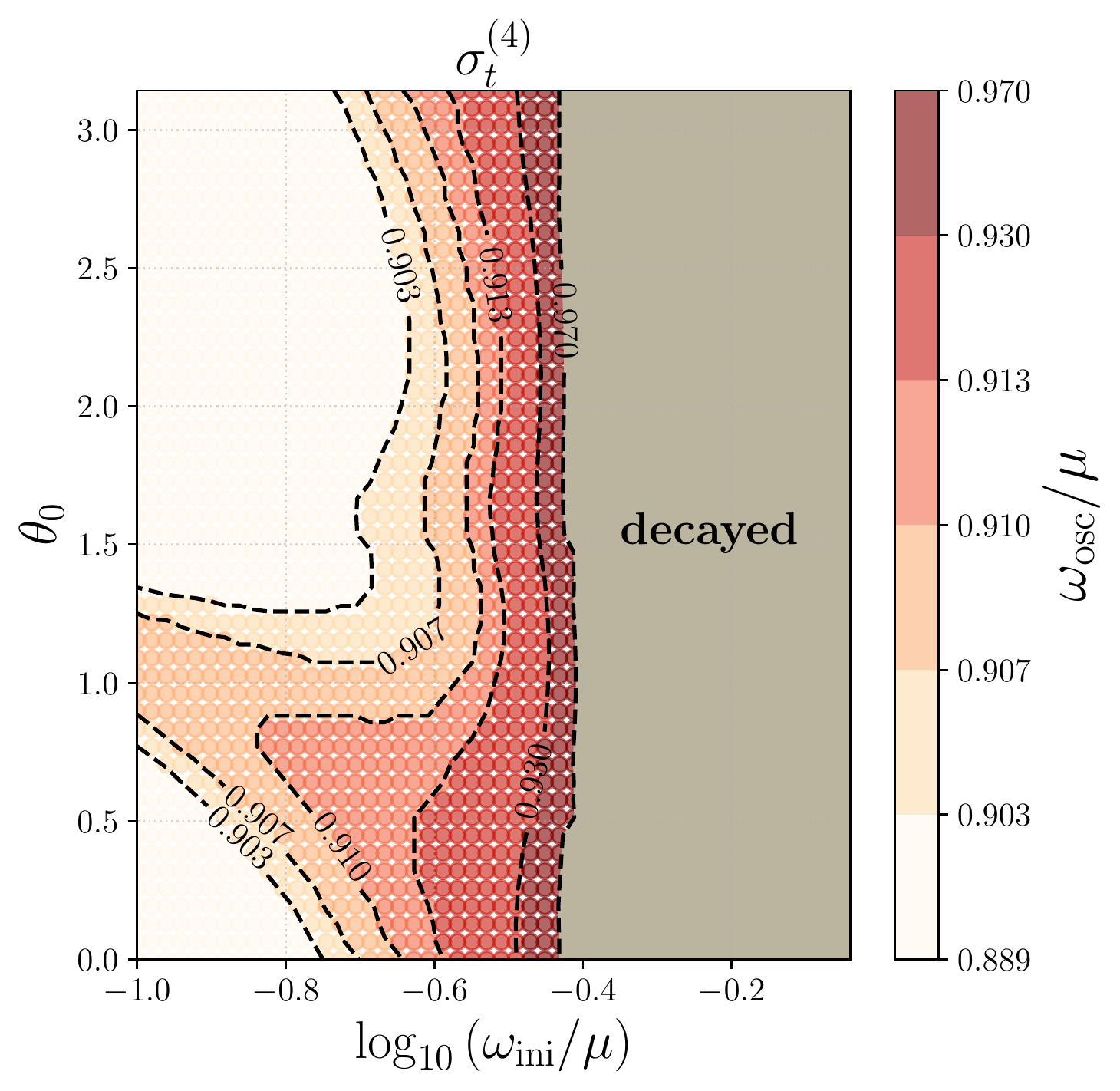}} \,
\subfigure{
\includegraphics[width=.4\textwidth]{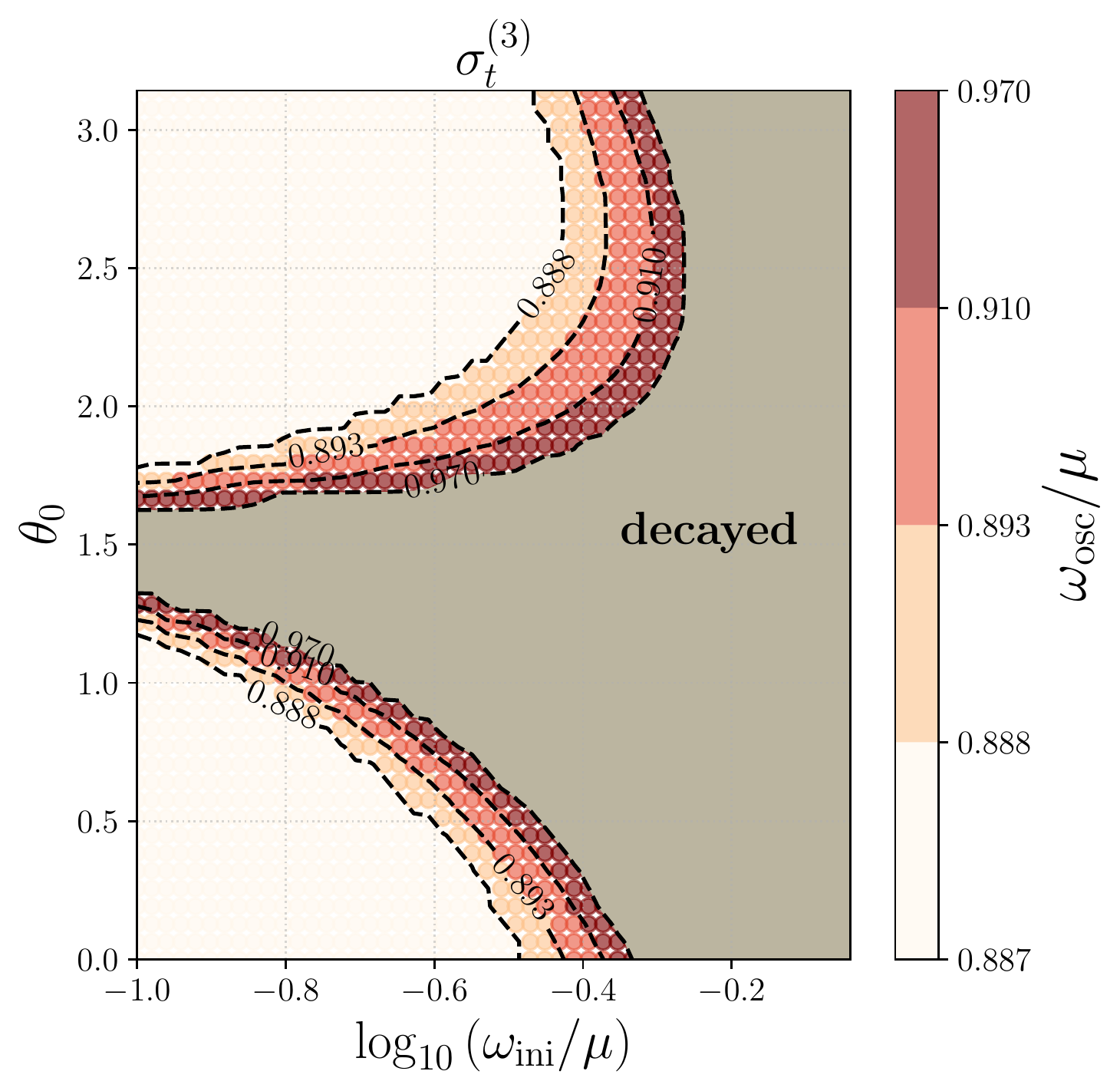}} \,
\subfigure{
\includegraphics[width=.4\textwidth]{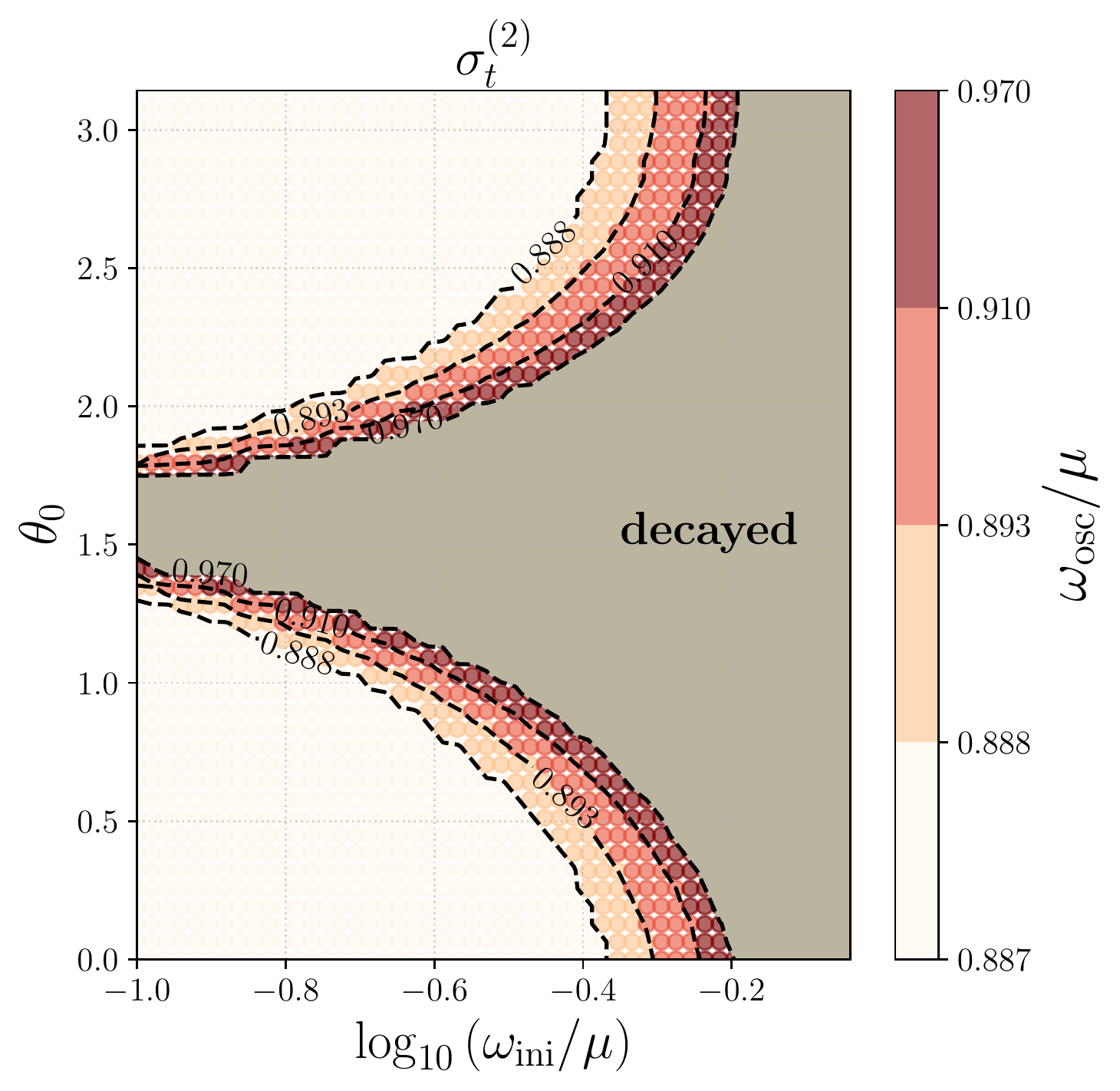}} \,
\subfigure{
\includegraphics[width=.4\textwidth]{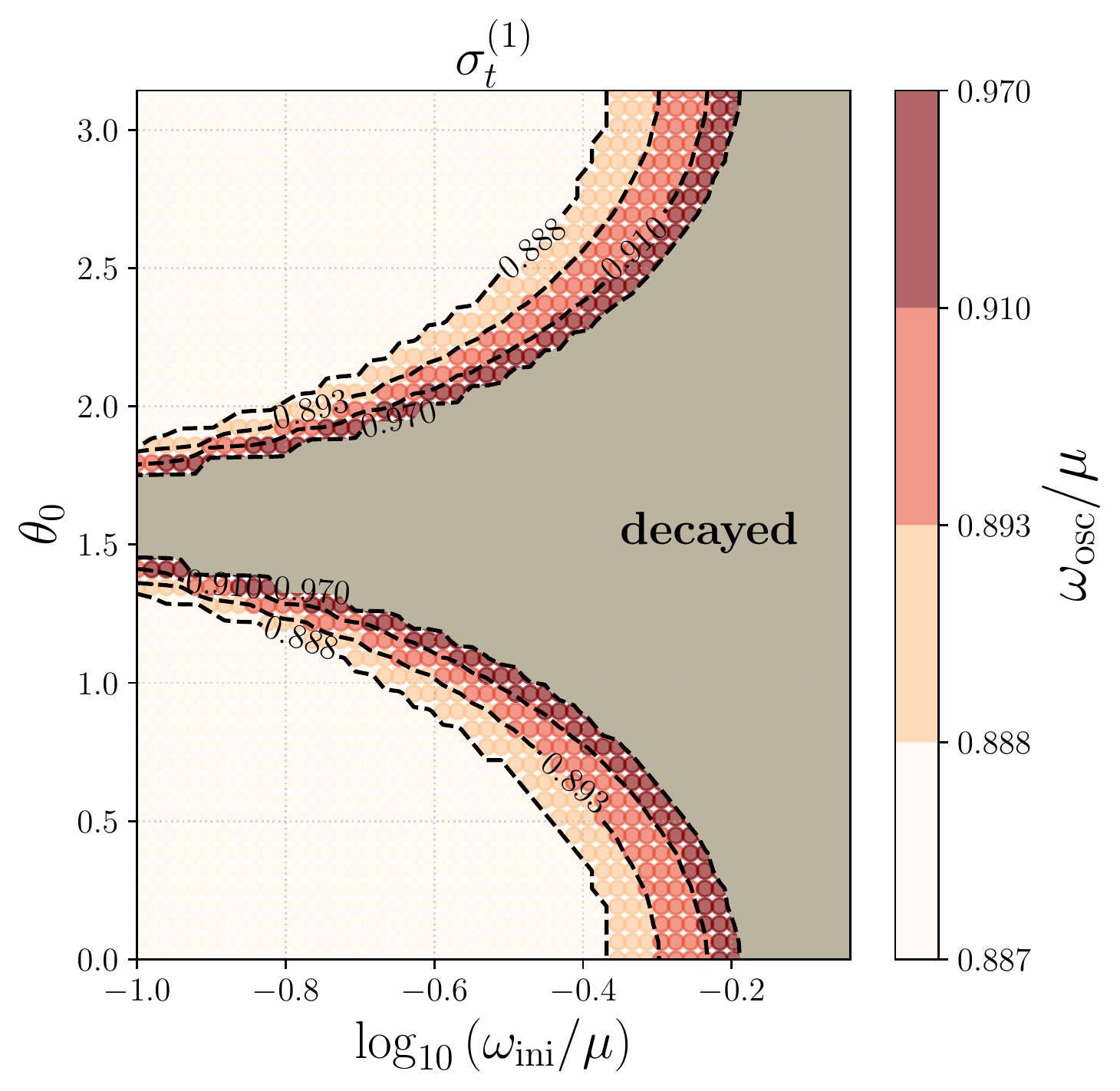}}
\caption{\label{fig:var_eps_freq} Oscillon frequency as a function of the phase and initial breather frequency for the 1D to 2D transitions described in Table~\ref{tab:duration}. 
We use a grid of $50\times 50$ initial configurations of frequencies $(\omegaini)$ and phases $(\theta_0)$ (which is the same as in \Figref{fig:ct_eps_freq}) distributed uniformily in 
$\log_{10}(\omegaini/\mu)\in [-1,-0.02]$ and in $\theta_0\in [0,\pi)$. The cusps reported in Fig.~\ref{fig:ct_eps_freq} emerge in the limit of abrupt dimensional transitions, such as the 
two panels at the bottom, and conversely become less sharp as the transition slows down (\ie the two panels at the top). In all scenarios, decayed states (in gray contours) represent 
40\% (approx.) of the 2500 solutions evolved. Symmetry around $\theta_0=\pi/2$ is only restored in the limit $\sigma_t\rightarrow 0$. We chose the range of oscillation frequencies to 
coincide with the interval of oscillon frequencies for the flow line $\varepsilon=1.0$ (orange dots) in \Figref{fig:pert_npert}.}
\end{figure*} 
}

\newcommand{\tabduration}{%
\begin{table}[t!]
\centering
\scalebox{1.1}{
  \setlength{\tabcolsep}{0.3em}
  \begin{tabular}{|c|c|c|c|}
    \hline\hline \noalign{\smallskip}
  {~Cases~} & \head{1.5cm} {~duration $\sigma_t [\mu^{-1}]$~} & \head{1.75cm}{~\# of oscillations~} & \head{1.5cm}{~\# of decaying states~}\\
  \noalign{\smallskip} \hline \noalign{\smallskip}
    $\sigma^{(1)}_t$ & ${0.1}$ &$[0.02;0.16]$  & $1080$\\
    $\sigma^{(2)}_t$ & ${0.5}$ & $[0.1;0.8]$  & $1080$\\
    $\sigma^{(3)}_t$ & ${2.5}$ & $[0.5;4.0]$  & $1183$\\
    $\sigma^{(4)}_t$ & ${12.5}$ & $[2.5;20.0]$  & $984$\\
  \noalign{\smallskip} \hline\hline
\end{tabular}}
\caption{\label{tab:duration} Cases and duration (range of the number of oscillations, given the span in $\omegaini$) of the transitions from 
one to two spatial dimensions. The right column contains the number of rapidly decaying solutions obtained in the maps of Fig.~\ref{fig:var_eps_freq}, 
contained by the contours colored in gray.}
\end{table}
}

\newcommand{\enconsbr}{%
\begin{figure}[t!]
\centering
\includegraphics[width=.45\textwidth]{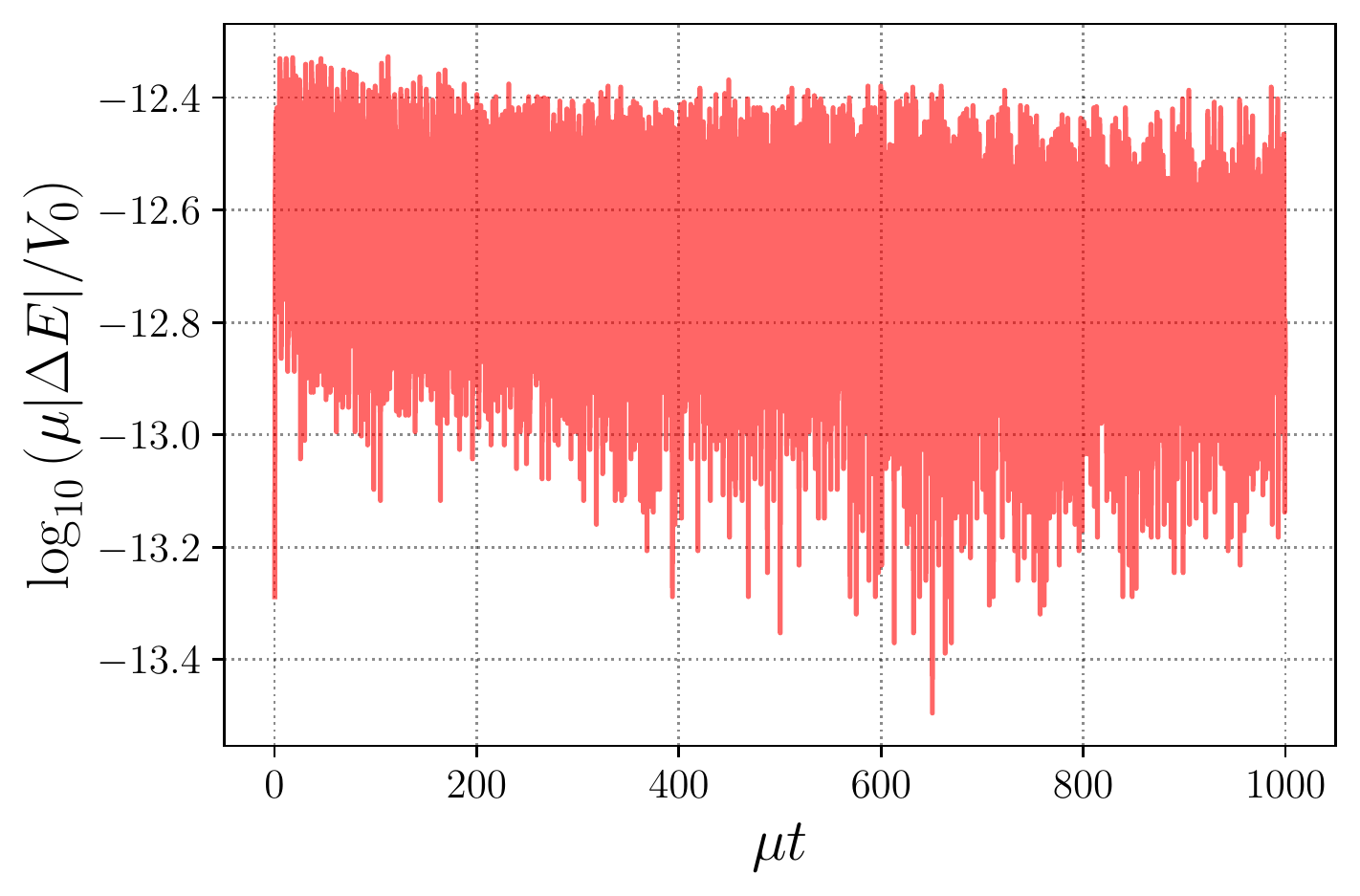}
\caption{\label{fig:en_cons} Energy conservation for a standing sine-Gordon breather with $\omegaini=0.794\mu$ and $\theta_0=0$, 
evolved from the equations of motion (\ref{eq:dphidt_rad_fin}--\ref{eq:cons_2_fin}) in the case $\varepsilon=0$. Showing that perfectly 
matched layers do not interfere with the solution inside the simulation length.}
\end{figure}
}

\newcommand{\hitpml}{%
\begin{figure}[t!]
\centering
\includegraphics[width=.4\textwidth]{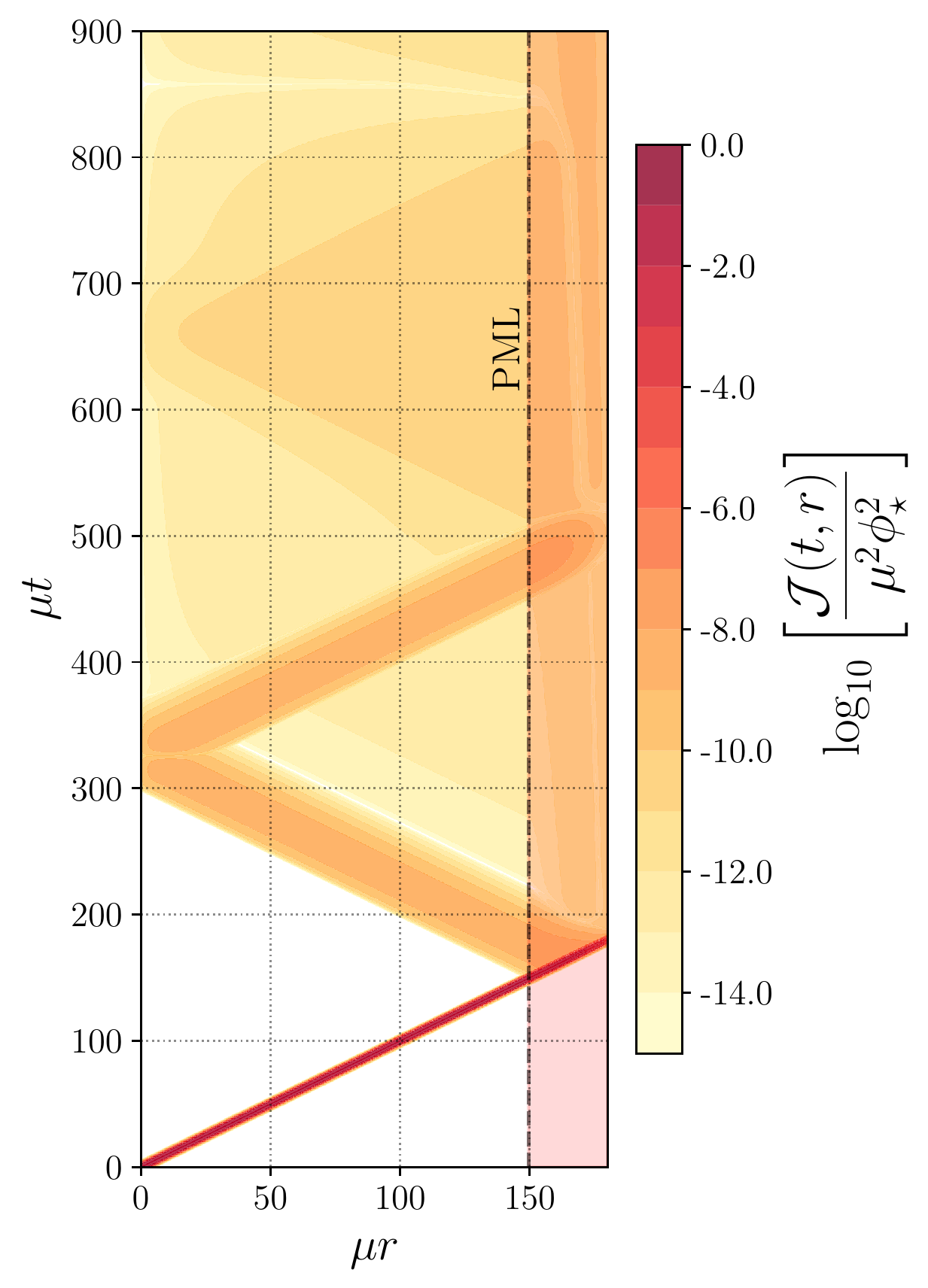}
\caption{\label{fig:hit_pml} Impact of a Gaussian wave packet on a perfectly matched layer (PML). Scalar flux reduces in 9--10 orders of magnitude after 
hitting the absorbing layer for the first time, which suggests that the setup is operational. With the purpose of showing the action of the PML on the ingoing flux, here we included 
some of the grid nodes within the plot. Nevertheless, the width of the layer is not considered within the simulation box.}
\end{figure}
}

\newcommand{\ampmodtwod}{%
\begin{figure*}
\centering
\subfigure{
\includegraphics[width=1.0\textwidth]{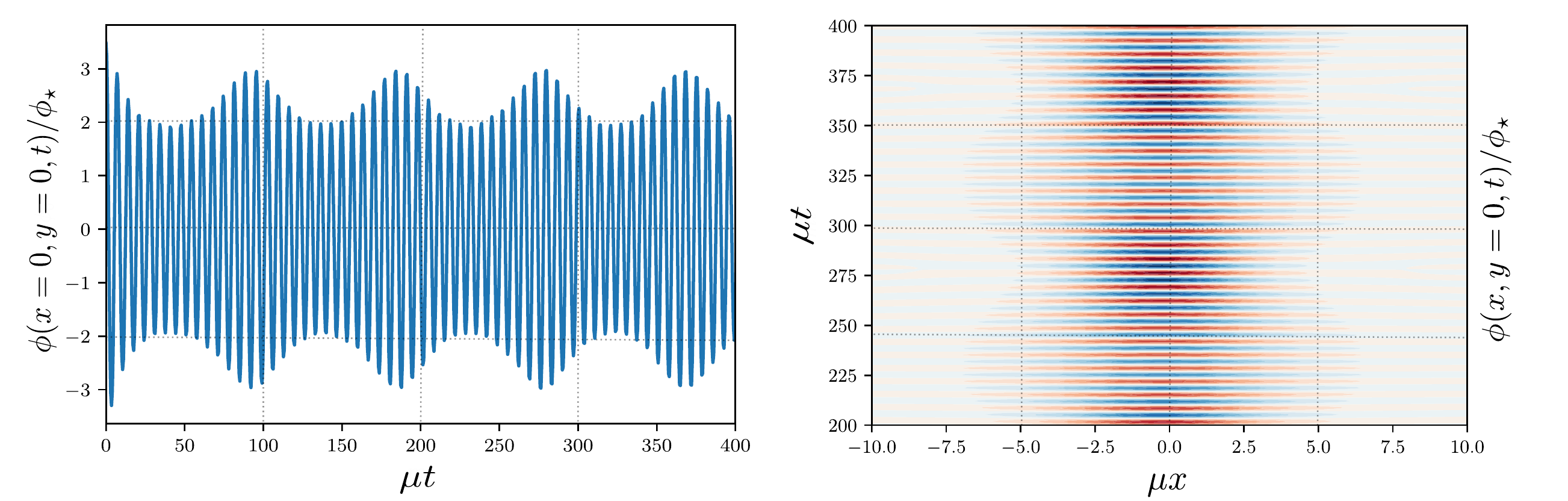}} \,
\caption{\label{fig:osc_2d_amp_mod} Evolution of a breather-like initial condition in two-dimensional 
sine-Gordon model for $\omegaini=0.63\mu$. Left panel: Projection of the solution in the $y=0$ plane, 
showing quasi-periodic phases of contraction and expansion of the oscillon core. Right panel: Evolution 
of the solution in the origin, consistent with modulations in the amplitude described for the perturbative 
regime in subection \ref{subsec:pert_dim_def}.}
\end{figure*} 
}

\newcommand{\multires}{%
\begin{figure*}
\centering
\subfigure{
\includegraphics[width=.44\textwidth]{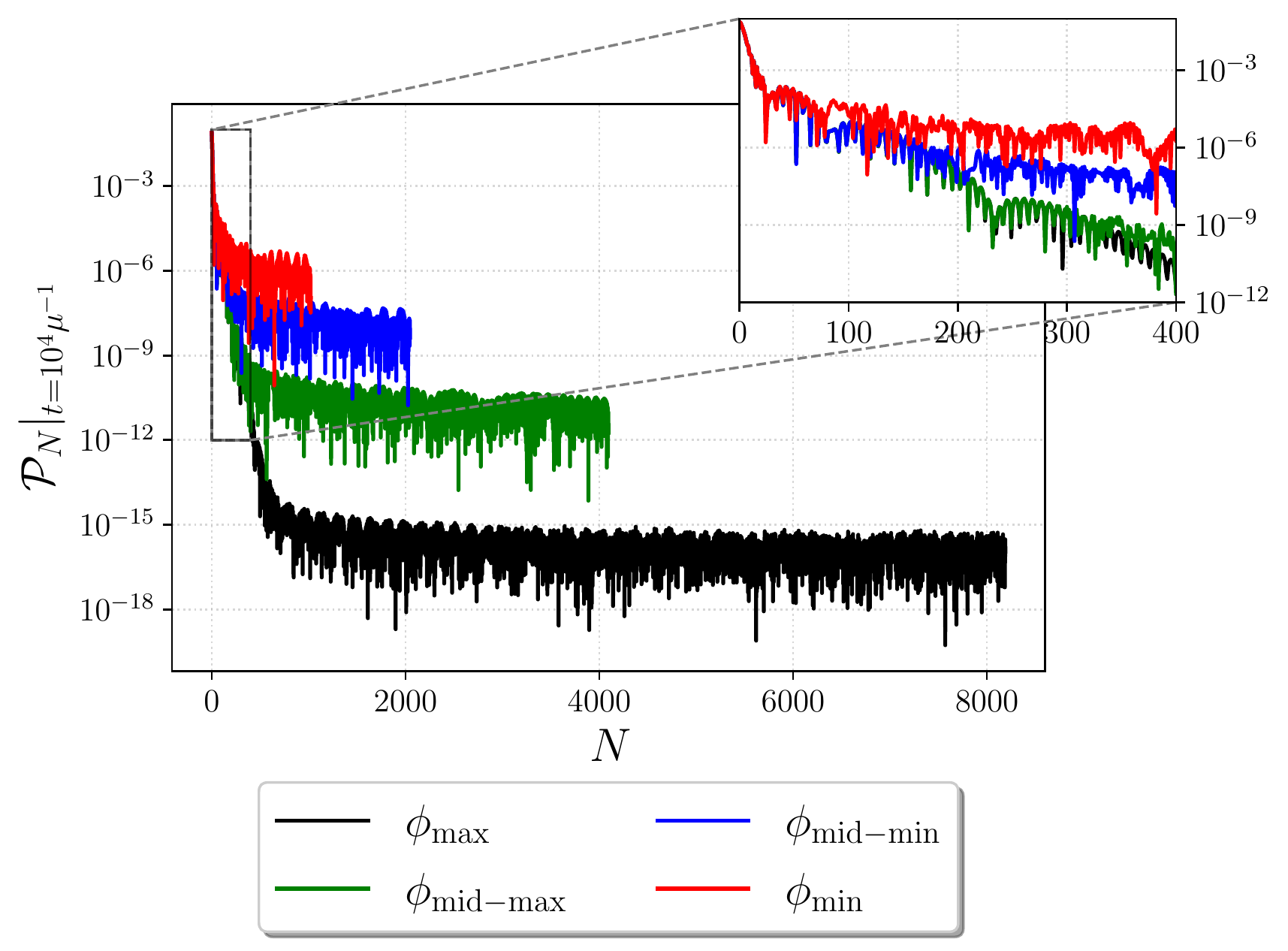}} \,
\subfigure{
\includegraphics[width=.44\textwidth]{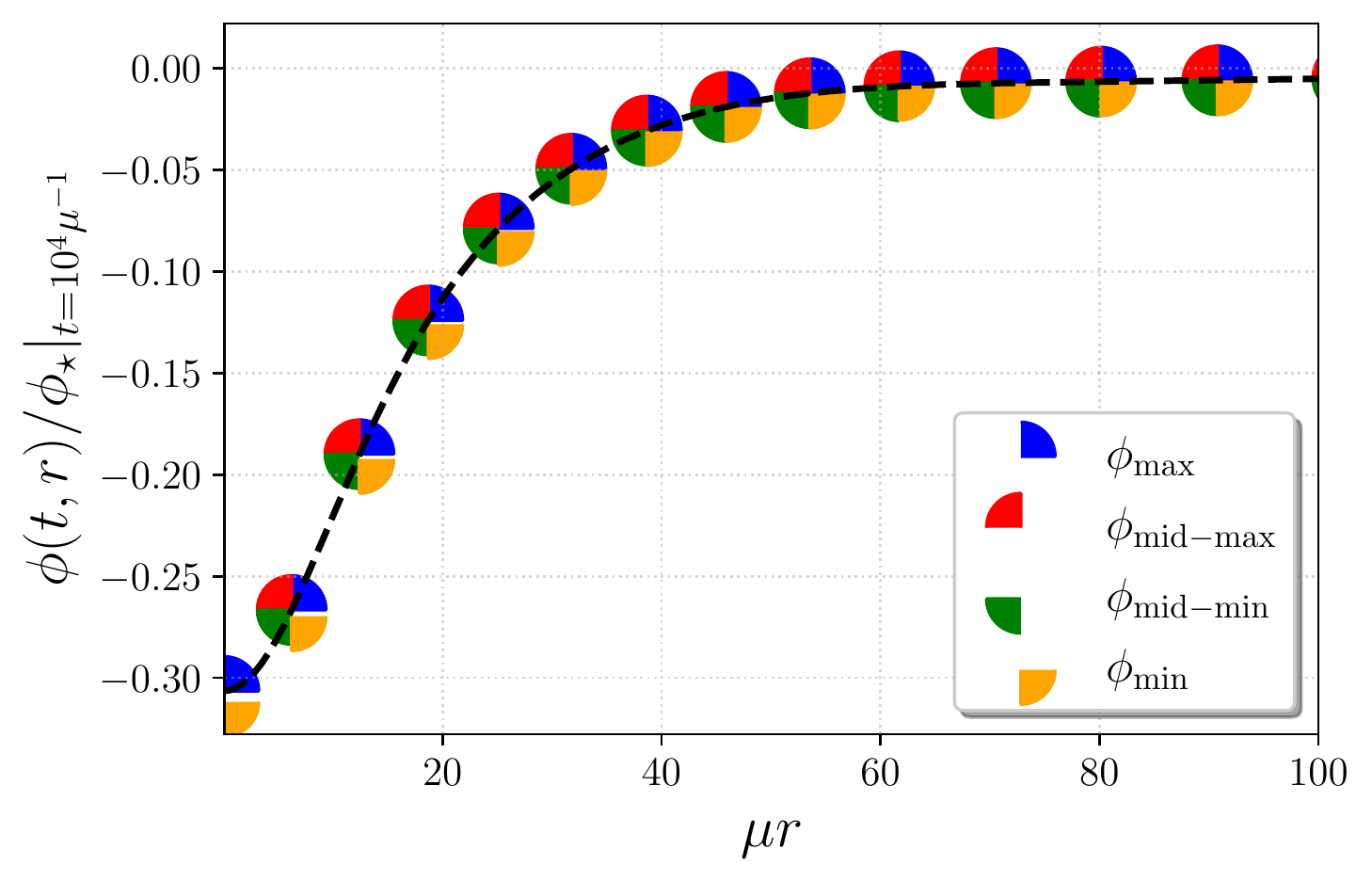}} 
\caption{\label{fig:multi_res} Left panel: Spectral coefficients as functions of the number of collocation points, corresponding to the solution 
in the middle and right panels of Fig.~\ref{fig:diag_n_mod} for $\varepsilon=0.75$. Initial condition is the breather parameterized by 
$\omegaini=0.398\mu$ and $\theta_0=0.6\pi$, we plot the solution at fixed $t=10^4\mu^{-1}$ for the resolutions reported in Table 
\ref{tab:resolutions}. Results at all resolutions coincide for the first hundred nodes, which are sufficient to produce the oscillon core. 
Right panel: Oscillating field as a function of the radius for the same resolutions. The origin is where the difference between solutions 
is the clearest; however, all of the solutions coincide approximately. Dashed black curve is a snapshot (at the same instant) 
of the solution resolved at the highest resolution. 
}
\end{figure*} 
}

\newcommand{\tabresolutions}{%
\begin{table}[t!]
\centering
\scalebox{1.1}{
  \setlength{\tabcolsep}{0.3em}
  \begin{tabular}{|c|c|c|}
    \hline\hline \noalign{\smallskip}
  {~Resolutions~} & {~\# of nodes~} & \head{1.5cm}{~time step $[\Delta t^{(\mathrm{res})}_{\mathrm{CFL}}]$}\\
  \noalign{\smallskip} \hline \noalign{\smallskip}
    max & $8192$ & $1/16$\\
    mid-max & $4096$ & $1/8$\\
    mid-min & $2048$ & $1/4$\\
    min & $1024$ & $1/4$\\
  \noalign{\smallskip} \hline\hline
\end{tabular}}
\caption{\label{tab:resolutions} Resolutions used to solve Eqns.~\eqref{eq:dphidt_rad_fin}--\eqref{eq:cons_2_fin} for 
$\varepsilon=0.75$. Considering $\ell=100\mu^{-1}$ and the absorbing boundary layer centered at the end of the simulation 
box, we perform convergence tests for the field configuration with intermittent phases of expansion and contraction shown in 
Fig.~\ref{fig:diag_n_mod}.}
\end{table}
}

\newcommand{\convrmax}{%
\begin{figure}[t!]
\centering
\includegraphics[width=.44\textwidth]{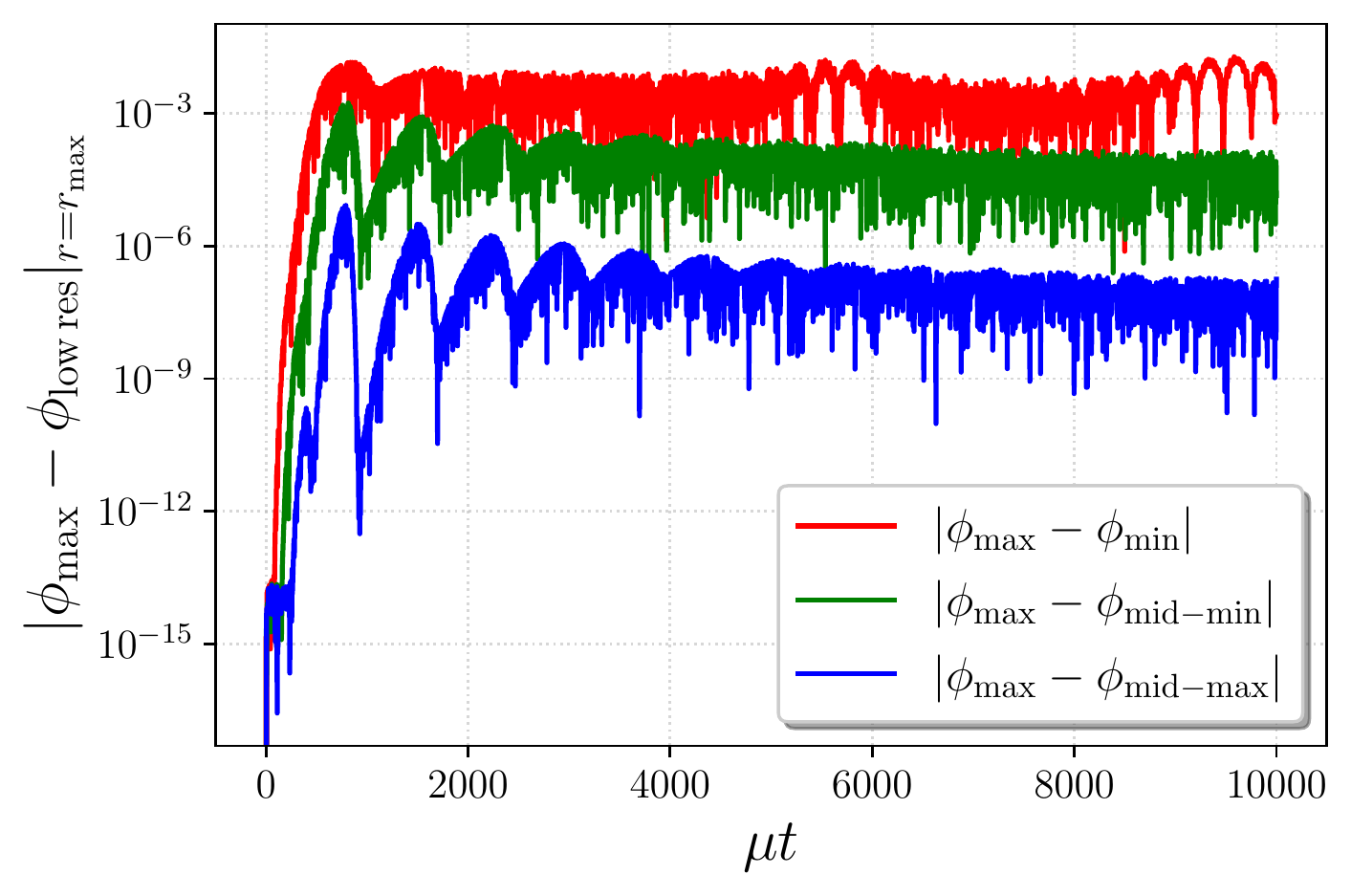}
\caption{\label{fig:conv_r_max} Subtracting field configuration at $r=r_{\mathrm{max}}$, where the differences between field 
configurations are the largest (this is near the origin, as depicted in the right panel of Fig.~\ref{fig:multi_res}). Numerical error 
decreases as the solutions are resolved with more collocation modes.}
\end{figure}
}

\newcommand{\brprof}{%
\begin{figure}
  \includegraphics[width=246pt]{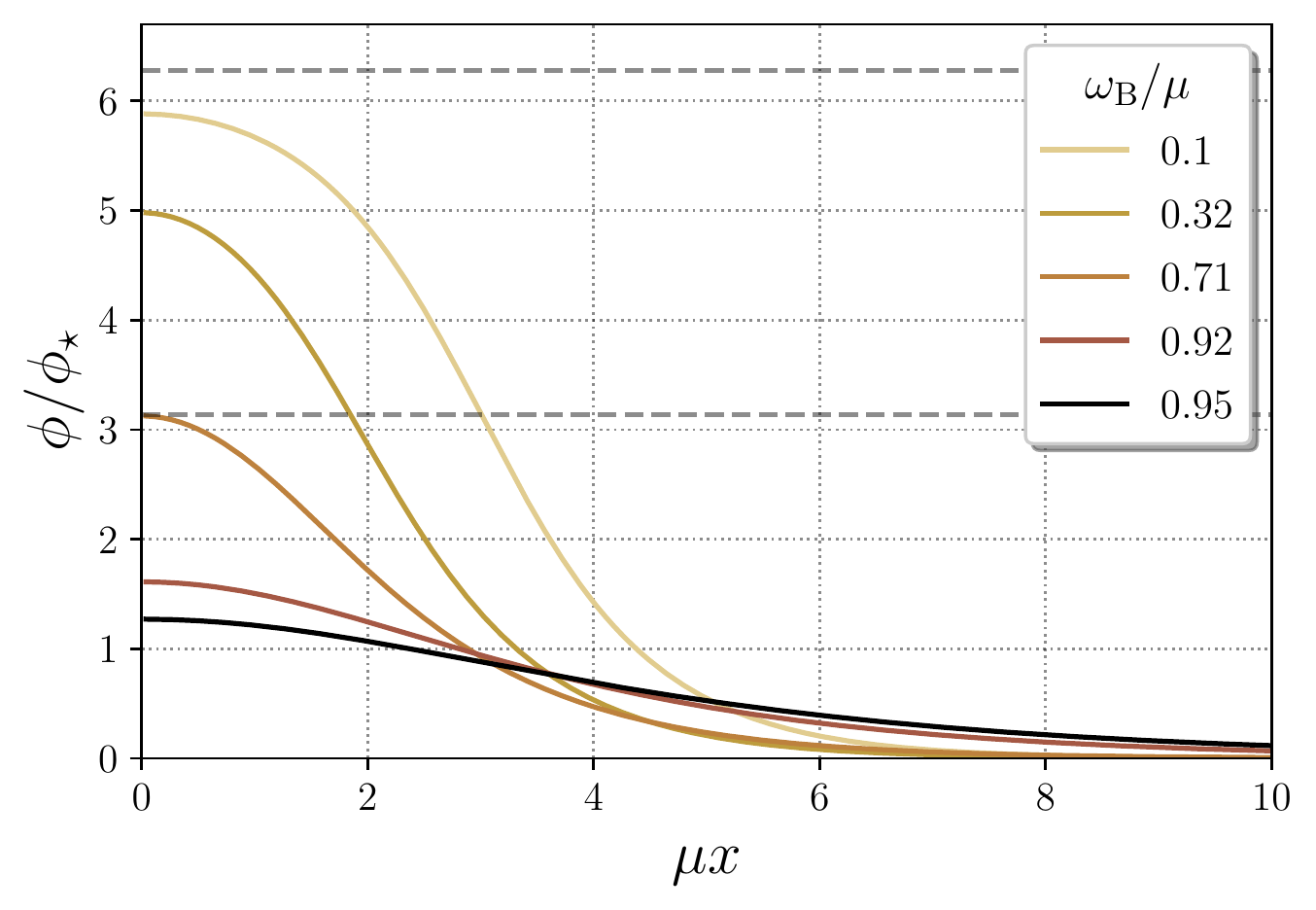}
  \caption{Radial breather profiles~\eqref{eqn:breather-profile} for several choices of the parameter $\omegaB$ describing the breather's frequency.  For $0 < \mu-\omegaB \ll 1$, 
 the profiles have small amplitude at the origin and damp slowly as $\mu r\to\infty$.  As $\omegaB$ is increased, the breathers become more peaked at the origin and damp more rapidly 
 at large radii.  We explicitly illustrate the breathers that just probe the inflection point of the potential ($\omegaB = 0.92\mu$) and the nearest maximum of the potential 
($\omegaB = 0.71\mu$).  For reference, we also plot the minimum ($\omegaB = 10^{-1}\mu$) and maximum ($\omegaB = 0.95\mu$) frequency breathers used as initial 
conditions for our simulations.}
  \label{fig:breather-profiles}
\end{figure}
}

\newcommand{\deltapotential}{%
\begin{figure}
\includegraphics[width=246pt]{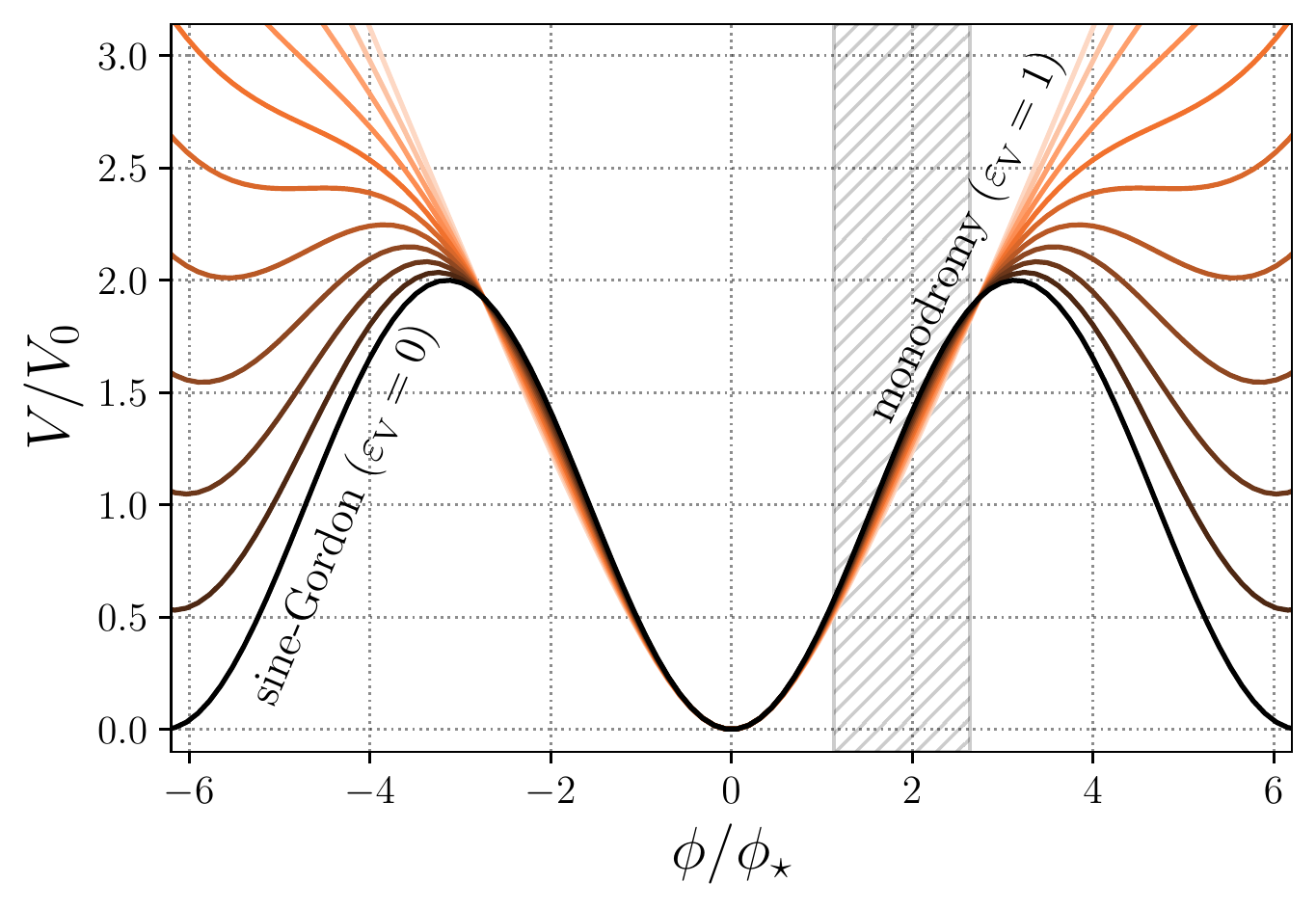}
\caption{An illustration of the potentials~\eqref{eqn:pot-deformed} for $\varepsilon_{\rm V} \in [0,1]$ deformed in incremental steps, and plotted with respect to the energy density 
scale $V_0=\mu^2\phiScl^2$.  For reference, the monodromy ($\varepsilon_{\rm V} = 1$) and sine-Gordon ($\varepsilon_{\rm V} = 0$) potentials are shown as a solid salmon and 
a solid black line, respectively.  The peak amplitude of the maximum ($\omegaini =0.96\mu$) and minimum ($\omegaini = 0.79\mu$) frequency breather profiles used in our 
parameter scans span the area hatched with gray lines. Within this region, the monodromy and the sine-Gordon potentials are nearly the same.}
\label{fig:mono-potential}
\end{figure}
}